\documentclass[twocolumn,aps,prd,showpacs,nofootinbib,floatfix,eqsecnum,10pt]{revtex4-1}  
\usepackage{amsmath,latexsym,multirow,longtable}
\usepackage{graphicx}

\usepackage{amssymb}

\begin{document}

\title{Gyroscopes orbiting black holes: A frequency-domain approach to
  precession and spin-curvature coupling for spinning bodies on
  generic Kerr orbits}

\author{Uchupol Ruangsri}
\affiliation{Department of Physics and MIT Kavli Institute, MIT,
  Cambridge, MA 02139, USA}

\author{Sarah J.\ Vigeland}
\affiliation{Jet Propulsion Laboratory, California Institute of
  Technology, Pasadena, CA 91109, USA}
\affiliation{Center for Gravitation, Cosmology and Astrophysics,
  University of Wisconsin Milwaukee, Milwaukee WI, 53211}

\author{Scott A.\ Hughes}
\affiliation{$\!\,$Department of Physics and MIT Kavli Institute, MIT,
  Cambridge, MA 02139, USA}

\date{\today}
\begin{abstract}
A small body orbiting a black hole follows a trajectory that, at
leading order, is a geodesic of the black hole spacetime.  Much effort
has gone into computing ``self force'' corrections to this motion,
arising from the small body's own contributions to the system's
spacetime.  Another correction to the motion arises from coupling of
the small body's spin to the black hole's spacetime curvature.
Spin-curvature coupling drives a precession of the small body, and
introduces a ``force'' (relative to the geodesic) which shifts the
small body's worldline.  These effects scale with the small body's
spin at leading order.  In this paper, we show that the equations
which govern spin-curvature coupling can be analyzed with a
frequency-domain decomposition, at least to leading order in the small
body's spin.  We show how to compute the frequency of precession along
generic orbits, and how to describe the small body's precession and
motion in the frequency domain.  We illustrate this approach with a
number of examples.  This approach is likely to be useful for
understanding spin coupling effects in the extreme mass ratio limit,
and may provide insight into modeling spin effects in the strong field
for non-extreme mass ratios.
\end{abstract}

\pacs{04.25.-g, 04.25.Nx, 04.30.Db}

\maketitle

\section{Introduction}
\label{sec:intro}

\subsection{The motion of spinning bodies in curved spacetime}
\label{sec:motivation}

Perturbative techniques have proven to be very fruitful for
understanding the two-body problem of general relativity.  Typically
one uses the mass ratio of the system as the small parameter defining
the perturbation.  An example of such a system is an extreme mass
ratio inspiral, or ``EMRI,'' a small compact object (mass $\mu \sim
1-100\,M_\odot$) whose orbit about a massive black hole (mass $M \sim
10^6\,M_\odot$) evolves (at lowest order) due to
gravitational-wave-driven backreaction.  Such systems are regarded as
important sources of gravitational waves for low-frequency,
space-based gravitational-wave detectors such as eLISA
{\cite{elisa2013}}.  Modeling EMRI systems has been a major motivator
for much of the progress this field has seen.  More fundamentally,
such a system represents a clean limit of the general two-body
problem.  Many quantities which describe the evolution of binaries can
be solved in this limit quite precisely.  Although formally applicable
only to systems with large mass ratios, perturbation-theory-based
analyses have proven to be useful for understanding binary systems
even for mass ratios of order unity {\cite{letiec2011, abds2012,
    letiec2013}}.

At zeroth order in the mass ratio, the small body moves on a geodesic
of the background spacetime:
\begin{equation}
	\frac{Dp^\alpha}{d\tau} = 0 \;.
\label{eq:geodesic}
\end{equation}
The operator $D/d\tau$ is a covariant derivative evaluated along the
small body's trajectory, with $\tau$ proper time along that
trajectory.  Corrections to geodesic motion enter at higher order in
the mass ratio.  Schematically, we write
\begin{equation}
	\frac{Dp^\alpha}{d\tau} = f^\alpha \;.
\end{equation}
The force $f^\alpha$ that pushes the smaller body away from a geodesic
can arise from a number of physical effects.  A particularly important
and interesting example is the gravitational self force,
$f^\alpha_\mathrm{sf}$.  It arises from the small body's own
contribution to the binary's spacetime, and enters at order
$(\mu/M)^2$.  This force encodes the loss of energy and angular
momentum from the binary due to gravitational waves, as well as
smaller scale but observationally important shifts to observables like
orbital frequencies.  Many groups have put considerable effort into
computing the gravitational self-force; see Refs.~\cite{ppv2011} and
{\cite{pound2015}} for a very detailed overview and discussion of the
self-force problem in general relativity, and Refs.\ {\cite{dvwd2012,
    wabgs2012, pmb2014, dwhlwb2014, ibdlnstw2014, albsw2015,
    warburton2015, vdms2015}} for some examples of recent progress in
the field.

Other forces arise when we consider that the small body is not a point
mass, but has internal structure.  The leading example of such
structure is spin.  Spin couples to the curvature of the background
spacetime, modifying the smaller body's motion
{\cite{mathisson1,mathisson2,papa51,dixon70}}:
\begin{eqnarray}
\frac{Dp^\alpha}{d\tau} &=& -\frac{1}{2} {R^\alpha}_{\nu\lambda\sigma}
u^\nu S^{\lambda\sigma}\,,
\label{eq:papapetrou1}
\\
\frac{DS^{\alpha\beta}}{d\tau} &=& p^\alpha u^\beta - p^\beta u^\alpha
\;.
\label{eq:papapetrou2}
\end{eqnarray}
These are the Mathisson-Papapetrou equations, commonly called the
Papapetrou equations.  Here, ${R^\alpha}_{\mu\lambda\sigma}$ is the
Riemann curvature of the background spacetime (which we'll take to be
a Kerr black hole), $u^\alpha = dx^\alpha/d\tau$ is the orbiting
body's four-velocity (tangent to its orbit worldline), and $p^\alpha$
is its four-momentum.  The tensor $S^{\lambda\sigma}$ describes the
spin of the small body; we define it precisely in
Sec.\ {\ref{sec:spincurv}}.  If the small body is itself a Kerr black
hole, then $S^{\lambda\sigma} \propto s\mu^2$, where $s \le 1$ is the
small body's dimensionless spin parameter, and where the small body's
mass $\mu$ will be defined precisely in
Sec.\ {\ref{sec:spincurvgeneral}}.  In this case, typical components
of the spin will have magnitude $S \equiv s\mu^2 \le \mu^2$, so the
leading effect of the spin-curvature force will be of order $\mu^2$.

A major motivation for our analysis is that non-geodesic effects
associated with the self force and with spin-curvature coupling both
lead to forces that scale at leading order as $\mu^2$ (at least if the
small body is considered to be a Kerr black hole, in the case of
spin-curvature coupling).  This very crudely suggests that the effects
of these two forces might be comparable.  However, it should be
emphasized that this scaling argument can only motivate a more careful
analysis which ultimately examines gauge invariant, observable
quantities.  It is only through such an analysis that one can
ascertain whether these forces drive secularly accumulating phase
shifts (for example), or produce periodic signatures that may be
searched for in astrophysical data.  The work in this paper is the
first step in a program to develop such an analysis, laying the
foundations for an examination of such effects for generic orbits of
spinning bodies with arbitrary spin-orbit orientations.

Equations (\ref{eq:papapetrou1}) and (\ref{eq:papapetrou2}) must be
supplemented by an additional constraint, reflecting an internal
degree of freedom associated with the small body's structure.  One can
reference an extended body's motion to an infinite number of
worldlines which pass through it.  The additional constraint which
must be imposed can be understood as specifying which of these
worldlines one uses.  This constraint is called the {\it
  spin-supplementary condition}, or SSC, in the case of a body
described by its mass and spin.

Figure 1 of Ref.\ {\cite{costanatario2015}} provides a very clear and
simple illustration of why the SSC is needed.  In non-relativistic
dynamics, a natural choice for the worldline that one uses to describe
a body's motion is the one which passes through its ``center of
mass,'' or COM.  However, in relativistic dynamics, the COM is
observer dependent.  An observer who is at rest with respect to the
axis of rotation of a uniform spinning sphere will place the COM at
the sphere's geometric center.  An observer moving relative to the
rotation axis will not: thanks to the relativistic transformation of
mass currents, and thanks also to the fact that different mass
elements of the sphere have different relative velocities to this
observer, the moving observer places the spinning sphere's COM away
from the rotation axis.

In this language, the SSC can be regarded as choosing a particular
observer and tying the body's worldline to their choice of COM.  Costa
and Nat\'ario {\cite{costanatario2015}} provide excellent discussion
and many further details elaborating on this point.  An SSC which is
commonly used in the literature is the Tulczyjew SSC
{\cite{tulczyjew}},
\begin{equation}
p_\alpha S^{\alpha\beta} = 0\;.
\label{eq:spinsupplementary1}
\end{equation}
Several variations on Eq.\ (\ref{eq:spinsupplementary1}) have been
formulated, and are compared by Kyrian and Semer\'ak in
Ref.\ {\cite{semerakII}}.  Many SSCs\footnote{Though not all SSCs: as
  emphasized by Costa and Nat\'ario {\cite{costanatario2015}}, some
  SSCs differ even at linear order in $S$.  The motion one finds using
  these different SSCs is equivalent; see
  Ref.\ {\cite{costanatario2015}} for detailed discussion.  We are
  grateful to Costa and Nat\'ario for correspondence which alerted us
  to this point.} differ from one another at $O(S^2)$; if the smaller
body is itself a Kerr black hole, this difference is $O(\mu^4)$.
These differences will be extremely small for large mass ratio
systems, suggesting that it would be useful to linearize our system in
spin.

Related to the internal degree of freedom is the fact that the
four-momentum and four-velocity are in general not parallel to one
another when the small body's spin is taken into account.  Instead, we
have
\begin{equation}
p^\alpha = \mu u^\alpha - u_\gamma\frac{DS^{\alpha\gamma}}{d\tau}\;;
\end{equation}
the momentum and velocity are only parallel if the second term
vanishes (which in fact it does for some SSCs; see Sec.\ 3.4 of
Ref.\ {\cite{semerakII}} for an example).  For several cases
(including the Tulczyjew SSC, which we will use),
\begin{equation}
\frac{DS^{\alpha\beta}}{d\tau} = O(S^2)\;.
\label{eq:papapetrou2smalls}
\end{equation}
In such a case, $p^\alpha$ and $u^\alpha$ are parallel at order $S$.

Given an initial position, velocity, and spin configuration,
Eqs.\ (\ref{eq:papapetrou1}), (\ref{eq:papapetrou2}), and
(\ref{eq:spinsupplementary1}) can be integrated to describe the motion
of the small body as it orbits the larger black hole.  When $S = 0$,
Eq.\ (\ref{eq:papapetrou1}) reduces to the geodesic equation
(\ref{eq:geodesic}), and Eqs.\ (\ref{eq:papapetrou2}) and
(\ref{eq:spinsupplementary1}) can be ignored.  The trajectories they
describe reduce to Kerr geodesic orbits.  For $S$ small, these
equations describe a nearly geodesic orbit, modified by a
spin-curvature force of order $S = s \mu^2$.  As already mentioned,
the scaling of the spin-curvature force with mass is naively similar
to the scaling for the gravitational self force, motivating a more
careful analysis to ascertain whether its effects are of observational
relevance.  Examples of such analyses have been done for special
cases.  Burko {\cite{burko2004}} and Burko and Khanna
{\cite{burkokhanna2015}} find that the spin-curvature force may in
fact dominate over certain aspects of the self force for circular and
equatorial orbits, at least when the small body's spin is large and
aligned with the orbit; similar results were found by Steinhoff and
Puetzfeld {\cite{sp2012}}.

\subsection{Past work}
\label{sec:past}

Because all astrophysical objects spin, it has long been known that it
is important to include the impact of spin when modeling the evolution
of and gravitational-wave emission from compact binaries.  This has
long been done in post-Newtonian studies of comparable mass coalescing
compact objects (see Refs.\ {\cite{bo79, kww93, mbfb2013, bmfb2013,
    bmb2013, blanchet2014, bfh2013, mbfb2013, mbbf2013, prr2011,
    kgobs2015, gksbo2015}} and references therein for some examples of
historical and current work on this problem), and has been a major
focus of work in numerical relativity (see
Refs.\ {\cite{marronetti2008, hannam2010, lousto2010, hemberger2013,
    boyle2014, ossokine2015, loustohealy2015_1, loustohealy2015_2}}
for some examples of work on this problem in the past half decade).
Active work is now ongoing to expand the effective one-body framework
{\cite{bd99, damour2014}} to include the effect of spin
{\cite{djs2008_1, djs2008_2, brb2009, baraussebuonanno2010,
    baraussebuonanno2011, taracchini2012, pan2014, damournagar2014,
    taracchini2014, djs2015}}.

Although there has been a great deal of work studying motion under the
influence of the Papapetrou equations (see Kyrian and Semar\'ak
{\cite{semerakII}} and references therein for an outstanding review),
the impact that the small body's spin has for potentially relevant
EMRI sources has not yet been carefully quantified.  It appears that,
to date, the most careful and rigorous analyses of spin-enhanced,
large-mass ratio dynamics have focused upon simple orbits and simple
spin-orbit configurations.  Most work on this problem has focused upon
circular orbits with spin parallel to the orbital angular momentum
{\cite{mst96, smsm98, han2010, sp2012, hlops2014}}; at least one paper
{\cite{tmss96}} considered the small body's spin to be slightly
misaligned from parallel, leading to precession kinematics which
influenced the system's emitted gravitational waves.  A very recent
analysis using a Hamiltonian formulation of spin-curvature coupling
has the promise of analyzing motions for general orbits and spin
orientations {\cite{dsvv2015}}, though at present the authors confine
their analysis to spherical background spacetimes and planar orbits.
With one exception, analyses which focus on issues of
gravitational-wave measurement and data analysis likewise use the
circular, equatorial, parallel spin limit
{\cite{burko2004,huertagair2011}}.  The one exception is Barack and
Cutler {\cite{barackcutler2004}}, who developed a ``kludge'' model for
the inspiral and waveforms by pushing the post-Newtonian expansion
beyond its formal range of validity in order to assess parameter
measurement accuracy with space-based gravtiational-wave detectors.
Barack and Cutler found that the small body's spin has a negligible
impact on parameter accuracy, except when the small body has maximal
spin.  It will be important to revisit this analysis using techniques
that accurately describe the small body's motion and precession deep
into the strong field.

\subsection{Our approach to spin-enhanced orbital motion}
\label{sec:thispaper}

A major goal of the work we present here is to develop tools that can
be used to efficiently compute orbits that include the impact of
spin-curvature forces for EMRI systems with arbitrary spin-orbit
configurations.  In this paper, we focus on computation of the
precession and forces for EMRIs; later work will examine how to evolve
orbits using the resulting force.  We assume $\mu \ll M$, where $M$ is
the mass of the large black hole, and $\mu$ is the mass of the
orbiting body.  We assume that the smaller body has a spin $S$, and
define the dimensionless spin parameter $s = S/\mu^2$.  If the small
body's spin obeys the Kerr bound, then $0 \le s \le 1$.

Our approach to modeling spin-enhanced orbital motion is based on the
fact that, for large mass ratio systems, the motion will be ``close
to'' a Kerr geodesic, with the spin-curvature force introducing
corrections to the orbiting body's worldline of order $\mu/M \ll 1$.
This suggests that a useful way to compute the force's effect is with
the method of {\it osculating geodesics}
{\cite{poundpoisson08,gfdhb11}}, in which the body is regarded as
evolving through a sequence of geodesic orbits.  This sequence of
geodesics shares the tangent to the body's true worldline at each
moment.  Osculating geodesics generalize the technique of osculating
orbits that are commonly used in astrophysical dynamics, and are a
powerful way to describe motion which is perturbed away from an exact,
integrable trajectory.

It should be emphasized at this point that the equations of motion for
a spinning body on a generic orbit about a Kerr black hole are not
integrable\footnote{We are very grateful to Georgios
  Loukes-Gerakopoulos for useful discussions on this point.}.  One can
in fact find configurations that lead to chaotic orbital evolution
{\cite{suzukimaeda96, hartl03, levin06}}.  Our proposed method of
treating the motion as a sequence of osculating geodesics presumes
that the motion is ``nearly'' integrable, in the sense that the
non-integrable spin-curvature coupling acts as a weak perturbation to
the integrable geodesic motion.  This is presumably reasonable
provided that the small body's spin is ``small enough,'' though it
remains to be quantified what ``small enough'' really means.

In our formulation of this problem, we use the fact that functions
arising from motion on a Kerr orbit can be expanded in a Fourier
series {\cite{dh04}}.  A Kerr black hole orbit is triperiodic, with
separate frequencies describing its radial, polar, and axial motions.
Many functions which arise from motion along such an orbit, including
those governing spin-curvature coupling, can be written
\begin{equation}
f({\bf z}) = \sum_{k,n = -\infty}^\infty f_{kn} e^{-i(k\Omega_\theta +
  n\Omega_r)t}\;,
\end{equation}
where ${\bf z}$ is the worldline followed by the orbiting body, $t$ is
time as measured by a distant observer, and $\Omega_{\theta,r}$ are
frequencies characterizing the orbit's motion in the $\theta$ and $r$
directions.  (A third frequency associated with axial motions,
$\Omega_\phi$, exists, but is not important in this analysis thanks to
the axisymmetry of the Kerr spacetime.)  Similar forms can be written
down using different time parameterizations.  Including the small
body's spin introduces a new frequency $\Omega_s$ which is associated
with the small body's precession as it moves along its orbit.  These
frequencies all vary smoothly with the parameters that describe bound
Kerr orbits, and so evolve in a simple way as the worldline moves
through its sequence of osculating geodesics.

We find that frequency-domain expansions of the quantities needed to
model spin-enhanced motion tend to converge rapidly, at least for
orbital eccentricities $e \lesssim 0.7$ or so\footnote{Based on
  experience studying gravitational waves from eccentric binary
  systems (e.g., Ref.\ {\cite{dh06}}), we expect that the
  frequency-domain expansion will converge very slowly as $e \to 1$.
  We have not, however, done a systematic study to determine the value
  of $e$ at which the frequency-domain approach becomes less useful.}.
This approach should thus facilitate efficient calculation of the
quantities needed to compute spin-enhanced orbits.  Describing these
quantities in the frequency domain can be useful for understanding how
harmonic features of the orbital and precessional motion are imprinted
on observationally important characteristics of the system.  We expect
that a frequency-domain description will prove useful for extending
codes which compute radiation from generic Kerr orbits (e.g.,
{\cite{dh06}}) to include the effects of spin.  Work of this nature
has been done so far for simpler orbit and spin configurations
{\cite{mst96,smsm98,tmss96,han2010,harmsetal2015}}.

\subsection{Organization of this paper}

In this paper, we show how to compute the new frequency associated
with the small body's spin and examine the Fourier decomposition of
all the relevant quantities needed to compute spin-enhanced EMRI
orbits.  Since our analysis is based on expanding about geodesics of
Kerr black holes, we begin in Sec.\ {\ref{sec:background}} by
reviewing the properties of these geodesics.  We introduce the
notation that we use, and describe briefly the parameterization and
properties of Kerr black hole orbits.  Chief among these properties is
the periodic structure of these orbits.  We review the triperiodic
structure of Kerr black hole orbits, and describe how any function
which arises from motion along such an orbit can be expanded in a
Fourier series of the orbit's fundamental frequencies.  In
Sec.\ {\ref{sec:spincurv}}, we move beyond geodesics, summarizing how
the small body's spin couples to spacetime curvature.  We first review
the general equations that govern spin-curvature coupling in
Sec.\ {\ref{sec:spincurvgeneral}}, and then examine in
Sec.\ {\ref{sec:spincurvleading}} how these equations simplify when we
linearize in the small body's spin.

In Sec.\ {\ref{sec:prec_FD}}, we then study the precession of the
small body's spin in the frequency domain.  We do this analysis in
what is often called ``Mino time,'' a time variable that fully
separates the radial and polar motions of Kerr black hole orbits, and
that has proven to be well suited for studies of strong-field motions.
We begin by studying precession along a circular equatorial orbit
(Sec.\ {\ref{sec:preccirceq}}).  This limit turns out to be very clean
when analyzed using Mino time.  Changing to the Boyer-Lindquist time
coordinate (appropriate for analyzing quantities that distant
observers would measure) and taking the weak-field limit, we recover
well-known results describing gyroscope precession in general
relativity.  We then generalize to more complicated orbit geometries
in Secs.\ {\ref{sec:precessonefreq}} and {\ref{sec:precessgeneric}}.
In all these cases, we find that the frequency characterizing
precession is the eigenvalue of a matrix representation of the
relevant equations.  This representation can become fairly complicated
for generic orbits, thanks to the complicated time-frequency structure
of the underlying orbits.  Nonetheless, we find it is not terribly
difficult in this framework to accurately construct the solution for
the small body's precession.

With the small body's precession in hand, we then examine the
spin-curvature force (\ref{eq:papapetrou1}) in
Sec.\ {\ref{sec:force}}.  We find that the frequency-domain expansion
converges quickly, and is an excellent way to describe these
quantities in the large mass ratio regime.  Interestingly, there exist
certain configurations in which the precession and orbit frequencies
are ``commensurate'' with one another: some linear combination of the
precession and orbit frequencies sums to zero.  Fourier components of
the force corresponding to such combinations would be constant in
time, and one might imagine that the system's evolution would be
substantially modified as compared to ``nearby'' orbit configurations
(as has been seen in studies of the self force's behavior at black
hole orbit resonances {\cite{fh12,fhr14}}).  In Sec.\ {\ref{sec:res}},
we show that, at least at linear order in spin, resonant
spin-curvature coupling does not have any special impact on the
system's evolution.  This is due to the existence of conserved
constants of the motion.  To ``protect'' these constants, certain
contributions to the spin-curvature force vanish when the precession
and orbit frequencies are commensurate.  We speculate that such
resonances (or an analog of such resonances) might have more
interesting effects if the analysis were taken to higher order in the
small body's spin, and might be connected to the onset of chaos that
has been seen in past analyses of the orbits of spinning bodies
{\cite{suzukimaeda96, hartl03, levin06}}.

We conclude in Sec.\ {\ref{sec:conclusion}} with a sketch of our plans
for future work on this problem.  Chief among these plans is to marry
this scheme for computing the motion of a small body to an osculating
geodesic integrator {\cite{poundpoisson08,gfdhb11}} in order to
compute the full worldline describing a small spinning body orbiting a
Kerr black hole.  This analysis is underway, and early results
indicate that it is a robust and accurate way to describe these
orbits.  We then hope to use it to broaden the class of sources that
can be incorporated into future EMRI models, and to model two-body
dynamics in general relativity more broadly.

\section{Background: Conventions, notation, and geodesic motion}
\label{sec:background}

We begin by laying out the conventions and notation that we will use,
and describe in some detail the Kerr geodesics that constitute the
background motion we use for orbits of spinning bodies.  This material
has appeared in great depth in quite a few other sources, so we simply
summarize what is relevant for us here and point the reader to
relevant literature.

\subsection{Generalities}

Our goal is to understand the motion of a small body of mass $\mu$
orbiting a Kerr black hole with mass $M$ and spin parameter $a$.  We
will use Boyer-Lindquist coordinates for our analysis, in which the
metric is
\begin{eqnarray}
ds^2 &=& -\left(1 - \frac{2Mr}{\Sigma}\right)dt^2 -
\frac{4Mar\sin^2\theta}{\Sigma}dt\,d\phi + \frac{\Sigma}{\Delta}dr^2
\nonumber\\
&+& \!\! \Sigma\,d\theta^2 + \frac{(r^2 + a^2)^2 -
  a^2\Delta\sin^2\theta}{\Sigma}\sin^2\theta\,d\phi^2\;,
\label{eq:Kerr_BL}
\end{eqnarray}
where
\begin{equation}
\Delta = r^2 - 2Mr + a^2\;,\qquad
\Sigma = r^2 + a^2\cos^2\theta\;.
\label{eq:DeltaSigma}
\end{equation}
(We use geometric units with $G = 1 = c$ throughout this paper.)  From
Eqs.\ (\ref{eq:Kerr_BL}) and (\ref{eq:DeltaSigma}), it is
straightforward to compute the Kerr metric's connection coefficients
and Riemann curvature components in these coordinates, both of which
we need in explicit form.  Semer\'ak provides a catalog of these
quantities (Ref.\ {\cite{semerakI}}, Appendix A; see also footnote 2
of Ref.\ {\cite{semerakII}}).

The Kerr spacetime admits two Killing vectors, $\xi^\alpha_t$ and
$\xi^\alpha_{\phi}$, corresponding to timelike and axial
diffeomorphisms.  It also admits a Killing tensor $K_{\mu\nu}$, given
by
\begin{equation}
K_{\mu\nu} = \Sigma\left(m_\mu\bar m_\nu + \bar m_\mu m_\nu\right)
-a^2\cos^2\theta g_{\mu\nu}\;,
\end{equation}
where
\begin{eqnarray}
m_\mu &\doteq& \left[m_t, m_r, m_\theta, m_\phi\right]
\nonumber\\
&\doteq&
\frac{1}{\sqrt{2}(r + ia\cos\theta)}\left[-ia\cos\theta, 0, \Sigma,
  i(r^2 + a^2)\sin\theta\right]
\nonumber\\
\end{eqnarray}
is a Newman-Penrose null tetrad leg\footnote{Here and below, we use
  ``$\doteq$'' to mean ``the indexed object on the left-hand side is
  represented by the components on the right-hand side in
  Boyer-Lindquist coordinates.''}, and ${\bar m}_\mu$ is the complex
conjugate of $m_\mu$.  This tensor satisfies a generalized form of
Killing's equation,
\begin{equation}
\nabla_\alpha K_{\beta\gamma} + \nabla_\beta K_{\gamma\alpha} +
\nabla_\gamma K_{\alpha\beta} = 0\;.
\end{equation}
To our knowledge, $K_{\alpha\beta}$ does not have a simple geometrical
interpretation as the Killing vectors do.  Note that $K_{\alpha\beta}$
is symmetric under exchange of indices.  For later discussion, it is
useful to note that we can write
\begin{equation}
K_{\alpha\beta} = f_{\alpha\gamma}{f_\beta}^\gamma\;,
\end{equation}
where the anti-symmetric Killing-Yano tensor $f_{\alpha\beta}$
is given by {\cite{tmss96}}
\begin{equation}
f_{\mu\nu} = a\cos\theta\left(e^1_\mu e^0_\nu - e^0_\mu e^1_\nu\right)
+ r\left(e^2_\mu e^3_\nu - e^3_\mu e^2_\nu\right)\;,
\label{eq:KYdef}
\end{equation}
and where
\begin{eqnarray}
e^0_\mu &\doteq& \left[\sqrt{\frac{\Delta}{\Sigma}}, 0, 0, -a\sin^2\theta
\sqrt{\frac{\Delta}{\Sigma}}\right]\;,
\\
e^1_\mu &\doteq& \left[0, \sqrt{\frac{\Sigma}{\Delta}}, 0, 0\right]\;,
\\
e^2_\mu &\doteq& \left[0, 0, \sqrt{\Sigma}, 0\right]\;,
\\
e^3_\mu &\doteq& \left[-\frac{a\sin\theta}{\sqrt{\Sigma}}, 0, 0,
\frac{(r^2 + a^2)\sin\theta}{\sqrt{\Sigma}}\right]\;.
\end{eqnarray}
The Killing-Yano tensor satisfies
\begin{equation}
\nabla_\gamma f_{\alpha\beta} + \nabla_\beta f_{\alpha\gamma} = 0\;.
\end{equation}

Associated with the Killing vectors and the Killing tensor are three
constants which are conserved along a geodesic worldline.  These are
the orbit's energy, axial angular momentum, and ``Carter constant'':
\begin{eqnarray}
E^{\rm G} &\equiv& -\xi^\mu_t p^{\rm G}_\mu = -p^{\rm G}_t\;,
\label{eq:EGdef}\\
L^{\rm G}_z &\equiv& \xi^\mu_\phi p^{\rm G}_\mu = p^{\rm G}_\phi\;,
\label{eq:LzGdef}\\
K^{\rm G} &\equiv& K_{\mu\nu} p_{\rm G}^\mu p_{\rm G}^\nu\;,
\label{eq:KGdef}\\
Q^{\rm G} &\equiv& K^{\rm G} - (L_z^{\rm G} - a E^{\rm G})^2
\nonumber\\
&=& (p^{\rm G}_\theta)^2 - a^2\cos^2\theta \left[(E^{\rm G})^2 -
  \mu^2\right] + \cot^2\theta (L^{\rm G}_z)^2\;.
\nonumber\\
\label{eq:QGdef}
\end{eqnarray}
The name ``Carter constant'' is commonly used for both $K^{\rm G}$ and
$Q^{\rm G}$.  In all of these quantities, the superscript ``G''
indicates that these are quantities associated with motion along a
geodesic.  This is to be contrasted with analogs we will introduce in
the next section which are associated with a more general class of
trajectories.  In much of our analysis, it is useful to normalize
these quantities to the rest mass of the orbiting body $\mu$.  We
indicate their values so normalized with a hat:
\begin{eqnarray}
\hat E^{\rm G} &=& E^{\rm G}/\mu\;,
\nonumber\\
\hat L_z^{\rm G} &=& L_z^{\rm G}/\mu\;,
\nonumber\\
\hat K^{\rm G} &=& K^{\rm G}/\mu^2\;,
\nonumber\\
\hat Q^{\rm G} &=& Q^{\rm G}/\mu^2\;.
\end{eqnarray}

The existence of these constants of motion makes it possible to
separate the equations governing Kerr geodesics.  In first order form,
they become [Ref.\ {\cite{mtw}}, Eqs.\ (33.32a)--(33.32d)]
\begin{eqnarray}
\Sigma^2\left(\frac{dr}{d\tau}\right)^2 &=& \left[\hat E^{\rm
    G}(r^2+a^2) - a \hat L^{\rm G}_z\right]^2
\nonumber\\
& & - \Delta\left[r^2 + (\hat L^{\rm G}_z - a \hat E^{\rm G})^2 + \hat
  Q^{\rm G}\right]
\nonumber\\
&\equiv& R(r)\;,\label{eq:rdot}\\
\Sigma^2\left(\frac{d\theta}{d\tau}\right)^2 &=& \hat Q^{\rm G} -
\cot^2\theta (\hat L^{\rm G}_z)^2 -a^2\cos^2\theta[1 - (\hat E^{\rm
    G})^2]\nonumber\\
&\equiv&\Theta(\theta)\;,\label{eq:thetadot}\\
\Sigma\left(\frac{d\phi}{d\tau}\right) &=& \csc^2\theta \hat L^{\rm
  G}_z + a \hat E^{\rm G}\left(\frac{r^2+a^2}{\Delta} - 1\right)
-\frac{a^2\hat L^{\rm G}_z}{\Delta}\nonumber\\
&\equiv&\Phi(r,\theta)\;,\label{eq:phidot}\\
\Sigma\left(\frac{dt}{d\tau}\right) &=&
\hat E^{\rm G}\left[\frac{(r^2+a^2)^2}{\Delta} - a^2\sin^2\theta\right]
\nonumber\\
& & + a\hat L^{\rm G}_z\left(1 - \frac{r^2+a^2}{\Delta}\right)
\nonumber\\
&\equiv& T(r,\theta)\;.\label{eq:tdot}
\end{eqnarray}
These equations use proper time $\tau$ as the independent parameter
along the geodesic.  Another time parameter which is very useful for
studying strong-field Kerr black hole orbits is $\lambda$, defined by
$d\lambda = d\tau/\Sigma$.  The geodesic equations parameterized in
this way are
\begin{eqnarray}
\left(\frac{dr}{d\lambda}\right)^2 = R(r)\;,
&\qquad&
\left(\frac{d\theta}{d\lambda}\right)^2 = \Theta(\theta)\;,
\label{eq:rdotthdotMino}\\
\frac{d\phi}{d\lambda} = \Phi(r,\theta)\;,
&\qquad&
\frac{dt}{d\lambda} = T(r,\theta)\;.
\label{eq:phidottdotMino}
\end{eqnarray}
By using $\lambda$ as our time parameter, the $r$ and $\theta$
coordinate motions completely separate.  The parameter $\lambda$ is
often called ``Mino time,'' following Mino's use of it to untangle
these coordinate motions {\cite{mino03}}.

\subsection{Parameterizing geodesics}

We have found it very useful to introduce the following
reparameterization of $r$ and $\theta$:
\begin{equation}
r = \frac{pM}{1 + e\cos(\psi + \psi_0)}\;,\qquad
\cos\theta = \cos\theta_{\rm m}\cos(\chi + \chi_0)\;.
\label{eq:rdefthdef}
\end{equation}
We further find it helpful to remap $\theta_{\rm m}$ to $\theta_{\rm
  inc}$, defined by
\begin{equation}
\theta_{\rm inc} = \pi/2 - {\rm sgn}(L_z)\theta_{\rm m}\;.
\label{eq:thincdef}
\end{equation}
This angle encodes whether the orbit is prograde ($L_z > 0$) or
retrograde ($L_z < 0$).  It smoothly varies from $\theta_{\rm inc} =
0^\circ$ for prograde equatorial ($\theta_{\rm m} = 90^\circ$, $L_z >
0$) to $\theta_{\rm inc} = 180^\circ$ for retrograde equatorial
($\theta_{\rm m} = 90^\circ$, $L_z < 0$).

There is a fairly simple mapping between the constants of motion
$(\hat E^{\rm G}, \hat L^{\rm G}_z, \hat Q^{\rm G})$ and the orbital
geometry parameters $(p,e,\theta_{\rm m})$.  See Appendix B of
Ref.\ {\cite{schmidt02}} for explicit formulas relating these two
parameterizations; Ref.\ {\cite{fujitahikida09}} also provides many
valuable results for studies of Kerr geodesic motion.  Notice that $r$
oscillates between $r_{\rm min}$ and $r_{\rm max}$, given by
\begin{equation}
r_{\rm min} = \frac{pM}{1 + e}\;,\qquad
r_{\rm max} = \frac{pM}{1 - e}\;.
\end{equation}
Likewise, $\theta$ oscillates from $\theta_{\rm min} = \theta_{\rm m}$
to $\theta_{\rm max} = \pi - \theta_{\rm m}$.

The transformations (\ref{eq:rdefthdef}) replace the variables $r$ and
$\theta$ with secularly accumulating angles $\psi$ and $\chi$.  As
$\psi$ and $\chi$ evolve from 0 to $2\pi$, $r$ and $\theta$ move
through their full ranges of motion.  By combining
Eq.\ (\ref{eq:rdefthdef}) with Eq.\ (\ref{eq:rdotthdotMino}), it is
straightforward to develop equations for $d\psi/d\lambda$, and
$d\chi/d\lambda$.  In this parameterization, a geodesic worldline is
defined by a set of four functions [$\psi(\lambda)$, $\chi(\lambda)$,
  $\phi(\lambda)$, $t(\lambda)$], three orbital parameters
($p,e,\theta_{\rm m}$), and four initial conditions.  We define $\chi
= \psi = 0$ at $\lambda = 0$; the value of $r$ and $\theta$ at
$\lambda = 0$ is then set by the angles $\psi_0$ and $\chi_0$ defined
in Eq.\ (\ref{eq:rdefthdef}).  We likewise define $t = t_0$ and $\phi
= \phi_0$ at $\lambda = 0$.

Let us define $\lambda_{r0} = \lambda_{r0}(\psi_0)$ to be the value of
$\lambda$ closest to zero\footnote{Due to periodicity, $\lambda =
  \lambda_{r0} \pm 2\pi j\Lambda_r$ will also satisfy this condition
  for any integer $j$ (where $\Lambda_r$ is the Mino-time radial
  period).  A similar statement holds for $\lambda_{\theta0}$.}  for
which $r(\lambda_{r0}) = r_{\rm min}$; likewise, $\lambda_{\theta0} =
\lambda_{\theta0}(\chi_0)$ is the value of $\lambda$ closest to zero
for which $\theta(\lambda_{\theta0}) = \theta_{\rm min}$.  We define
the ``fiducial geodesic'' as the geodesic for which $\lambda_{r0} =
\lambda_{\theta0} = \phi_0 = t_0 = 0$.  It is easy to show that if
$\psi_0 = 0$, then $\lambda_{r0} = 0$; likewise, if $\chi_0 = 0$, then
$\lambda_{\theta0} = 0$.

\subsection{Geodesics and functions of geodesics in the frequency
domain}

Bound Kerr orbits are triperiodic, as was first explicitly shown by
Schmidt {\cite{schmidt02}}.  It is simplest to compute the three
orbital periods in Mino time, which (as discussed above) separates the
radial and polar motions.  Using this time variable, it is not too
difficult to compute three periods $\Lambda_r$, $\Lambda_\theta$, and
$\Lambda_\phi$ associated with the radial, polar, and axial motions,
respectively, as well as associated frequencies
$\Upsilon_{r,\theta,\phi} = 2\pi/\Lambda_{r,\theta,\phi}$.  One can
also compute a quantity $\Gamma$ which converts these periods and
frequencies to their Boyer-Lindquist coordinate time analogues:
\begin{equation}
T_{r,\theta,\phi} = \Gamma\Lambda_{r,\theta,\phi}
\end{equation}
are the periods in Boyer-Lindquist time, and the associated frequencies
are
\begin{equation}
\Omega_{r,\theta,\phi} = \frac{\Upsilon_{r,\theta,\phi}}{\Gamma}\;.
\end{equation}
Since Boyer-Lindquist time corresponds to time as measured by distant
observers, this formulation is useful for describing the
time-frequency behavior of observable quantities.  Schmidt
{\cite{schmidt02}} first showed how to compute the frequencies
$\Omega_{r,\theta,\phi}$ using elegant but somewhat abstract
techniques.  Drasco and Hughes {\cite{dh04}} showed how to understand
these periodicities in simpler terms, with easy-to-compute quadratures
describing $\Lambda_{r,\theta,\phi}$ and $\Gamma$ (from which
computing $\Omega_{r,\theta,\phi}$ is straightforward).  Fujita and
Hikida {\cite{fujitahikida09}} then showed that these quadratures can
be evaluated analytically, yielding easy-to-use formulas for
$\Upsilon_{r,\theta,\phi}$ and $\Omega_{r,\theta,\phi}$ as functions
of $p$, $e$, and $\theta_{\rm m}$.

These frequency-domain expansions are excellent tools for
characterizing the behavior of functions associated with geodesic
black hole orbits.  The Mino-time periodic expansion is particularly
good for describing strong-field orbital dynamics, and will be used
extensively\footnote{Although not needed here, it's worth noting that
  converting from this form to the Boyer-Lindquist expansion is simple
  {\cite{dh04}}.} in this paper.  Let $f(\lambda) \equiv f[r(\lambda),
  \theta(\lambda)]$ be a function of $r$ and $\theta$ that is computed
along a geodesic worldline ${\bf z} = \left[t(\lambda), r(\lambda),
  \theta(\lambda), \phi(\lambda)\right]$.  Then,
\begin{equation}
f = \sum_{k,n = -\infty}^\infty f_{kn} e^{-i(k\Upsilon_\theta +
  n\Upsilon_r)\lambda}\;,
\label{eq:genfourier}
\end{equation}
where
\begin{eqnarray}
f_{kn} &=& \frac{4\pi^2}{\Lambda_r\Lambda_\theta}\times \nonumber\\
& & \int_0^{\Lambda_r} \int_0^{\Lambda_\theta}
f[r(\lambda_r),\theta(\lambda_\theta)] e^{i
  k\Upsilon_\theta\lambda_\theta}e^{i
  n\Upsilon_r\lambda_r}d\lambda_\theta\,d\lambda_r\;.
\nonumber\\
\label{eq:genfourier_amp}
\end{eqnarray}
A more complicated expansion must be used for functions which depend
in addition on $t$ and $\phi$; see Ref.\ {\cite{dh04}} for discussion.
We will not need this more complicated form for this paper.

Let us denote by $\tilde f_{kn}$ the amplitude
(\ref{eq:genfourier_amp}) computed along the fiducial geodesic,
$\chi_0 = \psi_0 = 0$.  Once $\tilde f_{kn}$ is known, it is simple to
compute the amplitude along an arbitrary geodesic {\cite{dfh05}}:
\begin{equation}
f_{kn} = e^{i\xi_{kn}(\psi_0,\chi_0)} \tilde f_{kn}\;,
\end{equation}
where
\begin{equation}
\xi_{kn}(\psi_0,\chi_0) = k\Upsilon_\theta\lambda_{\theta0}(\chi_0) +
n\Upsilon_r\lambda_{r0}(\psi_0)\;.
\end{equation}
This means that one only needs to compute $\tilde f_{kn}$; the Fourier
amplitudes along all other geodesics can be obtained from this quite
simply.

\section{Spin-curvature coupling}
\label{sec:spincurv}

Strictly speaking, geodesic motion only applies to a zero-size point
body moving through spacetime (and neglecting self force effects).  If
the body has any structure beyond the point particle description, that
structure will couple to the spacetime through which it moves.  This
coupling will push the small body away from the geodesic, appearing as
a force driving the body's motion.

Papapetrou {\cite{papa51}} pioneered a moment-based technique for
determining these couplings (see also Mathisson's discussion in
Ref.\ {\cite{mathisson2}}).  Following the discussion in
{\cite{wald72}}, the key idea is to choose some representative
worldline that passes through a body moving through spacetime.  One
computes moments of the body's stress-energy tensor $T^{\alpha\beta}$
about that worldline.  If the body is sufficiently compact that
moments beyond the $n$th can be taken to vanish, then one can find
equations of motion in terms of those non-vanishing moments by
enforcing $\nabla_\alpha T^{\alpha\beta} = 0$.  If all moments beyond
$n = 0$ vanish (meaning that the small body is a monopole point mass),
then the geodesic equations for motion in the background spacetime
emerge from this procedure.  The first moment to couple beyond this
describes the small body's spin angular momentum.  Spin couples to the
curvature of the background spacetime, producing a force which pushes
the small body away from the geodesic.  One can consider the coupling
of higher order moments as well {\cite{sp2010}}; for example, the
small body's quadrupole moment couples to the gradient of the
curvature {\cite{bfgo2008,bg2014}}.

Here we discuss the spin-curvature coupling and the force which
results from it.  Further detailed discussion and derivations can be
found in Refs.\ {\cite{mathisson1,mathisson2,papa51,dixon70,wald72}}.
We first summarize the equations which govern the motion of a small
spinning body through curved spacetime in full generality, making no
assumptions about the magnitude of the small body's spin or the
worldline that it follows.  We then introduce a perturbative
expansion, developing these equations at leading order in the small
body's spin, and treating its four-velocity as ``close to'' geodesic
motion (in a sense that is quantified precisely below).  A similar
approach was developed in Refs.\ {\cite{bgj2011,bg2011}}, though
without the goal of then connecting this perturbative force to an
osculating geodesic integrator.

\subsection{General form}
\label{sec:spincurvgeneral}

We begin by defining the spin tensor, which is central to our
analysis.  Let $\delta z^\alpha = x^\alpha - z^\alpha$, where
$x^\alpha$ is a general point in spacetime, and $z^\alpha$ is position
along the small body's worldline.  The spin tensor is defined as
\begin{eqnarray}
S^{\alpha\beta} &=& 2 \int_\Sigma \delta z^{[\alpha}\: T^{\beta]\gamma}\,
d\Sigma_\gamma
\label{eq:spindef_covar}\\
&=& 2\int_t \delta z^{[\alpha}\:T^{\beta]t}\,\sqrt{-g}\,d^3x\;.
\label{eq:spindef_coord}
\end{eqnarray}
In these equations, $T^{\alpha\beta}$ is the stress-energy tensor of
the orbiting body, and square brackets around indices denote
antisymmetrization: $A^{[\alpha\beta]} = (A^{\alpha\beta} -
A^{\beta\alpha})/2$.  In the covariant form (\ref{eq:spindef_covar}),
the spin is defined on an arbitrary spacelike hypersurface $\Sigma$,
and $x^\alpha$ are coordinates in that surface.  In
Eq.\ (\ref{eq:spindef_coord}), we have taken $\Sigma$ to be a
``slice'' of constant $t$, where $t$ (not necessarily the
Boyer-Lindquist time) is used to parameterize the orbiting body's
worldline; $g$ is the determinant of the metric.

By enforcing $\nabla_\alpha T^{\alpha\beta} = 0$, we find the
following equations governing the motion of the small body
{\cite{mathisson2,papa51,dixon70}}:
\begin{eqnarray}
\frac{Dp^\alpha}{d\tau} &=&
-\frac{1}{2}{R^\alpha}_{\nu\lambda\sigma}u^\nu S^{\lambda\sigma}\;,
\label{eq:force1}
\\
\frac{DS^{\alpha\beta}}{d\tau} &=& p^\alpha u^\beta - p^\beta u^\alpha\;.
\label{eq:precess1}
\end{eqnarray}
The operator $D/d\tau$ denotes a covariant derivative along the
worldline that the small body follows.  The four-velocity is defined,
as usual, by $u^\alpha = dx^\alpha/d\tau$.  However, it is not the
case that $p^\alpha = \mu u^\alpha$ (where $\mu$ is the small body's
rest mass).

As discussed in Sec.\ {\ref{sec:intro}}, Eq.\ (\ref{eq:force1}) and
(\ref{eq:precess1}) do not fully determine the small body's motion.
One must also impose a spin-supplementary condition (SSC), which
accounts for degrees of freedom which are implicit in its
non-point-like structure.  An outstanding review and discussion of
this condition and its physical meaning is given by Costa and
Nat\'ario {\cite{costanatario2015}}; further excellent review and
comparison is given in Kyrian and Semer\'ak {\cite{semerakII}}, and
additional comparison of spin-supplementary conditions is provided by
Ref.\ {\cite{lsk2014}}.

A commonly used choice is the Tulczyjew SSC
{\cite{tulczyjew}},
\begin{equation}
p_\alpha S^{\alpha\beta} = 0\;.
\label{eq:ss_p}
\end{equation}
We use Eq.\ (\ref{eq:ss_p}) for our analysis.  We emphasize that this
is an arbitrary choice, and is adopted primarily because it is often
used in literature that examines gravitational-wave generation from
spin-enhanced orbits (e.g.,
Refs.\ {\cite{mst96,smsm98,han2010,sp2012,hlops2014,tmss96}}).
Equivalent motion can be shown to follow from a range of SSCs, though
one must be careful to compare properly, as described in detail in
Ref.\ {\cite{costanatario2015}}.

Using Eq.\ (\ref{eq:ss_p}), one can show that
\begin{equation}
u^\mu = \frac{\cal M}{\mu^2}\left(p^\mu +
\frac{2S^{\mu\nu}R_{\nu\rho\sigma\tau}p^\rho S^{\sigma\tau}}{4{\cal
    \mu}^2 +
  R_{\alpha\beta\gamma\delta}S^{\alpha\beta}S^{\gamma\delta}}\right)\;.
\end{equation}
The parameters $\mu$ and ${\cal M}$ both have the dimensions of mass,
and are related\footnote{We have reversed the definitions of $\mu$ and
  ${\cal M}$ relative to how they are defined in
  Ref.\ {\cite{semerakII}}.  We do this so that $-p^\mu p_\mu$ has the
  same name in both the spinning and spinless cases.}  to $u^\alpha$
and $p^\alpha$ by
\begin{eqnarray}
\mu &\equiv& \sqrt{-p_\alpha p^\alpha}\;,
\label{eq:calMdef}
\\
{\cal M} &\equiv& -p_\alpha u^\alpha\;.
\label{eq:muMdef}
\end{eqnarray}
Note that $u_\alpha u^\alpha = -1$.  Using Eq.\ (\ref{eq:ss_p}), one
can show that $\mu$ is constant along the orbiting body's worldline,
but that $\cal M$ is not.  From these definitions, it is simple to see
that $\mu = {\cal M} + O(S^2)$, and that $u^\alpha = p^\alpha/\mu +
O(S^2)$.  The mass parameters $\mu$ and ${\cal M}$ are identical for
geodesic orbits (as they must be), and cannot be distinguished at
linear order in the small body's spin.

From the spin tensor, one finds the spin vector {\cite{semerakII}}
\begin{equation}
S^\mu = -\frac{1}{2\mu}{\epsilon^{\mu\nu}}_{\alpha\beta} p_\nu
S^{\alpha\beta}\;.
\label{eq:spinvec1}
\end{equation}
Here,
\begin{equation}
\epsilon_{\alpha\beta\gamma\delta} =
\sqrt{-g}\left[\alpha\beta\gamma\delta\right]
\end{equation}
where the metric determinant $\sqrt{-g} = \Sigma\sin\theta$ for Kerr,
and where $\left[\alpha\beta\gamma\delta\right]$ is the totally
antisymmetric symbol:
\begin{eqnarray}
\left[\alpha\beta\gamma\delta\right]
&=& +1\quad\mbox{if $\alpha\beta\gamma\delta$ is an even permutation
  of $0123$}
\nonumber\\
&=& -1\quad\mbox{if $\alpha\beta\gamma\delta$ is an odd permutation of
  $0123$}
\nonumber\\
&=& 0\quad\mbox{otherwise}\;.
\end{eqnarray}
Notice that $p_\mu S^\mu = 0$.  Less obviously,
\begin{equation}
S^2 \equiv S^\alpha S_\alpha = \frac{1}{2} S_{\alpha\beta} S^{\alpha\beta}
\label{eq:spinmag}
\end{equation}
is a constant.

Equations (\ref{eq:force1}), (\ref{eq:precess1}), and (\ref{eq:ss_p})
can be integrated to build the worldline of a spinning body.  The
worldline which one finds from such an integration admits a constant
of the motion for each of the spacetime's Killing vectors: If
$\xi^\alpha$ is one of these Killing vectors, then
\begin{equation}
{\cal C} = p_\alpha\xi^\alpha - \frac{1}{2}S^{\alpha\beta}\nabla_\beta\xi_\alpha
\end{equation}
is constant along the spinning body's worldline.  For the Kerr
spacetime, these constants are
\begin{eqnarray}
E^{\rm S} &=& E^{\rm G} + \frac{1}{2} \partial_\beta g_{t\alpha}
S^{\alpha\beta}\;,
\label{eq:ESdef}\\
L_z^{\rm S} &=& L_z^{\rm G} - \frac{1}{2} \partial_\beta g_{\phi\alpha}
S^{\alpha\beta}\;.
\label{eq:LzSdef}
\end{eqnarray}
These quantities define conserved energy and axial angular momentum
for a spinning body orbiting a Kerr black hole; the superscript ``S''
distinguishes them from the conserved quantities associated with
geodesics.  In general, there is no conserved quantity analogous to
the Carter constant for orbits of spinning bodies.

\subsection{Leading order in small body's spin}
\label{sec:spincurvleading}

We now imagine that the small body's spin can be taken to be ``small''
in some meaningful way.  Let us define the dimensionless spin
parameter $s$ by
\begin{equation}
S = s\mu^2\;,
\end{equation}
where $S$, defined in Eq.\ (\ref{eq:spinmag}), is the magnitude of the
spin vector.  If the small body is itself a Kerr black hole, we expect
that $s \le 1$.  Even for non-Kerr small bodies, this bound is likely
to be a useful guide as long as the small body is compact.  We thus
expect that spin effects will have an impact on the motion at order
$\mu^2$.  It's worth noting other literature uses different
conventions to normalize $S$.  For example,
Refs.\ {\cite{suzukimaeda96}} and {\cite{harmsetal2015}} define
\begin{equation}
S = \sigma \mu M\;.
\end{equation}
In these works, the authors are not restricting their analysis to the
large mass ratio limit, so $\mu$ is the reduced mass of the system,
and $M$ is its total mass.  If the binary's secondary is in fact a
Kerr black hole, then this condition implies that $\sigma \le
(\mu/M)$.  It should be noted that these conditions are somewhat
arbitrary, since it is the spin $S$ that matters in the equations of
motion, not the parameters $s$ and $\sigma$.  The definition of these
dimensionless parameters is simply a convenience to guide our
intuition.

The scaling of $S$ with $\mu$ for the case of the small body being a
Kerr black hole suggests that, especially for large mass ratio
systems, it would be fruitful to neglect terms that are of $O(S^2)$
and higher.  With our convention for the scaling of the small body's
spin, the terms we are neglecting would introduce effects at fourth
order in the small body's mass.  Truncating at $O(S)$, we find
\begin{equation}
u^\mu = p^\mu/\mu\;.
\end{equation}
It follows that
\begin{equation}
\frac{DS^{\mu\nu}}{d\tau} = 0\;,
\label{eq:precess2}
\end{equation}
so that the spin tensor is parallel transported along the small body's
worldline at $O(S)$.

We now rewrite the equation of motion Eq.\ (\ref{eq:force1}) to
leading order in $S$.  We emphasize again that the derivative operator
$D/d\tau$ is a derivative along the small body's worldline, and that
our goal is to develop a force term that can be used with an
osculating geodesic integrator.  As such, our small body's motion will
be regarded as tangent to some geodesic at every moment along its
worldline.  The geodesic to which the worldline is tangent will differ
at each step, but there will be some ``reference geodesic'' defining
$D/d\tau$ at each moment.

We begin by writing the small body's 4-velocity
\begin{equation}
u^\alpha = u^\alpha_{\rm G} + u^\alpha_{\rm S}\;,
\end{equation}
where $u^\alpha_{\rm G}$ satisfies the geodesic equation,
\begin{equation}
\frac{Du^\alpha_{\rm G}}{d\tau} = 0\;,
\end{equation}
and where $u^\alpha_{\rm S} = O(S)$.  Linearizing
Eq.\ (\ref{eq:force1}) in $S$, we have
\begin{equation}
\frac{Du^\alpha_{\rm S}}{d\tau} =
-\frac{1}{2\mu}
{R^\alpha}_{\nu\lambda\sigma}u^\nu_{\rm G}S^{\lambda\sigma}\;.
\label{eq:force_lin}
\end{equation}
Equation (\ref{eq:force_lin}) defines the force which, in the
osculating geodesic picture, carries the small body's worldline from
one reference geodesic to another.  It is worth emphasizing that
Eq.\ (\ref{eq:force_lin}) only makes sense in this picture, since the
derivative operator $D/d\tau$ is defined quite strictly with respect
to this geodesic.  The results that we present in
Sec.\ {\ref{sec:results}} for the spin-curvature force should thus be
regarded as the force tending to push a spinning body away from a
specified reference geodesic.

The SSC (\ref{eq:ss_p}) becomes
\begin{equation}
u^{\rm G}_\alpha S^{\alpha\beta} = 0\
\label{eq:ss_u}
\end{equation}
when we linearize in $S$.  The spin vector (\ref{eq:spinvec1}) becomes
\begin{equation}
S^\mu = -\frac{1}{2}{\epsilon^{\mu\nu}}_{\alpha\beta} u^{\rm G}_\nu
S^{\alpha\beta}\;.
\label{eq:spinvec}
\end{equation}
This equation can be inverted:
\begin{equation}
S^{\alpha\beta} = \epsilon^{\alpha\beta\mu\nu}u^{\rm G}_\mu S_\nu\;.
\label{eq:spinvecinvert}
\end{equation}
It follows from Eq.\ (\ref{eq:spinvec}) that
\begin{equation}
\frac{DS^\mu}{d\tau} = 0\;,
\label{eq:spinparallel}
\end{equation}
so that the spin vector is parallel transported along the body's
worldline at $O(S)$.  It also follows that
\begin{equation}
u^{\rm G}_\alpha S^\alpha = 0
\label{eq:Stransport}
\end{equation}
at this order.  This means that we only need to know three of the
components of $S^\alpha$; the fourth component is determined by the
constraint.  Equation (\ref{eq:Stransport}) tells us that, in the
orbiting body's rest frame, the spin vector is purely spatial.

At linear order in $S$, the motion acquires a new conserved constant
{\cite{rudiger83,tmss96}}:
\begin{equation}
K^{\rm S} = K^{\rm G} - 2 p_{\rm G}^\mu S^{\rho\sigma}
\left({f^\nu}_\sigma\nabla_\nu f_{\mu\rho} - {f^\nu}_\rho\nabla_\nu
f_{\sigma\rho}\right)\;,
\label{eq:KSdef}
\end{equation}
where $f_{\mu\nu}$ is the Killing-Yano tensor introduced in
Eq.\ (\ref{eq:KYdef}).  Although there is in general no analog of the
Carter constant for spinning bodies orbiting Kerr black holes, there
is such an analog at leading order in the smaller body's spin.

The leading-order-in-spin approach we have described is designed to be
used for integrating orbits using the technique of osculating
geodesics {\cite{poundpoisson08,gfdhb11}}.  The small body's true
worldline is, at every moment, tangent to some geodesic, the
``osculating'' geodesic.  Using Eq.\ (\ref{eq:force_lin}), we compute
the force which pushes the small body away from that osculating
geodesic.  We use this to step the system by some amount $\Delta\tau$
along its worldline, updating the spin via
Eq.\ (\ref{eq:spinparallel}).  At the end of this step, we update the
osculating geodesic, and then repeat.

Integrating orbits with this scheme, one will find that the geodesic
values for the energy, axial angular momentum, and Carter constant
will oscillate as the system moves between different osculating
geodesics.  However, the spin-enhanced energy, axial angular momentum,
and Carter constant, given by Eqs.\ (\ref{eq:ESdef}),
(\ref{eq:LzSdef}), and (\ref{eq:KSdef}), will all be constant.

\section{Frequency domain treatment of spin precession}
\label{sec:prec_FD}

We now expand on the details of our approach to analyzing
spin-curvature forces.  In addition to treating the small body's spin
as a perturbative parameter, we take advantage of the fact that Kerr
geodesics are triperiodic, so any function arising from these
geodesics can be computed in a Fourier expansion.  In the next
section, we will use this to show how the components of the spin
curvature force behave in the frequency domain.

We start by analyzing the precession of the small body's spin.
Precession introduces new frequencies into our analysis, which can be
represented as eigenvalues of a matrix representation of the master
precession equation.  Once the precession frequency is known, it is
straightforward to include it in a frequency-domain expansion of the
spin-curvature force.

Begin by expanding the covariant derivative in
Eq.\ (\ref{eq:spinparallel}), and changing the independent parameter
from proper time $\tau$ to Mino time $\lambda$.  The ``master
equation'' governing precession becomes
\begin{equation}
\frac{d S^\alpha}{d \lambda} = -\Gamma^\alpha_{\mu\nu} S^\mu U^\nu \;,
\label{eq:dSdlambda}
\end{equation}
where
\begin{equation}
U^\nu \equiv u^\nu \frac{d\tau}{d\lambda} = \frac{dx^\nu}{d\lambda}\;.
\end{equation}
This can be regarded as an eigenvector equation.  To expedite solving
it, write
\begin{equation}
\frac{d{\mathbf{S}}}{d\lambda} = \mathbf{P}\cdot{\mathbf{S}}\;,
\label{eq:masterprec}
\end{equation}
where ${\mathbf{S}} \doteq (S^r, S^\theta, S^\phi)$, and where
$\mathbf{P}$ is a $3 \times 3$ matrix which depends only on $r$ and
$\theta$.  Here and throughout the paper, boldface symbols denote
vectors and matrices whose indices are associated with the spatial
coordinates $r$, $\theta$, and $\phi$.  It should be clear from
context which quantities are two-index matrices, and which are
one-index vectors.

As discussed in Sec.\ {\ref{sec:spincurv}}, the timelike component
$S^t$ is determined by the constraint $u_\alpha^{\rm G}S^\alpha = 0$.
Taking this into account, the elements of $\mathbf{P}$ are given by
\begin{equation}
{P^i}_j = -\Gamma^i_{\nu j} U^\nu + \Gamma^i_{\nu 0} U^\nu
\frac{U_j}{U_0} \;.
\label{eq:precmatrix_gen}
\end{equation}
Each element of this matrix varies with time as the small body moves
through its orbit.  Because the matrix is itself a function of Kerr
geodesic motion, it can be expanded in the frequency domain, so we
write
\begin{equation}
\mathbf{P} = \sum_{k = -\infty}^\infty \sum_{n = -\infty}^\infty
\mathbf{P}_{kn} e^{-i(k\Upsilon_\theta + n\Upsilon_r)\lambda}\;.
\label{eq:precmatrix_gen_FD}
\end{equation}
Harmonics of $\Upsilon_\phi$ do not enter due to the spacetime's
axisymmetry.  Each Fourier component $\mathbf{P}_{kn}$ is itself a $3
\times 3$ matrix.  The indices $k$ and $n$ formally run from $-\infty$
to $\infty$, although in practice the sums converge at finite value.

We now discuss how to solve this.  There are three cases which we
discuss separately: circular equatorial orbits, for which we need only
the $k = n = 0$ terms; eccentric equatorial, circular inclined, or
``resonant'' orbits, for which we need one index; and generic orbits,
for which we need both.

\subsection{Circular and equatorial orbits}
\label{sec:preccirceq}

When the osculating geodesic is circular and equatorial, the small
body's precession has a particularly simple analytic solution.  We
develop this solution first in a straightforward way, and then
re-analyze this solution as an eigenvector problem.  This allows us to
begin developing the tools we will need to study more complicated
orbit geometries.

\subsubsection{Analytic solution for the spin}

Consider an orbit of constant radius $r$, with constant polar angle
$\theta = \pi/2$.  The precession matrix $\mathbf{P}$ is quite simple
in this limit:
\begin{eqnarray}
{P^\phi}_r &=& \frac{[a^2M\! - r(r - 2M)^2]\hat L^{\rm G}_z - aM(3r\!
  - 4M\!  +\! a^2)\hat E^{\rm G}}{\Delta}\;,
\nonumber\\
\label{eq:PCEphir}\\
{P^r}_\phi &=& -3M(\hat L_z^{\rm G} - a\hat E^{\rm G}) +
\frac{aM(\hat L_z^{\rm G} - a\hat E^{\rm G})^2}{r^2 \hat
  E^{\rm G}} +r \hat L_z^{\rm G}\;,
\nonumber\\
\label{eq:PCErphi}\\
{P^i}_j &=& 0\qquad\mbox{(all other indices)}\;.
\label{eq:PCEij}
\end{eqnarray}
Note that $\hat Q^{\rm G} = 0$ for equatorial orbits, so the Carter
constant does not appear in these expressions.

Using this, Eq.\ (\ref{eq:masterprec}) becomes
\begin{eqnarray}
\frac{dS^r}{d\lambda} &=& {P^r}_\phi S^{\phi}\;,
\label{eq:Sr} \\
\frac{dS^{\theta}}{d\lambda} &=& 0\;,
\label{eq:Stheta}\\
\frac{dS^{\phi}}{d\lambda} &=& {P^\phi}_rS^{r}\;.
\label{eq:Sphi}
\end{eqnarray}
For an equatorial orbit, $S^\theta$ is the component of the small
body's spin normal to the orbital plane.  Equation (\ref{eq:Stheta})
tells us that this spin component is constant for circular, equatorial
orbits.  As we'll see in Sec.\ {\ref{sec:results}}, the product $r
S^\theta$ is constant for all equatorial orbits, consistent with this
result.  For the other two components, we combine Eqs.\ (\ref{eq:Sr})
and (\ref{eq:Sphi}) yielding
\begin{equation}
\frac{d^2S^{r,\phi}}{d\lambda^2} - {P^r}_\phi{P^\phi}_r S^{r,\phi} = 0\;.
\label{eq:Srphi1}
\end{equation}
The values of $\hat E^{\rm G}$ and $\hat L^{\rm G}_z$ appearing in
${P^r}_\phi$ and ${P^\phi}_r$ are given by {\cite{bpt72}}
\begin{eqnarray}
\hat E^{\rm G} &=& \frac{r^{3/2} - 2Mr^{1/2} \pm aM^{1/2}}{\sqrt{r^3 -
    3Mr^2 \pm 2 a M^{1/2} r^{3/2}}}\;,
\label{eq:EGcirceq}\\
\hat L^{\rm G}_z &=& \pm\frac{M^{1/2}\left(r^2 \mp 2aM^{1/2}r^{1/2} +
  a^2\right)} {\sqrt{r^3 - 3Mr^2 \pm 2 a M^{1/2} r^{3/2}}}\;.
\label{eq:LzGcirceq}
\end{eqnarray}
The upper sign is for prograde orbits, the lower is for retrograde.

Combining Eqs.\ (\ref{eq:PCEphir}), (\ref{eq:PCErphi}),
(\ref{eq:EGcirceq}), and (\ref{eq:LzGcirceq}), we find the somewhat
remarkable simplification
\begin{equation}
{P^r}_\phi{P^\phi}_r = -r M \equiv -\left(\Upsilon^{\rm CE}_s\right)^2\;.
\end{equation}
In terms of Mino-time $\lambda$, this means that for circular
equatorial orbits, $S^{r,\phi}$ undergo simple harmonic oscillations
at the frequency
\begin{equation}
\Upsilon^{\rm CE}_s = \sqrt{M r}\;.
\label{eq:UpsilonCE}
\end{equation}
The ``CE'' superscript is a reminder that this quantity only applies
to circular and equatorial orbits.  By matching to initial conditions,
it is a straightforward exercise to construct $S^{r,\phi}(\lambda)$.

Using $d\lambda = d\tau/\Sigma$, with $\Sigma = r^2$ for an equatorial
orbit, we convert to frequency conjugate to proper time along the
orbit:
\begin{equation}
\omega^{\rm CE}_s = \frac{\Upsilon^{\rm CE}_s}{\Sigma} =
\sqrt{\frac{M}{r^3}}\;.
\end{equation}
Finally, using Eqs.\ (\ref{eq:tdot}), (\ref{eq:EGcirceq}), and
(\ref{eq:LzGcirceq}), we can convert this to a frequency conjugate
to observer time:
\begin{equation}
\Omega^{\rm CE}_S = \frac{M^{1/2}}{r^{3/2} \pm a M^{1/2}}
\left(1 \pm 2a\sqrt{\frac{M}{r^3}} - \frac{3M}{r}\right)^{1/2}\;.
\end{equation}
As before, upper sign labels prograde orbits, and lower labels
retrograde.

The rate at which the spin vector is seen to precess is given by the
difference between this frequency and the orbital frequency.  Using
\begin{equation}
\Omega_{\rm orb} = \frac{M^{1/2}}{r^{3/2} \pm a M^{1/2}}\;,
\end{equation}
we have
\begin{eqnarray}
\Omega_{\rm prec} &=& \Omega_{\rm orb} - \Omega^{\rm CE}_s
\nonumber\\
&=& \frac{3}{2}\frac{M^{3/2}}{r^{5/2}} \mp \frac{aM}{r^3} +
O\left(r^{-7/2}\right)\;.
\label{eq:CEprec_expand}
\end{eqnarray}
The term at $O(r^{-5/2})$ reproduces well known results for the
geodetic precession of an orbiting gyroscope; the term at $O(a/r^3)$
reproduces the ``gravitomagnetic'' or Lense-Thirring correction to
this precession arising from the larger body's spin.  See, for
example, Eqs.\ (3.3) and (3.4) of Ref.\ {\cite{bo79}} (noting that,
for an equatorial Kerr orbit, Barker and O'Connell's ${\bf n}$ and
${\bf n}^{(2)}$ are parallel to one another, and point along the black
hole's spin axis).

\subsubsection{The spin as an eigenvector}

Let us now reorganize the above analysis, introducing notation that
will generalize to more complicated orbit geometries.  We begin by
writing
\begin{equation}
{\mathbf{S}} = {\mathbf{S}}^a\, e^{-i \Upsilon^a_s \lambda}\;.
\end{equation}
Using this, Eq.\ (\ref{eq:masterprec}) becomes
\begin{equation}
-i\Upsilon^a_s{\mathbf{S}}^a = \mathbf{P}\cdot {\mathbf{S}}^a\;.
\label{eq:eigencase1}
\end{equation}
In other words, $\mathbf{S}^a$ is an eigenvector of $\mathbf{P}$ with
eigenvalue $-i\Upsilon^a_s$.  Using Eqs.\ (\ref{eq:PCEphir}),
(\ref{eq:PCErphi}), and (\ref{eq:PCEij}), we find the following three
eigenvectors and eigenvalues:
 \begin{eqnarray}
{\mathbf{S}}^0 &=& 
\left (
\begin{array}{ccc}
0 \\ 1 \\ 0
\end{array}
\right )\;,\quad \Upsilon^0_s = 0\;,
\label{eq:CEeigen0}\\
{\mathbf{S}}^{\pm1} &=& 
\left (
\begin{array}{ccc}
\pm {P^r}_\phi/{P^\phi}_r \\ 0 \\ 1
\end{array}
\right )\;,\quad \Upsilon^{\pm1}_s = \pm\Upsilon^{\rm CE}_s\;,
\label{eq:CEeigenpm1}
\end{eqnarray}
where $\Upsilon^{\rm CE}_S = \sqrt{M r}$ as before.

A general spin vector can then be written
\begin{equation}
\mathbf{S} = c_0\mathbf{S}^0 + c_{-1}
\mathbf{S}^{-1}e^{i\Upsilon^{\rm CE}_s\lambda} + c_1 \mathbf{S}^1
e^{-i\Upsilon^{\rm CE}_s\lambda}\;,
\end{equation}  
with $c_{-1,0,1}$ determined by initial conditions.  As we saw above,
the $\theta$ component of ${\bf S}$ is constant, while the components
$S^r$ and $S^\phi$ undergo simple harmonic oscillations [see
  Eqs.\ (\ref{eq:Sr})--(\ref{eq:UpsilonCE})].  A generalization of
this behavior holds for all equatorial orbits.

\subsection{Circular inclined, eccentric equatorial, and resonant generic
orbits}
\label{sec:precessonefreq}

In the circular equatorial case, the matrix $\mathbf{P}$ does not vary
with time, so the frequency-domain expansion
(\ref{eq:precmatrix_gen_FD}) is trivial.  We now consider cases in
which the matrix varies, but is characterized by one dynamically
important frequency.

If the orbit is circular but inclined, then the angle $\theta$ varies
over the orbit.  Only the polar frequency matters in this case, and
(\ref{eq:precmatrix_gen_FD}) simplifies to
\begin{equation}
\mathbf{P} = \sum_{k = -\infty}^\infty \mathbf{P}_k e^{-ik\Upsilon_\theta\lambda}\;.
\label{eq:Pmatrix_circinc}
\end{equation}
If the orbit is equatorial but eccentric, then the radius $r$ varies
over the orbit, and Eq.\ (\ref{eq:precmatrix_gen_FD}) becomes
\begin{equation}
\mathbf{P} = \sum_{n = -\infty}^\infty \mathbf{P}_n
e^{-in\Upsilon_r\lambda}\;.
\label{eq:Pmatrix_eqecc}
\end{equation}
In the general case, we expect both radial and polar frequencies to be
important, and we must use Eq.\ (\ref{eq:precmatrix_gen_FD}) with no
simplifications.  However, for the special case of resonant orbits
{\cite{fh12,fhr14}}, the two frequencies are simply related to one
another:
\begin{equation}
\Upsilon_\theta = \beta_\theta\Upsilon\;,\quad
\Upsilon_r = \beta_r\Upsilon\;,
\end{equation}
with $\beta_r$ and $\beta_\theta$ both integers.  A harmonic
$k\Upsilon_\theta + n\Upsilon_r = N\Upsilon$, with $N = k\beta_\theta
+ n\beta_r$.  In this case, the expansion takes the form
\begin{equation}
\mathbf{P} = \sum_N \mathbf{P}_N e^{-iN\Upsilon\lambda}\;.
\end{equation}
This sum runs from $-\infty$ to $\infty$, but the spacing between
frequencies will depend in detail on the nature of the resonance.

In these three cases, the expansion can be written\footnote{With
  perhaps an adjustment to the spacing for resonant orbits.}
\begin{equation}
\mathbf{P} = \sum_{j = -\infty}^\infty \mathbf{P}_j
e^{-ij\Upsilon_x\lambda}
\end{equation}
for an appropriate choice of $\Upsilon_x$.  To solve
Eq.\ (\ref{eq:masterprec}) with this form of $\mathbf{P}$, we assume
that there exist solutions $\mathbf{S}^a$ of the form
\begin{equation}
\mathbf{S}^a = \sum_{j' = -\infty}^\infty
e^{-i\Upsilon^a_s\lambda}{\mathbf{S}^a}_{j'}
e^{-ij'\Upsilon_x\lambda}\;.
\label{eq:eigenspin_1freq}
\end{equation}
The index $a$ labels different eigenvectors and eigenvalues. Equation
(\ref{eq:masterprec}) then becomes
\begin{widetext}
\begin{equation}
-i\sum_{j' = -\infty}^\infty(\Upsilon^a_s +
j'\Upsilon_x){\mathbf{S}^a}_{j'} e^{-i\Upsilon^a_s\lambda}
e^{-ij'\Upsilon_x\lambda} = \sum_{j,j' =
  -\infty}^\infty\mathbf{P}_{j}\cdot{\mathbf{S}^a}_{j'}
e^{-i\Upsilon^a_s\lambda} e^{-i(j+j')\Upsilon_x\lambda}\;.
\end{equation}
The common factor of $e^{-i\Upsilon^a_s\lambda}$ on both sides
cancels.  Multiply both sides by $e^{iq\Upsilon_x\lambda}$, and
integrate from $\lambda = 0$ to $2\pi/\Upsilon_x$:
\begin{equation}
-i\sum_{j' = -\infty}^\infty(\Upsilon^a_s +
j'\Upsilon_x){\mathbf{S}^a}_{j'}\delta_{j',q} = \sum_{j,j' =
  -\infty}^\infty\mathbf{P}_{j}\cdot{\mathbf{S}^a}_{j'}\delta_{(j+j'),q}\;.
\end{equation}
Performing the sums over $j'$, this simplifies further:
\begin{equation}
-i(\Upsilon^a_s + q\Upsilon_x){\mathbf{S}^a}_q = \sum_{j =
  -\infty}^\infty \mathbf{P}_{j}\cdot{\mathbf{S}^a}_{q-j}\;.
\end{equation}
Let us truncate the sum over $j$ at some finite value $\pm j_{\rm
  max}$.  Expand this equation and rearrange:
\begin{eqnarray}
\mathbf{P}_{-j_{\rm max}}\cdot {\mathbf{S}^a}_{q+j_{\rm max}} &+&
\mathbf{P}_{-j_{\rm max}+1}\cdot {\mathbf{S}^a}_{q+j_{\rm max}-1} +
\cdots
\nonumber\\
&+& \left(\mathbf{P}_{0} + iq\Upsilon_x\mathbf{I}\right)\cdot
{\mathbf{S}^a}_{q} +
\nonumber\\
\cdots &+& \mathbf{P}_{j_{\rm max}-1}\cdot {\mathbf{S}^a}_{q-j_{\rm
    max}+1} + \mathbf{P}_{j_{\rm max}}\cdot {\mathbf{S}^a}_{q-j_{\rm
    max}} = -i\Upsilon^a_s {\mathbf{S}^a}_{q}\;.
\label{eq:mastereigen}
\end{eqnarray}
The eigenvalue $\Upsilon^a_s$ must satisfy this for each of the
$2j_{\rm max} + 1$ Fourier components.

To solve Eq.\ (\ref{eq:mastereigen}), we rewrite it as a bigger matrix
equation.  We first assemble a ``vector of vectors'' by combining the
different Fourier components ${\mathbf{S}^a}_j$ as follows:
\begin{equation}
\mathbb{S}^a =
\begin{pmatrix}
{\mathbf{S}^a}_{-j_{\rm max}}\cr
{\mathbf{S}^a}_{-j_{\rm max} + 1}\cr
\vdots\cr
{\mathbf{S}^a}_0\cr
\vdots\cr
{\mathbf{S}^a}_{j_{\rm max} - 1}\cr
{\mathbf{S}^a}_{j_{\rm max}}\cr
\end{pmatrix}\;.
\end{equation}
There are $2j_{\rm max} + 1$ elements in $\mathbb{S}^a$, each of which
is itself a 3 element spin vector.  We likewise define the $(2j_{\rm
  max} + 1)\times(2j_{\rm max} + 1)$ matrix of matrices $\mathbb{P}$
whose elements are each $3 \times 3$ precession matrices:
\begin{equation}
\mathbb{P}_{gh} =
\left\{
\begin{matrix}
\mathbf{P}_{g - h}\qquad g \ne h\;,
\\
\mathbf{P}_0 + ig\Upsilon_x\mathbf{I}\qquad g = h\;.
\end{matrix}
\right.
\end{equation}
The indices $g,h \in [-j_{\rm max},j_{\rm max}]$, and $\mathbf{I}$ is
the $3\times3$ identity matrix.  As a concrete example, for $j_{\rm
  max} = 2$ this matrix is
\begin{equation}
\mathbb{P} =
\begin{pmatrix}
\left(\mathbf{P}_0 - 2i\Upsilon_x\mathbf{I}\right) & \mathbf{P}_{-1} &
\mathbf{P}_{-2} & \mathbf{P}_{-3} & \mathbf{P}_{-4} \cr
\mathbf{P}_1 & \left( \mathbf{P}_0 - i\Upsilon_x\mathbf{I}\right) &
\mathbf{P}_{-1} & \mathbf{P}_{-2} & \mathbf{P}_{-3} \cr
\mathbf{P}_2 & \mathbf{P}_1 & \mathbf{P}_0 &
\mathbf{P}_{-1} & \mathbf{P}_{-2} \cr
\mathbf{P}_3 & \mathbf{P}_2 & \mathbf{P}_1 & \left(\mathbf{P}_0 +
i\Upsilon_x\mathbf{I}\right) & \mathbf{P}_{-1} \cr
\mathbf{P}_4 & \mathbf{P}_3 & \mathbf{P}_2 & \mathbf{P}_1 &
\left(\mathbf{P}_0 + 2i\Upsilon_x\mathbf{I}\right) \cr
\end{pmatrix}\;.
\end{equation}

\end{widetext}

To find the solution for the precessional motion of the small body, we
then solve for the eigenvectors and eigenvalues of the system
\begin{equation}
\mathbb{P}\cdot \mathbb{S}^a = -i\Upsilon^a_s\mathbb{S}^a\;.
\label{eq:eigensystem_1freq}
\end{equation}
When we do this we find find $3 \times (2j_{\rm max} + 1)$ eigenvalues
and eigenvectors.  On physical grounds\footnote{The equation governing
  the spin vector ${\bf S}$ is first order, and we need three
  eigenvectors to set initial conditions for the three components of
  ${\bf S}$.  Furthermore, since the spin components are real numbers,
  we expect the three eigenvalues to take the form $\Upsilon_s^1 =
  -\Upsilon_s^{-1} \equiv \Upsilon_s$, and $\Upsilon_0 = 0$.}, we
expect only $3$ eigenvalues and eigenvectors, so we appear to have far
more solutions than are needed.

On careful analysis, we find that these many solutions are simply
related to one another: they fall into $3$ groups, each of which has
$2j_{\rm max} + 1$ members.  The eigenvalues in each group are simply
shifted from one another by some multiple of $\Upsilon_x$.  The
surfeit of solutions originates in a relabeling ambiguity in the
eigenvector expansion.  Take Eq.\ (\ref{eq:eigenspin_1freq}) and shift
the index $j'$ by some integer $\Delta j'$.  The resulting eigenvector
$\mathbf{S}^a$ is unchanged by this shift if we take
\begin{equation}
\Upsilon^a_s \to \Upsilon^a_s + \Delta j'\Upsilon_x\;,
\label{eq:eigenvalue_shift}
\end{equation}
and likewise shift the eigenvector components:
\begin{equation}
{\mathbf{S}^a}_{j'} \to {\mathbf{S}^a}_{j' + \Delta j'}\;.
\label{eq:eigenvec_shift}
\end{equation}
Strictly speaking, this shift leaves the system unchanged only in the
limit $j_{\rm max} \to \infty$.  However, provided $j_{\rm max}$ is
large and we confine ourselves to solutions with $\Delta j \ll j_{\rm
  max}$, the different precessional solutions that we construct with
these related eigensolutions will differ negligibly from one another.

We emphasize here that this relabeling ambiguity means that in
principle {\it any} eigenvalue/eigenvector set can be used to
construct a valid solution, provided one correctly accounts for the
index shift.  In practice, we find that the set which has
$\Upsilon^0_s = 0$ and $\Upsilon^{-1}_s = -\Upsilon^1_s$ is computed
most accurately, and is the one we have used in our analyses.

\begin{widetext}

An example usefully illustrates how this relabeling ambiguity works.
Consider an equatorial eccentric orbit around a black hole with $a =
0.9M$, $p = 10M$, and $e = 0.1$; for this orbit, $\Upsilon_r =
2.824M$.  Note that $e = 0.1$ is fairly small, so we expect the
expansion (\ref{eq:Pmatrix_eqecc}) to converge for a fairly small
value of $n_{\rm max}$.

Solving Eq.\ (\ref{eq:eigensystem_1freq}) with $n_{\rm max} = 3$, we
find 21 eigenvalues which we group as follows:
\begin{equation}
\Upsilon^{1,\Delta n}_s =
\left\{
\begin{matrix}
11.638M \cr
8.812M \cr
5.987M \cr
3.163M \cr
0.339M \cr
-2.485M \cr
-5.301M
\end{matrix}
\right.\;,\qquad \Upsilon^{0,\Delta n}_s =
\left\{
\begin{matrix}
8.456M \cr
5.648M \cr
2.824M \cr
0 \cr
-2.824M \cr
-5.648M \cr
-8.456M
\end{matrix}
\right.\;,\qquad \Upsilon^{-1,\Delta n}_s =
\left\{
\begin{matrix}
5.301M \cr
2.485M \cr
-0.339M \cr
-3.163M \cr
-5.987M \cr
-8.812M \cr
-11.638M
\end{matrix}
\right.\;.
\label{eq:ambigeigenvals}
\end{equation}
These eigenvalues are of the form
\begin{eqnarray}
\Upsilon^{1,\Delta n}_s &=& \Upsilon^1_s + \Delta n \Upsilon_r\;,
\nonumber\\
\Upsilon^{0,\Delta n}_s &=& \Upsilon^0_s + \Delta n \Upsilon_r\;,
\nonumber\\
\Upsilon^{-1,\Delta n}_s &=& \Upsilon^{-1}_s + \Delta n \Upsilon_r\;,
\label{eq:eigenshift}
\end{eqnarray}
with $\Delta n \in [-n_{\rm max}, n_{\rm max}]$, and with
$\Upsilon^1_s = 3.163M = -\Upsilon^{-1}_s$, and $\Upsilon^0_s = 0$.
(Note that $\Upsilon^{-1}_s \ne 1/\Upsilon_s$.)  The eigenvalues
corresponding to $\Delta n = \pm n_{\rm max}$ fall slightly off of the
trend given by Eq.\ (\ref{eq:eigenshift}).  As stated above, we expect
the shifted solutions to differ negligibly from one another when
$\Delta n \ll n_{\rm max}$.  The errors may be large when $\Delta n
\approx n_{\rm max}$.

Next examine one of the eigenvectors.  We focus on those corresponding
to the eigenvalues labeled $\Upsilon^{0,\Delta n}_s$ in
Eq.\ (\ref{eq:ambigeigenvals}).  Because data at $\Delta n = \pm
n_{\rm max}$ appears to be less accurate than the other solutions,
focus attention on the ``innermost'' 5 solutions (those corresponding
to $\Upsilon^0_s = \pm5.648M$, $\pm2.824M$, and $0$).  These five
eigenvectors are
\begin{eqnarray}
\mathbb{S}^{0} &=&
\begin{pmatrix}
0 \cr 2.157 \times 10^{-6} \cr 0 \cr
\hline
0 \cr 7.805 \times 10^{-7} \cr 0 \cr
\hline
0 \cr -0.000211 \cr 0 \cr
\hline
0 \cr 0.0499 \cr 0 \cr
\hline
0 \cr 0.998 \cr 0 \cr
\end{pmatrix}\;,
\begin{pmatrix}
0 \cr 7.806 \times 10^{-7} \cr 0 \cr
\hline
0 \cr -0.000211 \cr 0 \cr
\hline
0 \cr 0.0499 \cr 0 \cr
\hline
0 \cr 0.998 \cr 0 \cr
\hline
0 \cr 0.0499\cr 0 \cr
\end{pmatrix}\;,
\begin{pmatrix}
0 \cr -0.000211 \cr 0 \cr
\hline
0 \cr 0.0499 \cr 0 \cr
\hline
0 \cr 0.998 \cr 0 \cr
\hline
0 \cr 0.0499\cr 0 \cr
\hline
0 \cr -0.000211\cr 0 \cr
\end{pmatrix}\;,
\begin{pmatrix}
0 \cr 0.0499 \cr 0 \cr
\hline
0 \cr 0.998 \cr 0 \cr
\hline
0 \cr 0.0499\cr 0 \cr
\hline
0 \cr -0.000211\cr 0 \cr
\hline
0 \cr 7.806 \times 10^{-7}\cr 0 \cr
\end{pmatrix}\;,
\begin{pmatrix}
0 \cr 0.998 \cr 0 \cr
\hline
0 \cr 0.0499\cr 0 \cr
\hline
0 \cr -0.000211\cr 0 \cr
\hline
0 \cr 7.805 \times 10^{-7}\cr 0 \cr
\hline
0 \cr 2.157 \times 10^{-6}\cr 0 \cr
\end{pmatrix}\;.
\label{eq:eigenvec_example}
\end{eqnarray}
Recall that $\mathbb{S}$ is a ``vector of vectors,'' constructed by
combining the Fourier components ${\bf S}_j$ which are each associated
with a given eigenvalue.  These individual Fourier components are
indicated in the eigenvectors shown in
Eq.\ (\ref{eq:eigenvec_example}); five components are included.
Notice the cyclic nature of the eigenvectors' components.  This
demonstrates how these components must be shifted according to
Eq.\ (\ref{eq:eigenvec_shift}) when we shift the eigenvalues.  Notice
also the slight numerical differences between the smallest components:
Although they should be identical, we see both $7.806 \times 10^{-7}$
and $7.805 \times 10^{-7}$.  This difference goes away if we use a
larger value of $n_{\rm max}$, although similar errors then appear in
even smaller components.

Empirically, we find that the eigenvectors corresponding to larger
shifts tend to exhibit more numerical error.  We thus use the set
requiring no shift, for which the eigenvalues have the form
$\Upsilon^1_s = -\Upsilon^{-1}_s \equiv \Upsilon_s$, $\Upsilon^0_s =
0$.  Once these eigenvalues are identified, it is not difficult to
isolate the associated eigenvectors, $(\mathbb{S}^1, \mathbb{S}^0,
\mathbb{S}^{-1})$.  We then extract the coordinate-domain spin
eigenvector Fourier components ${\mathbf{S}^a}_j$ and assemble the
solution for spin precession along an orbit:
\begin{equation}
\mathbf{S} = \sum_{j = -j_{\rm max}}^{j_{\rm max}} \left(c_0
       {\mathbf{S}^0}_j + c_{-1} {\mathbf{S}^{-1}}_j
       e^{i\Upsilon_s\lambda} + c_1 {\mathbf{S}^1}_j
       e^{-i\Upsilon_s\lambda}\right) e^{-ij\Upsilon_x\lambda}\;.
\label{eq:spinprecconstrained}
\end{equation}
This sum runs from $-j_{\rm max}$ to $j_{\rm max}$.  For the example
discussed above, the coordinate-domain eigenvector Fourier components
are given by
\begin{equation}
{S^0}_{0} =
\begin{pmatrix}
0 \cr 0.998 \cr 0
\end{pmatrix}\;,\qquad
{S^0}_{\pm1} =
\begin{pmatrix}
0 \cr 0.0499 \cr 0
\end{pmatrix}\;,\qquad
{S^0}_{\pm2} =
\begin{pmatrix}
0 \cr -0.000211 \cr 0
\end{pmatrix}\;.
\end{equation}
Notice that only the $\theta$ components of ${\bf S}^0$ are non zero;
if we had examined the ${\bf S}^{\pm1}$ solutions, we would find by
contrast that their $\theta$ components were all zero.  Recall that
for circular equatorial orbits, $S^\theta$ is constant, and
$S^{r,\phi}$ undergo simple harmonic oscillation.  For an eccentric
equatorial orbit, we find a similar behavior: the components
$S^{r,\phi}$ are strongly coupled and oscillatory, whereas $S^\theta$
varies independent of the other two components.  We will revisit this
behavior in Sec.\ {\ref{sec:results}}.

\end{widetext}

\begin{figure}[ht]
\includegraphics[width = 0.48\textwidth]{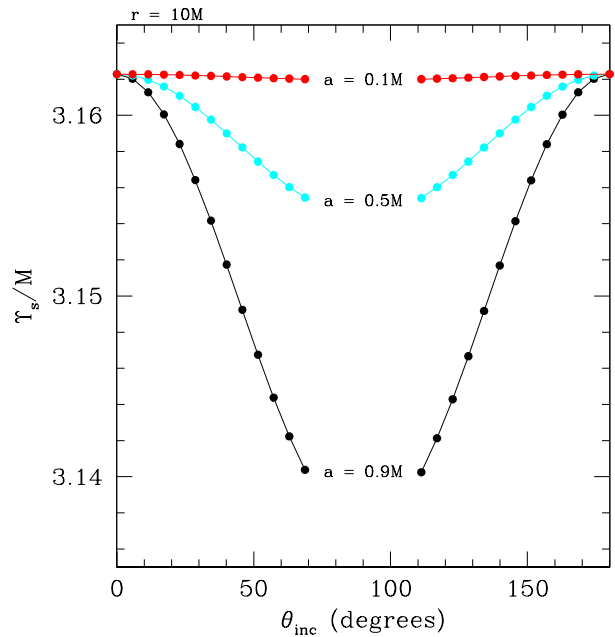}
\caption{Spin precession frequency $\Upsilon_s$ for circular inclined
  orbits of radius $r = 10M$ as a function of $\theta_{\rm inc}$.  The
  gap near $\theta_{\rm inc} = 90^\circ$ is because we truncate the
  expansion (\ref{eq:Pmatrix_circinc}) at $k_{\rm max} = 12$.  More
  terms than this must be kept in order for the precession matrix to
  be accurately represented in the frequency domain when the orbit is
  nearly polar.  Two interesting features we see here are that
  $\Upsilon_s$ varies very little as a function of $\theta_{\rm inc}$,
  and that there is a near symmetry between the prograde ($\theta_{\rm
    inc} < 90^\circ$) and retrograde ($\theta_{\rm inc} > 90^\circ$)
  branches: $\Upsilon_s(\theta_{\rm inc}) \simeq \Upsilon_s(180^\circ
  - \theta_{\rm inc})$.  These features are particularly pronounced
  for small $a$, and arise because the Kerr spacetime deviates from
  sphericity only slightly at $r = 10M$.}
\label{fig:r10_circ}
\end{figure}

Figures {\ref{fig:r10_circ}} -- {\ref{fig:p5_ecc_pro}} show
representative examples of how $\Upsilon_s$ computed by this procedure
varies with respect to orbital parameters for circular inclined and
equatorial eccentric orbits.  We omit the case of generic but resonant
orbits since they are more complicated to compute, and do not add much
to these results.  Since the amount of data involved is significant,
we do not show examples of the eigenvectors, though of course they are
found by this procedure as well.

\begin{figure}[ht]
\includegraphics[width = 0.48\textwidth]{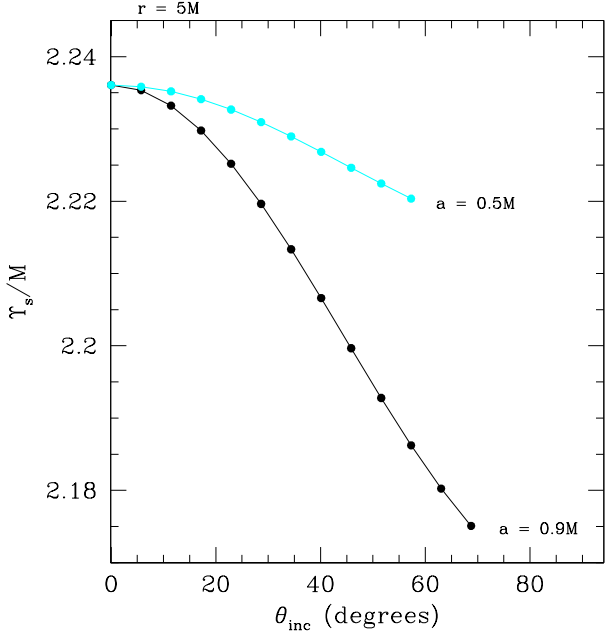}
\caption{Spin precession frequency $\Upsilon_s$ for circular inclined
  orbits of radius $r = 5M$ as a function of $\theta_{\rm inc}$.  The
  maximum allowed inclination at this radius is $\theta_{\rm inc}
  \simeq 59.9^\circ$ for $a = 0.5M$, and $\theta_{\rm inc} \simeq
  82.6^\circ$ for $a = 0.9M$; no stable orbits exist at this radius
  for $a = 0.1M$.  The variation of $\Upsilon_s$ with inclination is
  stronger at this radius than it was at $r = 10M$.  This is not
  surprising, since this is a stronger field region of the spacetime,
  and the deviations from spherical symmetry are be much larger here.
  Aside from the stronger variation and the cutoffs associated with
  the lack of stable orbits at large $\theta_{\rm inc}$, the trends we
  see are qualitatively similar to those seen in
  Fig.\ {\ref{fig:r10_circ}}.}
\label{fig:r5_circ}
\end{figure}

Figures {\ref{fig:r10_circ}} and {\ref{fig:r5_circ}} show how
$\Upsilon_s$ varies with inclination $\theta_{\rm inc}$ for orbits of
constant radius $r$ for several values of black hole spin.  The finite
span of data used in Fig.\ {\ref{fig:r5_circ}} is simply because no
stable orbits exist at $r = 5M$ beyond some inclination $\theta_{\rm
  inc}^{\rm max}$ for the spins we include here.  The gap near
$\theta_{\rm inc} = 90^\circ$ in Fig.\ {\ref{fig:r10_circ}} is because
we need a large value of $k_{\rm max}$ in
Eq.\ (\ref{eq:Pmatrix_circinc}) to accurately model highly inclined
orbits in the frequency domain.  We have fixed $k_{\rm max} = 12$ for
this initial analysis.  Empirically, we find that the expansion
converges to nine or ten digits of precision for $k_{\rm max} = 12$ on
the range $\theta_{\rm inc} \le 70^\circ$ and $\theta_{\rm inc} \ge
110^\circ$.  This holds for all values of $r$ and $a$ that we consider
here; a larger value of $k_{\rm max}$ would be needed if we examined
inclinations $\theta_{\rm inc}$ outside this range.  In other words,
the value of $k_{\rm max}$ needed for the expansion to converge
depends quite a bit on orbital inclination, but only weakly on black
hole spin and orbital radius.

One trend we see for the circular orbits is that $\Upsilon_s$ does not
vary by much with inclination angle at constant $r$, especially for
large radius and small spin.  This reflects the fact that the
spacetime is nearly spherically symmetric for these orbits, and so
$\Upsilon_s \simeq \sqrt{rM}$ at all inclinations.  Figure
{\ref{fig:r10_circ}} also shows a near symmetry between prograde and
retrograde orbits: $\Upsilon_s(\theta_{\rm inc}) \simeq
\Upsilon_s(180^\circ - \theta_{\rm inc})$.  This also follows from the
fact that the spacetime is nearly spherically symmetric at $r = 10M$.
In all cases, we find that $\Upsilon_s$ decreases from $\sqrt{rM}$ as
$\theta_{\rm inc}$ increases toward $90^\circ$ (presumably reaching a
minimum at $\theta_{\rm inc} = 90^\circ$), and increases back to
$\sqrt{rM}$ as $\theta_{\rm inc}$ increases toward $180^\circ$.

\begin{figure*}[ht]
\includegraphics[width = 0.48\textwidth]{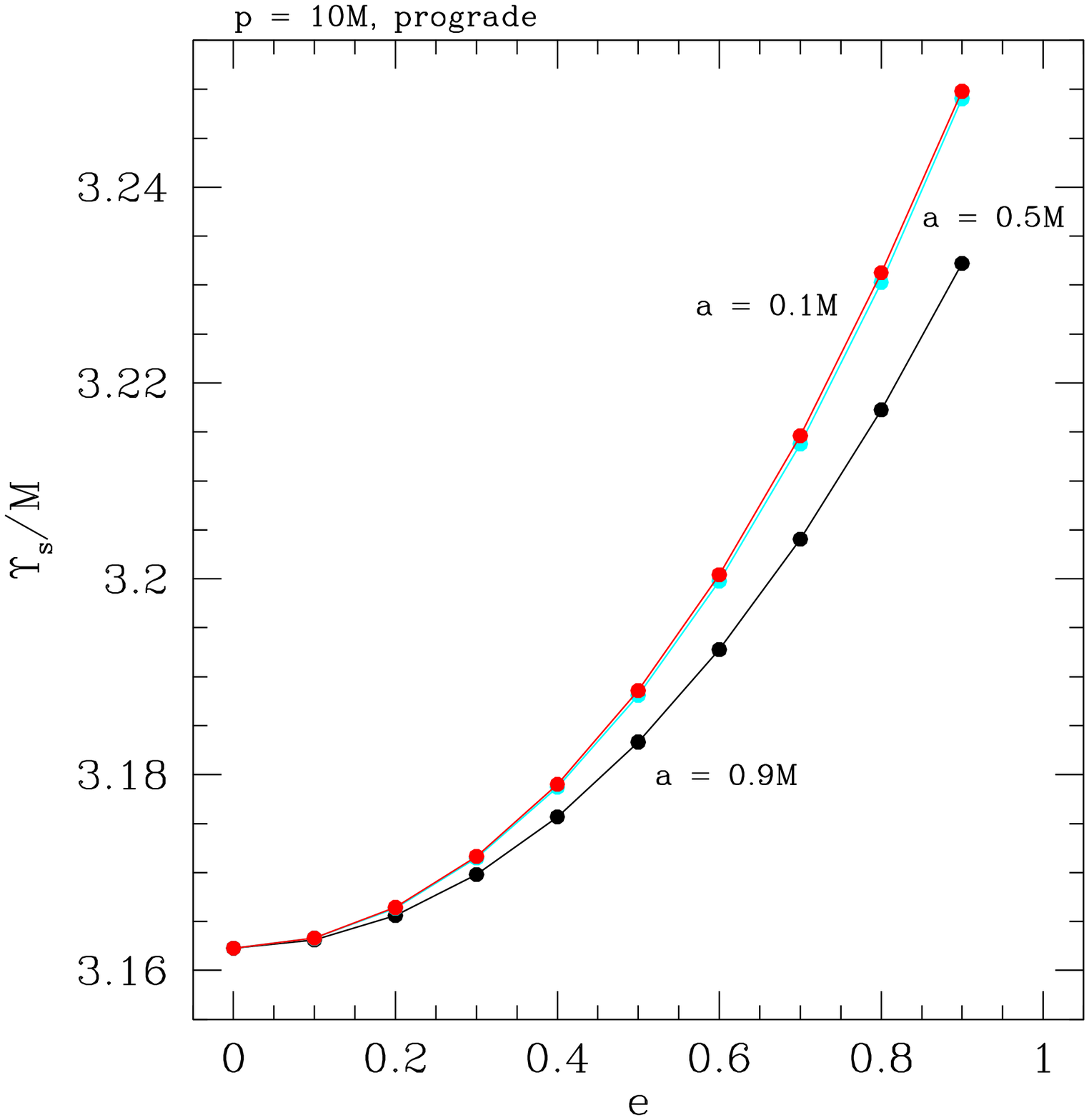}
\includegraphics[width = 0.48\textwidth]{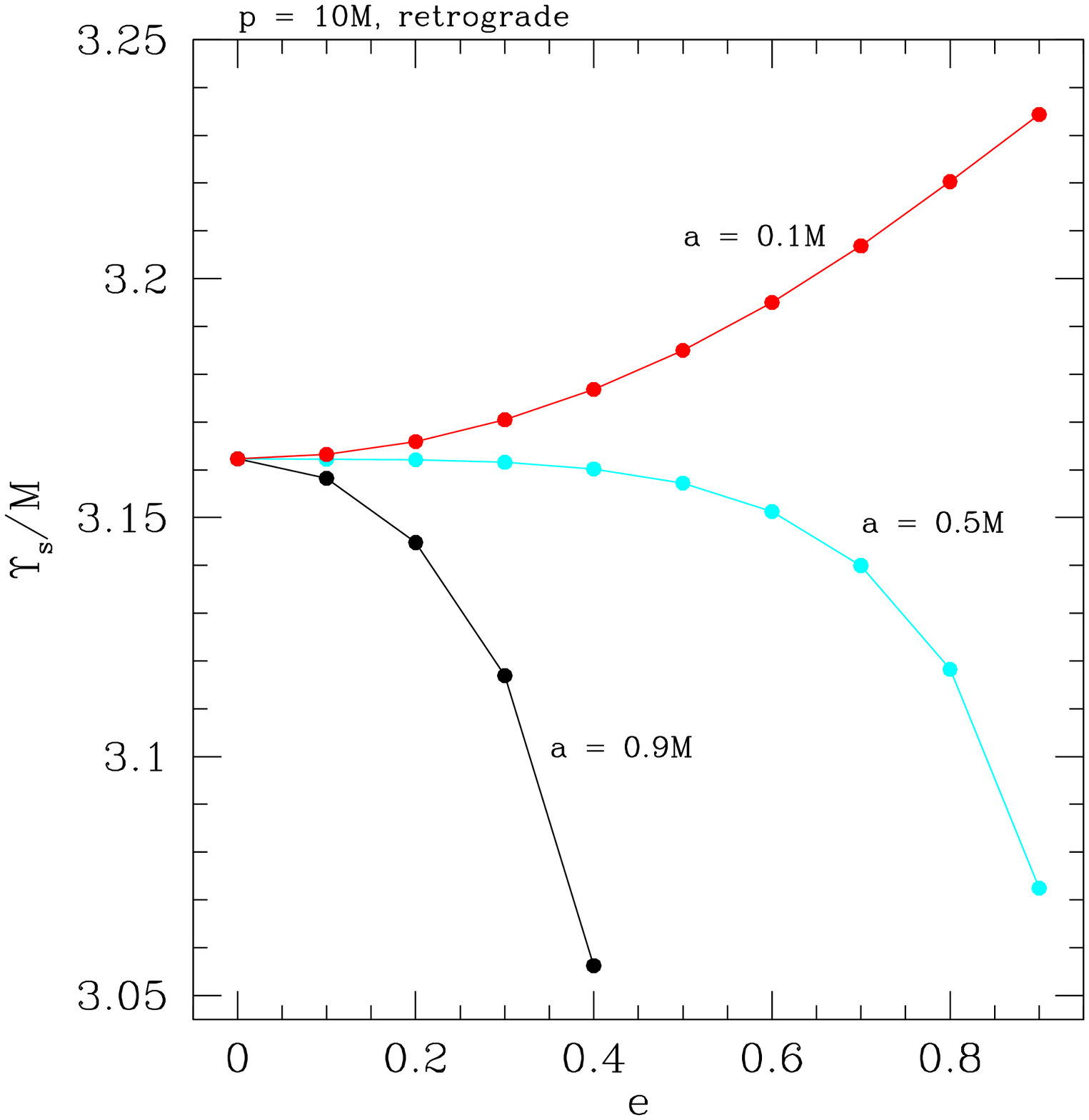}
\caption{Spin precession frequency $\Upsilon_s$ for eccentric
  equatorial orbits with $p = 10M$ as a function of eccentricity $e$.
  Left-hand panel shows prograde orbits ($L_z > 0$), right-hand panel
  is retrograde ($L_z < 0$).  In the prograde cases, $\Upsilon_s$
  smoothly and monotically increases with $e$, varying only slightly
  with spin.  (Indeed, the curves for $a = 0.1M$ and $a = 0.5M$ lie
  practically on top of each other.)  The retrograde cases show
  considerably more variation.  We find that $\Upsilon_s$ tends to
  decrease as orbits approach the last stable orbit: For retrograde
  orbits at $p = 10M$, $e_{\rm LSO} \simeq 0.4694$ at $a = 0.9M$.
  Orbits with $p = 10M$ are outside the separatrix for all $e$ at $a =
  0.5M$, although this value of $p$ is close to the separatrix as $e
  \to 1$.}
\label{fig:p10_ecc}
\end{figure*}

\begin{figure}[ht]
\includegraphics[width = 0.48\textwidth]{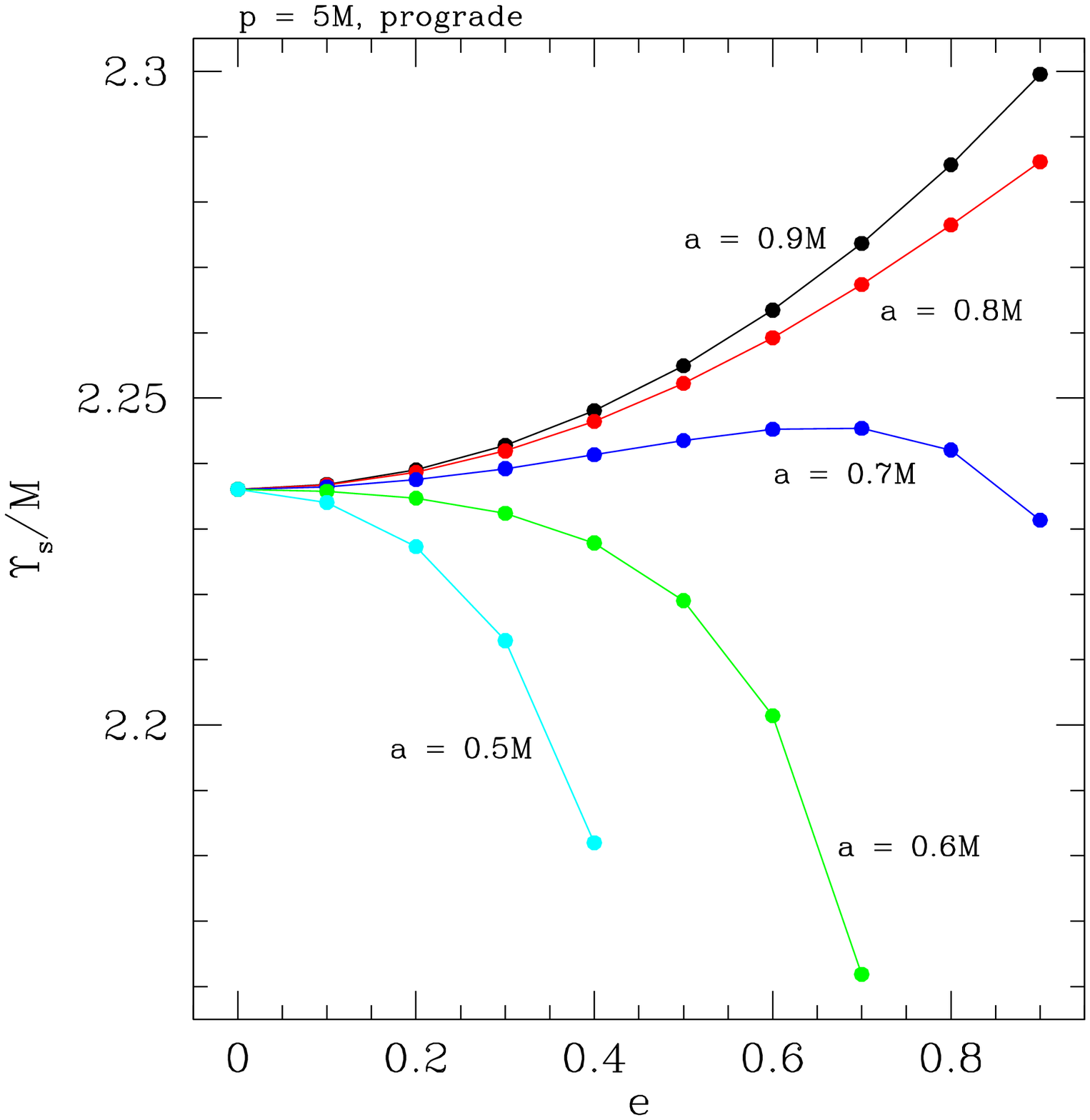}
\caption{Spin precession frequency $\Upsilon_s$ for prograde eccentric
  equatorial orbits with $p = 5M$ as a function of eccentricity $e$.
  We see considerably more variation in the behavior of $\Upsilon_s$
  here than we did for prograde orbits at $p = 10M$ (left-hand panel
  of Fig.\ {\ref{fig:p10_ecc}}).  The trend we find supports the idea
  that $\Upsilon_s$ decreases for orbits that come close to the last
  stable orbit: For prograde orbits at $p = 5M$ $e_{\rm LSO} \simeq
  0.796$ at $a = 0.6M$, and $e_{\rm LSO} \simeq 0.503$ at $a = 0.5M$.
  (For $a = 0.7M$, $ p = 5M$ is outside the separatrix between stable
  and unstable orbits for all $e$, though it is close to the
  separatrix as $e \to 1$.)}
\label{fig:p5_ecc_pro}
\end{figure}

Figures {\ref{fig:p10_ecc}} and {\ref{fig:p5_ecc_pro}} show the
variation of $\Upsilon_s$ with eccentricity $e$ for several examples
of orbits of constant semi-latus rectum $p$, for several spins.  The
general trend we see is that $\Upsilon_s$ monotonically increases from
$\Upsilon_s(e = 0) = \sqrt{pM}$ when $p$ is large.  This is apparent
in the left-hand (prograde) panel of Fig.\ {\ref{fig:p10_ecc}}, the $a
= 0.1M$ curve in the right-hand (retrograde) panel of
Fig.\ {\ref{fig:p10_ecc}}, and in the $a = 0.9M$ and $a = 0.8M$ curves
of Fig.\ {\ref{fig:p5_ecc_pro}}.

The behavior we find deviates from this tendency as orbits approach
the last stable orbit (LSO), which marks the boundary between stable
and unstable orbits.  For prograde orbits at $p = 5M$, the LSO is at
$e_{\rm LSO} = 0.503$ for $a = 0.5M$, and at $e_{\rm LSO} = 0.796$ for
$a = 0.6M$.  All prograde $p = 5M$ orbits are stable for $a = 0.7M$,
though the LSO is close to these orbits: $p_{\rm LSO} \to 4.79M$ as $e
\to 1$, not far beyond the large $e$ portion of the $a = 0.7M$ curve.
The turnaround we see in the $a = 0.9M$ and $a = 0.5M$ curves in the
right-hand panel of Fig.\ {\ref{fig:p10_ecc}} appears to be the same
phenomenon.  For retrograde orbits at $p = 10M$, $e_{\rm LSO} = 0.469$
at $a = 0.9M$.  All retrograde orbits at $p = 10M$ are stable for $a =
0.5M$, though this curve lies close to the LSO: $p_{\rm LSO} = 9.90M$
as $e \to 1$.

\subsection{Generic orbits}
\label{sec:precessgeneric}

Finally, consider generic orbits, which include both radial and polar
motions.  We now need to solve Eq.\ (\ref{eq:masterprec}) including
both $r$ and $\theta$ frequencies.  To do so, we begin by assuming
that there exist solutions $\mathbf{S}^a$ of the form
\begin{equation}
\mathbf{S}^a = \sum_{k'=-\infty}^\infty\sum_{n' = -\infty}^\infty
e^{-i\Upsilon^a_s\lambda}{\mathbf{S}^a}_{k'n'} e^{-i(k'\Upsilon_\theta
  + n'\Upsilon_r)\lambda}\;.
\label{eq:eigenspin_2freqs}
\end{equation}
\begin{widetext}
With this, Eq.\ (\ref{eq:masterprec}) becomes
\begin{equation}
-i\sum_{k';n' = -\infty}^\infty(\Upsilon^a_s + k'\Upsilon_\theta +
n'\Upsilon_r){\mathbf{S}^a}_{k'n'} e^{-i\Upsilon^a_s\lambda}
e^{-i(k'\Upsilon_\theta + n'\Upsilon_r)\lambda} = \sum_{k,k';n,n' =
  -\infty}^\infty\mathbf{P}_{kn}\cdot{\mathbf{S}^a}_{k';n'}
e^{-i\Upsilon^a_s\lambda} e^{-i[(k+k')\Upsilon_\theta +
    (n+n')\Upsilon_r]\lambda}\;.
\end{equation}
As in Sec.\ {\ref{sec:precessonefreq}}, we find a common factor of
$e^{-i\Upsilon^a_s\lambda}$ which cancels.  To go further, we use a
trick adapted from classical mechanics {\cite{dh04}} for analyzing
biperiodic functions: on both sides, change $\Upsilon_\theta\lambda$
to $\Upsilon_\theta\lambda_\theta$ and $\Upsilon_r\lambda$ to
$\Upsilon_r\lambda_r$; multiply by
$e^{ip\Upsilon_\theta\lambda_\theta} e^{iq\Upsilon_r\lambda_r}$;
integrate both $\lambda_\theta$ and $\lambda_r$ over one full period.
The result is
\begin{equation}
-i\sum_{k'n' = -\infty}^\infty(\Upsilon^a_s + k'\Upsilon_\theta +
n'\Upsilon_r) {\mathbf{S}^a}_{k'n'}\delta_{k',p}\,\delta_{n',q} =
\sum_{k,k';n,n' =
  -\infty}^\infty\mathbf{P}_{kn}\cdot{\mathbf{S}^a}_{k'n'}\delta_{(k+k'),p}\,
\delta_{(n+n'),q}\;.
\end{equation}


Performing the sums over $k'$ and $n'$, this simplifies to
\begin{equation}
-i(\Upsilon^a_s + p\Upsilon_\theta + q\Upsilon_r){\mathbf{S}^a}_{pq} =
\sum_{k;n = -\infty}^\infty
\mathbf{P}_{kn}\cdot{\mathbf{S}^a}_{(p-k)(q-n)}\;.
\end{equation}

As in the previous analysis, truncate these sums at $\pm k_{\rm max}$
and $\pm n_{\rm max}$, then expand the equation and rearrange.  If
written out directly, the rather cumbersome result would be difficult
to analyze.  We again clean things up substantially by nesting
matrices, though we will need to take things one layer ``deeper''
since we have a second dynamical frequency in our problem.  Let us
first define
\begin{equation}
{\mathbb{S}^a}_k =
\begin{pmatrix}
{\mathbf{S}^a}_{k,-n_{\rm max}} \cr
{\mathbf{S}^a}_{k,-n_{\rm max} + 1} \cr
\vdots \cr
{\mathbf{S}^a}_{k,-1} \cr
{\mathbf{S}^a}_{k,0} \cr
{\mathbf{S}^a}_{k,1} \cr
\vdots \cr
{\mathbf{S}^a}_{k,n_{\rm max} - 1} \cr
{\mathbf{S}^a}_{k,n_{\rm max}}
\end{pmatrix}\;.
\end{equation}
Next define the matrix $\mathbb{P}_k$ whose elements are given by
\begin{equation}
\mathbb{P}_{k,gh} =
\left\{
\begin{matrix}
\mathbf{P}_{k,g - h}\qquad g \ne h\;,
\\
\mathbf{P}_{k,0} + ig\Upsilon_r\mathbf{I}\qquad g = h\;.
\end{matrix}
\right.
\end{equation}
Recall that $\mathbf{I}$ is the $3\times3$ identity matrix.

We then define
\begin{equation}
\mathcal{S}^a =
\begin{pmatrix}
{\mathbb{S}^a}_{-k_{\rm max}} \cr
{\mathbb{S}^a}_{-k_{\rm max} + 1} \cr
\vdots \cr
{\mathbb{S}^a}_{-1} \cr
{\mathbb{S}^a}_{0} \cr
{\mathbb{S}^a}_{1} \cr
\vdots \cr
{\mathbb{S}^a}_{k_{\rm max} - 1} \cr
{\mathbb{S}^a}_{k_{\rm max}}
\end{pmatrix}\;,
\end{equation}
and a matrix ${\cal P}$ whose elements are given by
\begin{equation}
\mathcal{P}_{cd} =
\left\{
\begin{matrix}
\mathbb{P}_{c - d}\qquad c \ne d\;,
\\
\mathbb{P}_{0} + id\Upsilon_\theta\mathbb{I}\qquad c = d\;.
\end{matrix}
\right.
\end{equation}
The matrix $\mathbb{I}$ is the $(2n_{\rm max} + 1)\times(2n_{\rm
  max}+1)$ identity matrix.

As a concrete example, imagine that we truncate our sums at $k_{\rm
  max} = 2$, $n_{\rm max} = 3$.  The vector ${\mathbb{S}^a}_k$ has
$2n_{\rm max} + 1 = 7$ elements, and is given by
\begin{equation}
{\mathbb{S}^a}_k =
\begin{pmatrix}
{\mathbf{S}^a}_{k,-3} \cr
{\mathbf{S}^a}_{k,-2} \cr
{\mathbf{S}^a}_{k,-1} \cr
{\mathbf{S}^a}_{k,0} \cr
{\mathbf{S}^a}_{k,1} \cr
{\mathbf{S}^a}_{k,2} \cr
{\mathbf{S}^a}_{k,3}
\end{pmatrix}\;.
\end{equation}


The matrix $\mathbb{P}_k$ has $(2n_{\rm max} + 1)^2 = 49$ elements,
and is given by
\begin{equation}
\mathbb{P}_k = 
\begin{pmatrix}
\left(\mathbf{P}_{k,0} - 3i\Upsilon_r\mathbf{I}\right) &
\mathbf{P}_{k,-1} & \mathbf{P}_{k,-2} & \mathbf{P}_{k,-3} 
& \mathbf{P}_{k,-4} & \mathbf{P}_{k,-5} & \mathbf{P}_{k,-6} \cr
\mathbf{P}_{k,1} & \left(\mathbf{P}_{k,0} -
2i\Upsilon_r\mathbf{I}\right) & \mathbf{P}_{k,-1} & \mathbf{P}_{k,-2}
& \mathbf{P}_{k,-3} & \mathbf{P}_{k,-4} & \mathbf{P}_{k,-5} \cr
\mathbf{P}_{k,2} & \mathbf{P}_{k,1} & \left(\mathbf{P}_{k,0} -
i\Upsilon_r\mathbf{I}\right) & \mathbf{P}_{k,-1} &
\mathbf{P}_{k,-2} & \mathbf{P}_{k,-3} & \mathbf{P}_{k,-4} \cr
\mathbf{P}_{k,3} & \mathbf{P}_{k,2} & \mathbf{P}_{k,1} &
\mathbf{P}_{k,0} & \mathbf{P}_{k,-1} & \mathbf{P}_{k,-2} &
\mathbf{P}_{k,-3}\cr
\mathbf{P}_{k,4} & \mathbf{P}_{k,3} & \mathbf{P}_{k,2} &
\mathbf{P}_{k,1} & \left(\mathbf{P}_{k,0} +
i\Upsilon_r\mathbf{I}\right) & \mathbf{P}_{k,-1} &
\mathbf{P}_{k,-2}\cr
\mathbf{P}_{k,5} & \mathbf{P}_{k,4} & \mathbf{P}_{k,3} &
\mathbf{P}_{k,2} & \mathbf{P}_{k,1} & \left(\mathbf{P}_{k,0} +
2i\Upsilon_r\mathbf{I}\right) & \mathbf{P}_{k,-1}\cr
\mathbf{P}_{k,6} & \mathbf{P}_{k,5} & \mathbf{P}_{k,4} &
\mathbf{P}_{k,3} & \mathbf{P}_{k,2} & \mathbf{P}_{k,1} &
 \left(\mathbf{P}_{k,0} + 3i\Upsilon_r\mathbf{I}\right)\cr
\end{pmatrix}\;.
\end{equation}

Next, the vector $\mathcal{S}^a$ has $2k_{\rm max} + 1 = 5$ elements:
\begin{equation}
\mathcal{S}^a =
\begin{pmatrix}
{\mathbb{S}^a}_{-2} \cr
{\mathbb{S}^a}_{-1} \cr
{\mathbb{S}^a}_{0} \cr
{\mathbb{S}^a}_{1} \cr
{\mathbb{S}^a}_{2} \cr
\end{pmatrix}\;.
\end{equation}
The matrix $\mathcal{P}$ has $(2k_{\rm max} + 1)^2 = 25$ elements:
\begin{equation}
\mathcal{P} =
\begin{pmatrix}
\left(\mathbb{P}_0 - 2i\Upsilon_\theta\mathbb{I}\right) &
\mathbb{P}_{-1} & \mathbb{P}_{-2} & \mathbb{P}_{-3} &\mathbb{P}_{-4}
\cr
\mathbb{P}_1 & \left(\mathbb{P}_0 - i\Upsilon_\theta\mathbb{I}\right)
& \mathbb{P}_{-1} & \mathbb{P}_{-2} &\mathbb{P}_{-3}\cr
\mathbb{P}_2 & \mathbb{P}_1 & \mathbb{P}_0 &
\mathbb{P}_{-1} & \mathbb{P}_{-2}\cr
\mathbb{P}_3 & \mathbb{P}_2 & \mathbb{P}_1 & \left(\mathbb{P}_0 +
i\Upsilon_\theta\mathbb{I}\right) & \mathbb{P}_{-1}\cr
\mathbb{P}_4 & \mathbb{P}_3 & \mathbb{P}_2 & \mathbb{P}_1 &
\left(\mathbb{P}_0 + 2i\Upsilon_\theta\mathbb{I}\right)
\end{pmatrix}\;.
\end{equation}
\end{widetext}

Finally, to solve for the precessional motion of our spinning body, we
must find the eigenvalues and eigenvectors of the matrix equation
\begin{equation}
\mathcal{P}\cdot\mathcal{S}^a = -i\Upsilon^a_s\mathcal{S}^a\;.
\label{eq:eigensystem_2freqs}
\end{equation}
Doing so, we now find an even greater abundance of eigensolutions:
Eq.\ (\ref{eq:eigensystem_2freqs}) yields $3\times(2k_{\rm max} +
1)\times(2n_{\rm max} + 1)$ eigenvalues and eigenvectors.  This
originates from essentially the same relabeling ambiguity that was
responsible for the extra eigensolutions we found for
Eq.\ (\ref{eq:eigensystem_1freq}), but with an additional order of
``extraness'' arising from the additional frequency associated with
the underlying geodesic.  Starting with
Eq.\ (\ref{eq:eigenspin_2freqs}), we can shift the sum over $k'$ by
$\Delta k'$ and the sum over $n'$ by $\Delta n'$, provided we shift
the eigenvalue by
\begin{equation}
\Upsilon^a_s \to \Upsilon^a_s + \Delta k'\Upsilon_\theta + \Delta
n'\Upsilon_r\;,
\end{equation}
and correspondingly shift the components of the eigenvectors.

Bearing this in mind, we find that the eigensolutions can be organized
into three groups.  As we discussed in
Sec.\ {\ref{sec:precessonefreq}}, the relabeling ambiguity means that
the multiple solutions are in principle equivalent to one another,
provided that the eigenvector indices are properly shifted.  In
practice, we find the most accurate solution corresponds to the
eigenvector set $\Upsilon^1_s = -\Upsilon^{-1}_s \equiv \Upsilon_s$,
$\Upsilon_0 = 0$.  From the associated eigenvectors
$(\mathcal{S}^{-1},\mathcal{S}^0,\mathcal{S}^1)$, we extract the
Fourier components $\mathbf{S}_{kn}$, and finally assemble the
solution for spin precession along a generic orbit:
\begin{eqnarray}
\mathbf{S} &=& \sum_{k,n}\left(c_0{\mathbf{S}^0}_{kn} +
c_{-1}{\mathbf{S}^{-1}}_{kn}e^{i\Upsilon_s\lambda} +
c_0{\mathbf{S}^0}_{kn}e^{-i\Upsilon_s\lambda}\right)
\nonumber\\
& &\qquad \times e^{-i(k\Upsilon_\theta + n\Upsilon_r)\lambda}\;.
\label{eq:spinprecgeneric}
\end{eqnarray}
These sums run from $-k_{\rm max}$ to $k_{\rm max}$, and from $-n_{\rm
  max}$ to $n_{\rm max}$.

\begin{figure*}[ht]
\includegraphics[width = 0.48\textwidth]{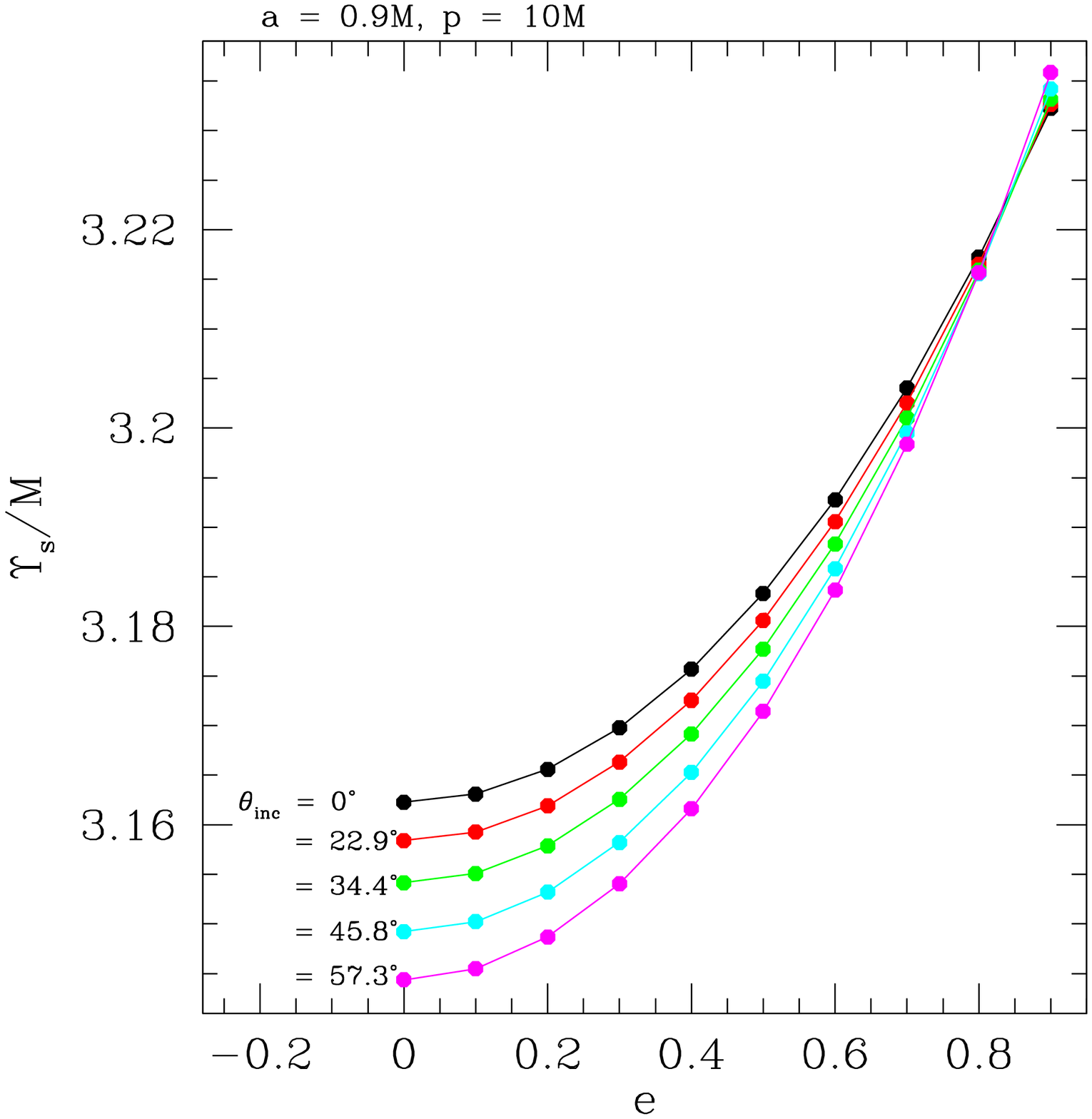}
\includegraphics[width = 0.48\textwidth]{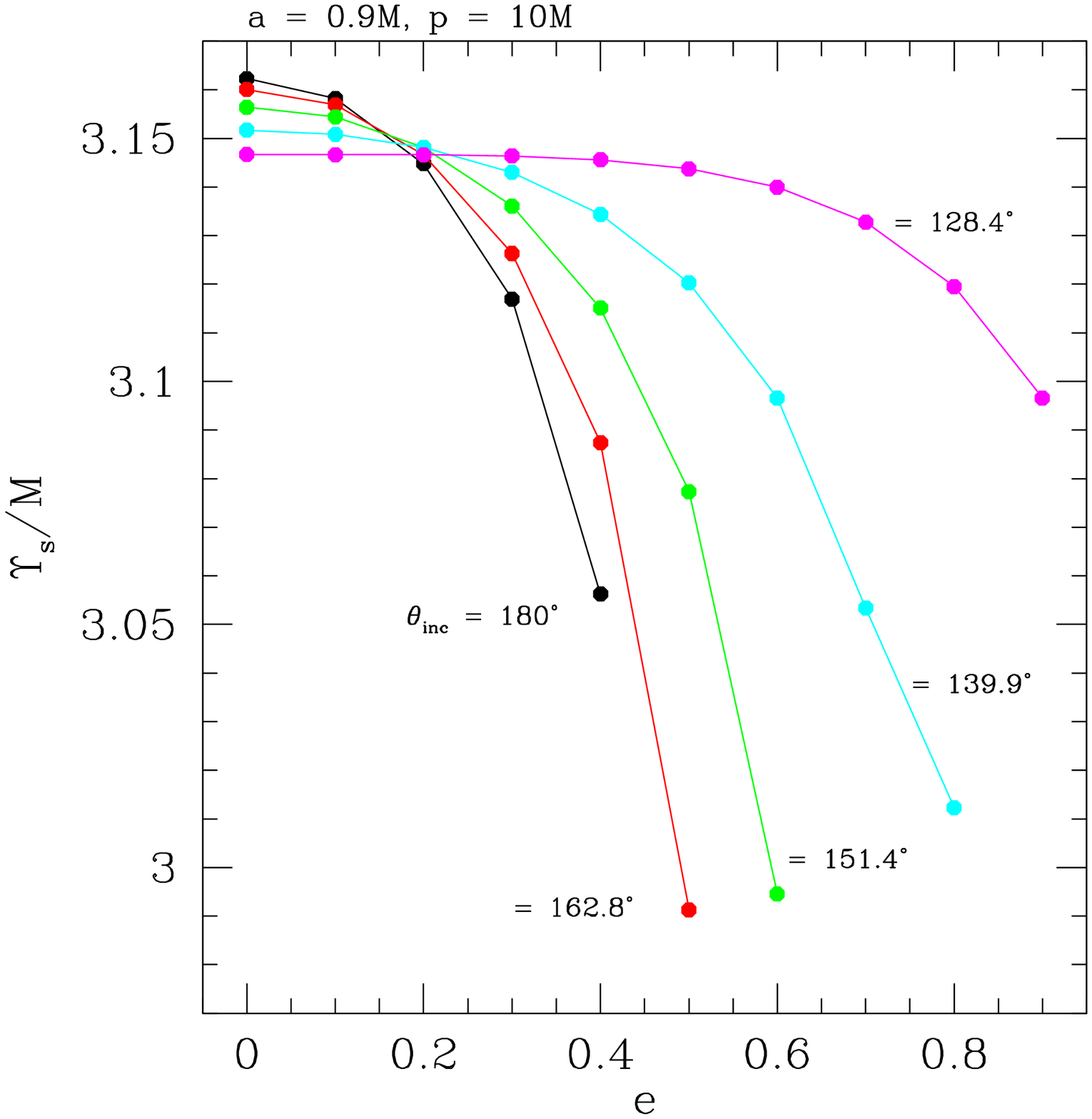}
\caption{Example of the spin precession frequency $\Upsilon_s$ for
  generic orbits.  All examples are for orbits of a black hole with $a
  = 0.9M$, and have $p = 10M$.  We plot $\Upsilon_s$ versus
  eccentricity at specified inclinations; left-hand panel shows
  prograde orbits ($\theta_{\rm inc} < 90^\circ$), right shows
  retrograde ($\theta_{\rm inc} > 90^\circ$).  We fixed $k_{\rm max} =
  5$ for this initial analysis.  With this choice of $k_{\rm max}$, we
  find empirically that we can accurately compute $\Upsilon_s$ for
  $\theta_{\rm inc} \lesssim 60^\circ$ and $\theta_{\rm inc} \gtrsim
  120^\circ$.  The behavior we see combines trends we see in both the
  equatorial eccentric and the inclined circular cases.  For the
  prograde orbits, we see that $\Upsilon_s$ decreases with
  $\theta_{\rm inc}$ at fixed $e$ (similar to what we see in
  Fig.\ {\ref{fig:r10_circ}}), but increases with $e$ at fixed
  $\theta_{\rm inc}$ (similar to what we see in the left-hand panel of
  Fig.\ {\ref{fig:p10_ecc}}).  Interestingly, for prograde orbits
  $\Upsilon_s$ varies much less with $\theta_{\rm inc}$ as $e$ becomes
  large.  For the retrograde cases, we see the strong influence of
  proximity to the last stable orbit (LSO), much as we saw in the
  right-hand panel of Fig.\ {\ref{fig:p10_ecc}} and in
  Fig.\ {\ref{fig:p5_ecc_pro}}.}
\label{fig:gen_UpS}
\end{figure*}

Figure {\ref{fig:gen_UpS}} shows an example of how $\Upsilon_s$ varies
with $e$ and $\theta_{\rm inc}$ for a sample of generic orbits.  We
consider orbits with $p = 10M$ about a black hole with spin $a =
0.9M$.  We have fixed $k_{\rm max} = 5$ and $n_{\rm max} = 5$.  The
restriction on $k$ means that we can only examine orbits with
$\theta_{\rm inc} \lesssim 60^\circ$ and $\theta_{\rm inc} \gtrsim
120^\circ$.  Although no issue of principle prevents us from
increasing $k_{\rm max}$ and examining more highly inclined orbits, we
hold to $k_{\rm max} = 5$ in order to keep the precession matrix small
and our analyses simple for this initial exploration.

The trends we see in Fig.\ {\ref{fig:gen_UpS}} combine the trends we
see for circular inclined and equatorial eccentric orbits.  In
particular, we see that $\Upsilon_s$ tends to increase with $e$ as
other parameters are held fixed; it tends to decrease with
$\theta_{\rm inc}$ with all other parameters fixed; and tends to
decrease sharply as the LSO is approached.  This last point is
particularly visible in the right-hand panel of
Fig.\ {\ref{fig:gen_UpS}}, which shows data for retrograde orbits
($\theta_{\rm inc} > 90^\circ$).  Each trend line we plot is
terminated near\footnote{For $p = 10M$ and $a = 0.9M$, we have $e_{\rm
    LSO} = 0.469$ at $\theta_{\rm inc} = 180^\circ$; $e_{\rm LSO} =
  0.534$ at $\theta_{\rm inc} = 162.8^\circ$; $e_{\rm LSO} = 0.654$ at
  $\theta_{\rm inc} = 151.4^\circ$; and $e_{\rm LSO} = 0.850$ at
  $\theta_{\rm inc} = 139.9^\circ$.  Stable orbits exist all the way
  to $e = 1$ at $\theta_{\rm inc} = 128.4^\circ$, the final data set
  included in this plot.} the LSO.  We see $\Upsilon_s$ sharply
decreasing in all of the cases we show here as we move to large $e$
and approach the LSO.

One new feature of the general case is apparent in the left-hand panel
of Fig.\ {\ref{fig:gen_UpS}}, which shows data for prograde orbits
($\theta_{\rm inc} < 90^\circ$): $\Upsilon_s$ varies much more weakly
with $\theta_{\rm inc}$ as $e \to 1$.  For $e$ near $1$, much of the
orbit is spent at large radius, where the spacetime is nearly
spherical, and $\Upsilon_s$ does not depend on $\theta_{\rm inc}$ in
the spherically symmetric limit.  This behavior suggests that it may
be useful to examine the limit $e \to 1$ (for which $\hat E^{\rm G} =
1$).  One might find a simple solution for $\Upsilon_s$ in this case,
much as we found $\Upsilon_s = \sqrt{pM}$ for circular and equatorial
orbits.  If such a solution exists, it may be a useful constraint for
computing $\Upsilon_s$ more generally.

\section{The spin-curvature force in the frequency domain}
\label{sec:force}

We now move to a discussion of the spin curvature force.  We begin
with some notation: we denote by $f^\alpha$ forces defined with
respect a trajectory's proper time $\tau$.  The force corresponding to
the spin-curvature interaction is given by
\begin{equation}
f_{\rm S}^\alpha = -\frac{1}{2}{R^\alpha}_{\nu\lambda\sigma}u^\nu_{\rm
  G} S^{\lambda\sigma}\;.
\label{eq:spinforcetau}
\end{equation}
This is just a rewriting of Eq.\ (\ref{eq:force_lin}).

Since we find it very useful to use Mino time $\lambda$ as our
independent parameter, it is very useful to also define a force
defined with respect to $\lambda$.  Let us put
\begin{equation}
F^\alpha \equiv \left(\frac{d\tau}{d\lambda}\right)f^\alpha
= \Sigma f^\alpha\;.
\label{eq:forcedeflambda}
\end{equation}
Combining Eqs.\ (\ref{eq:spinforcetau}) and (\ref{eq:forcedeflambda}), we
have
\begin{equation}
F^\alpha_{\rm S} = -\frac{1}{2}{R^\alpha}_{\nu\lambda\sigma}U^\nu_{\rm
  G}S^{\lambda\sigma}\;,
\label{eq:spinforcelambda}
\end{equation}
where $U^\alpha_{\rm G} \equiv dx^\alpha/d\lambda = \Sigma
u^\alpha_{\rm G}$.  Using Eq.\ (\ref{eq:spinvecinvert}), which relates
the spin tensor $S^{\alpha\beta}$ to the spin vector $S^\alpha$, plus
the rule
\begin{equation}
S^t = -\frac{u^{\rm G}_j}{u^{\rm G}_t}S^j = -\frac{U^{\rm G}_j}{U^{\rm
    G}_t}S^j\;,
\end{equation}
which follows from the constraint $u^{\rm G}_\alpha S^\alpha = 0$ and
from Eq.\ (\ref{eq:spinvecinvert}), Eq.\ (\ref{eq:spinforcelambda}) can be
written
\begin{equation}
F^\alpha_{\rm S} = {C^\alpha}_j S^j\;.
\end{equation}
The components of the $4 \times 3$ curvature coupling matrix
$\mathbf{C}$ are given by
\begin{equation}
{C^\alpha}_j = -\frac{1}{2\Sigma}{R^\alpha}_{\nu\lambda\sigma}
\epsilon^{\lambda\sigma\beta\gamma}p_{\gamma j}U^{\rm G}_\beta U^\nu_{\rm G}\;,
\end{equation}
with
\begin{equation}
p_{\gamma j} = g_{\gamma j} - g_{\gamma t}\frac{U^{\rm G}_j}{U^{\rm G}_t}\;.
\end{equation}

We now introduce the Fourier expansion.  Each matrix element
${C^\alpha}_j$ can be expanded using Eqs.\ (\ref{eq:genfourier}) and
(\ref{eq:genfourier_amp}).  We further know that the spin vector can
be written as (\ref{eq:spinprecgeneric}).  Putting all of this
together, the force components can be written
\begin{equation}
F^\alpha_{\rm S} = \sum_{j = -1}^1\sum_{k,n = -\infty}^\infty
\left(F^\alpha_{\rm S}\right)_{jkn}
e^{-ij\Upsilon_s\lambda}e^{-i(k\Upsilon_\theta +
  n\Upsilon_r)\lambda}\;.
\end{equation}
For presenting our results, it is useful to project $F^\alpha_{\rm S}$
onto the Killing vectors and tensors: we define
\begin{eqnarray}
\frac{dE^{\rm G}}{d\lambda} &\equiv& -F^\alpha_{\rm S}\xi^t_\alpha =
-F_t^{\rm S}\;,
\label{eq:dEdlambda}
\\
\frac{dL_z^{\rm G}}{d\lambda} &\equiv& F^\alpha_{\rm S}\xi^\phi_\alpha =
F_\phi^{\rm S}\;,
\label{eq:dLdlambda}
\\
\frac{dK^{\rm G}}{d\lambda} &\equiv& 2 K_{\alpha\beta}u^\alpha_{\rm G}
F^\beta_{\rm S}\;.
\label{eq:dKdlambda}
\end{eqnarray}
As the small body moves along its orbit, its geodesic energy, axial
angular momentum, and Carter constant will each vary according to
Eqs.\ (\ref{eq:dEdlambda})--(\ref{eq:dKdlambda}) due to the
spin-curvature force.

These quantities can likewise be Fourier expanded: for $\mathcal{C}
\in (E^{\rm G}, L^{\rm G}_z, K^{\rm G})$, we have
\begin{equation}
\frac{d\mathcal{C}}{d\lambda} = \sum_{j = -1}^1 \sum_{k,n =
  -\infty}^\infty \left(\frac{d\mathcal{C}}{d\lambda}\right)_{jkn}
e^{-ij\Upsilon_s\lambda} e^{-i(k\Upsilon_\theta + n\Upsilon_r)}\;.
\label{eq:forcefourier}
\end{equation}
With one exception, we use the expansion (\ref{eq:forcefourier}) of
the components (\ref{eq:dEdlambda})--(\ref{eq:dKdlambda}) in our
detailed discussion of the spin force components that we study in
Secs.\ {\ref{sec:results}} and {\ref{sec:converge}}.  The exception is
for the radial component of the spin force, $F^r_{\rm S}$.  It is not
difficult to see that $F^r_{\rm S}$ completely\footnote{There are two
  ways to write the tensor $K_{\alpha\beta}$.  In one way, $dK^{\rm
    G}/d\lambda$ is independent of $F^r_{\rm S}$, but involves the
  other three components; the other way, it is independent of
  $F^\theta_{\rm S}$ and involves the other three.  In either case,
  one of the force components decouples from the projection of
  $F^\alpha_{\rm S}$ onto the Kerr metric's Killing quantities.}
decouples from Eqs.\ (\ref{eq:dEdlambda})--(\ref{eq:dKdlambda}).
Since it nonetheless includes important information, we examine it
along with the components $dE^{\rm G}/d\lambda$, $dL_z^{\rm
  G}/d\lambda$, and $dK^{\rm G}/d\lambda$.

Before beginning our analysis, it is useful to examine how the
quantities we will study scale with the small body's mass $\mu$ and
the black hole's mass $M$.  Begin with the spin components themselves.
As already discussed, the spin magnitude $S \equiv s\mu^2$, with $s
\le 1$ a dimensionless parameter.  We also know that
\begin{equation}
S^2 = g_{\alpha\beta} S^\alpha S^\beta\;.
\end{equation}
Consider now a point at some coordinate $r = A M$ (where $A$ is some
positive number).  Noting how the different metric components scale
with black hole mass $M$ at this coordinate, we infer that
\begin{eqnarray}
S^r &\sim& \mu^2\;,
\nonumber\\
S^{\theta,\phi} &\sim& \mu^2/M\;.
\end{eqnarray}
Next examine the scaling of the various quantities which enter into
the two forms of the spin-curvature force,
Eqs.\ (\ref{eq:spinforcetau}) and (\ref{eq:spinforcelambda}).  From
the behavior of the Riemann tensor, the components of the 4-velocity,
the spin tensor, and the factor $d\tau/d\lambda$, we see that
\begin{equation}
f^{t,r}_S \sim \frac{s\mu^2}{M^2}\;,\quad
f^{\theta,\phi}_S \sim \frac{s\mu^2}{M^3}\;\;,
\end{equation}
\begin{equation}
F^{t,r}_S \sim s \mu^2\;,\quad
F^{\theta,\phi}_S \sim \frac{s\mu^2}{M}\;.
\end{equation}
Projecting onto the Killing vectors and Killing tensor to assemble the
rates of change of geodesic energy, angular momentum, and Carter
constant associated with the spin force, we find
\begin{equation}
\frac{dE^{\rm G}}{d\lambda} \sim s\mu^2\;,\quad
\frac{dL^{\rm G}_z}{d\lambda} \sim s\mu^2M\;,\quad
\frac{dK^{\rm G}}{d\lambda} \sim s\mu^3M^2\;.
\end{equation}
In the figures that follow, we divide all of the quantities we plot by
these scaling rules, so that one can easily assess the impact of our
analysis for different masses and different small body spins.

\subsection{Initial conditions for the spin vector}
\label{sec:initialconditions}

Our first step is to select initial conditions for the small body's
spin vector.  We begin by picking initial spatial components as
measured in an orthonormal frame, and then convert the components to a
Boyer-Lindquist coordinate frame.  We emphasize that the orthonormal
frame components are merely a convenient tool for visualizing the
initial spin vector.  An ensemble of spins that have the same
magnitude $S$ but differing orientations will have orthonormal frame
components of similar magnitudes.  By contrast, in the Boyer-Lindquist
coordinate frame, the components' magnitudes will vary strongly as a
function of the orbit's radius $r$ and angle $\theta$.

Let $S^{\hat\jmath}$ be an orthonormal frame component, and let $S^j$
be a coordinate frame component.  Then,
\begin{equation}
S^{\hat r} = \sqrt{g_{rr}} S^r\;,
\quad
S^{\hat\theta} = \sqrt{g_{\theta\theta}}\, S^\theta\;,
\quad
S^{\hat\phi} = \sqrt{g_{\phi\phi}} S^\phi\;.
\label{eq:SjhatSj}
\end{equation}

We must next fix the timelike spin component.  To do so, we enforce
$S^\alpha u^{\rm G}_\alpha = 0$, and find
\begin{equation}
S^t = \frac{u^{\rm G}_i S^i}{\hat E^{\rm G}}\;.
\label{eq:St}
\end{equation}
We used $u^{\rm G}_t = -\hat E^{\rm G}$.  Finally, we have the rule
that $S \equiv \sqrt{S^\alpha S_\alpha} = {\rm constant}$ along the
small body's worldline.

With all this in mind, we use the following algorithm to set initial
conditions on the spin vector:
\begin{itemize}

\item Select initial components in the orthonormal frame.

\item Convert to Boyer-Lindquist coordinate frame components using
  Eq.\ (\ref{eq:SjhatSj}).

\item Compute the initial $S^t$ using Eq.\ (\ref{eq:St}).

\item Increase or decrease all components by whatever factor is needed
  that so that $S$ is some prescribed value.

\end{itemize}

In the results we discuss below, we set $S/\mu^2 = 1$, and choose our
initial components so that $S^{\hat r} = S^{\hat\theta} =
S^{\hat\phi}$.  Orbits that are eccentric begin at peripasis ($\psi =
0$); orbits that are inclined begin at $\theta = \theta_{\rm min}$
($\chi = 0$).

\subsection{Results}
\label{sec:results}

Here we show results for three representative cases.  All are for
orbits about a black hole with spin parameter $a = 0.9M$, but we
consider different orbit geometries in order to explore how precession
and the spin-curvature force behave in these different cases.

\begin{itemize}

\item {\it Equatorial eccentric}: We examine an orbit with $p = 5M$,
  $e = 0.7$, $\theta_{\rm inc} = 0^\circ$.  This orbit is
  characterized by a radial frequency $\Upsilon_r = 1.7842M$, and a
  precession frequency $\Upsilon_s = 2.2737M$.

\item {\it Inclined circular}: We examine an orbit with $p = 5M$, $e =
  0$, $\theta_{\rm inc} = 60^\circ$.  This orbit is characterized by a
  polar frequency $\Upsilon_\theta = 2.9230M$, and a precession
  frequency $\Upsilon_s = 2.1833M$.

\item {\it Generic}: We examine an orbit with $p = 8M$, $e = 0.5$,
  $\theta_{\rm inc} = 30^\circ$.  This orbit is characterized by a
  radial frequency $\Upsilon_r = 2.4304M$, a polar frequency
  $\Upsilon_\theta = 3.2367M$, and a precession frequency $\Upsilon_s
  = 2.8429M$.

\end{itemize}

To present our results, we compare frequency-domain expansions for the
evolution of the spin vectors [computed using
  Eq.\ (\ref{eq:spinprecconstrained})] and the spin-curvature force
[computed with Eq.\ (\ref{eq:forcefourier})] to time-domain solutions
computed by directly integrating Eq.\ (\ref{eq:dSdlambda}), and
combining with Eq.\ (\ref{eq:spinforcelambda}).  The time-domain
solutions were found using the {\sc Mathematica} function {\tt
  NDSolve}, which adaptively chooses integration methods to optimize
accuracy.  As we'll discuss below and in Sec.\ {\ref{sec:converge}},
our time- and frequency-domain solutions agree with one another to
within $10^{-6}$ or $10^{-7}$ over many hundred of orbital periods, an
indication that the solutions are quite accurate.  It is worth noting
that in the frequency-domain approach, most of the computational
effort is spent solving for the frequencies, the precession
eigenvectors, and the Fourier-modes of the force components.  Although
we have not yet done a careful study of computational cost, our
present studies indicate that the frequency-domain expansion is likely
to be useful for large-scale studies of spin-enhanced binary motion,
especially if it is practical to precompute and store such quantities
in order for them to be read into a code and used as needed.

\subsubsection{Equatorial orbits}

Equatorial orbits have two particularly nice properties.  First, the
spin component $S^{\hat\theta} = rS^\theta$ is constant along the
background geodesic.  To see this, examine the $\theta$ component of
the precession equation (\ref{eq:precmatrix_gen}):
\begin{equation}
\frac{dS^\theta}{d\lambda} = {P^\theta}_j S^j
= -\frac{U^r_{\rm G}}{r}S^\theta\;.
\end{equation}
We used the equatorial condition ($\theta = \pi/2$, $u^\theta = 0$) to
simplify the general expression for ${P^\theta}_j$.  Use $U_{\rm G}^r
= dr/d\lambda$ and rearrange:
\begin{equation}
S^\theta\frac{dr}{d\lambda} + r\frac{dS^\theta}{d\lambda} = 0\;,
\end{equation}
or
\begin{equation}
\frac{d(rS^\theta)}{d\lambda} = \frac{dS^{\hat\theta}}{d\lambda} =
0\;,
\end{equation}
so $S^{\hat\theta} = \mbox{constant}$.

The second useful property of these orbits is that $K^{\rm G} =
\mbox{constant}$, at least at leading order in $S$.  We initially saw
this empirically, finding $dK^{\rm G}/d\lambda = 0$ in our numerics
for all equatorial configurations.  We then realized it is simple to
prove this analytically, which we do in App.\ {\ref{app:eqCarter}}.

\begin{figure*}[ht]
\includegraphics[width = 0.48\textwidth]{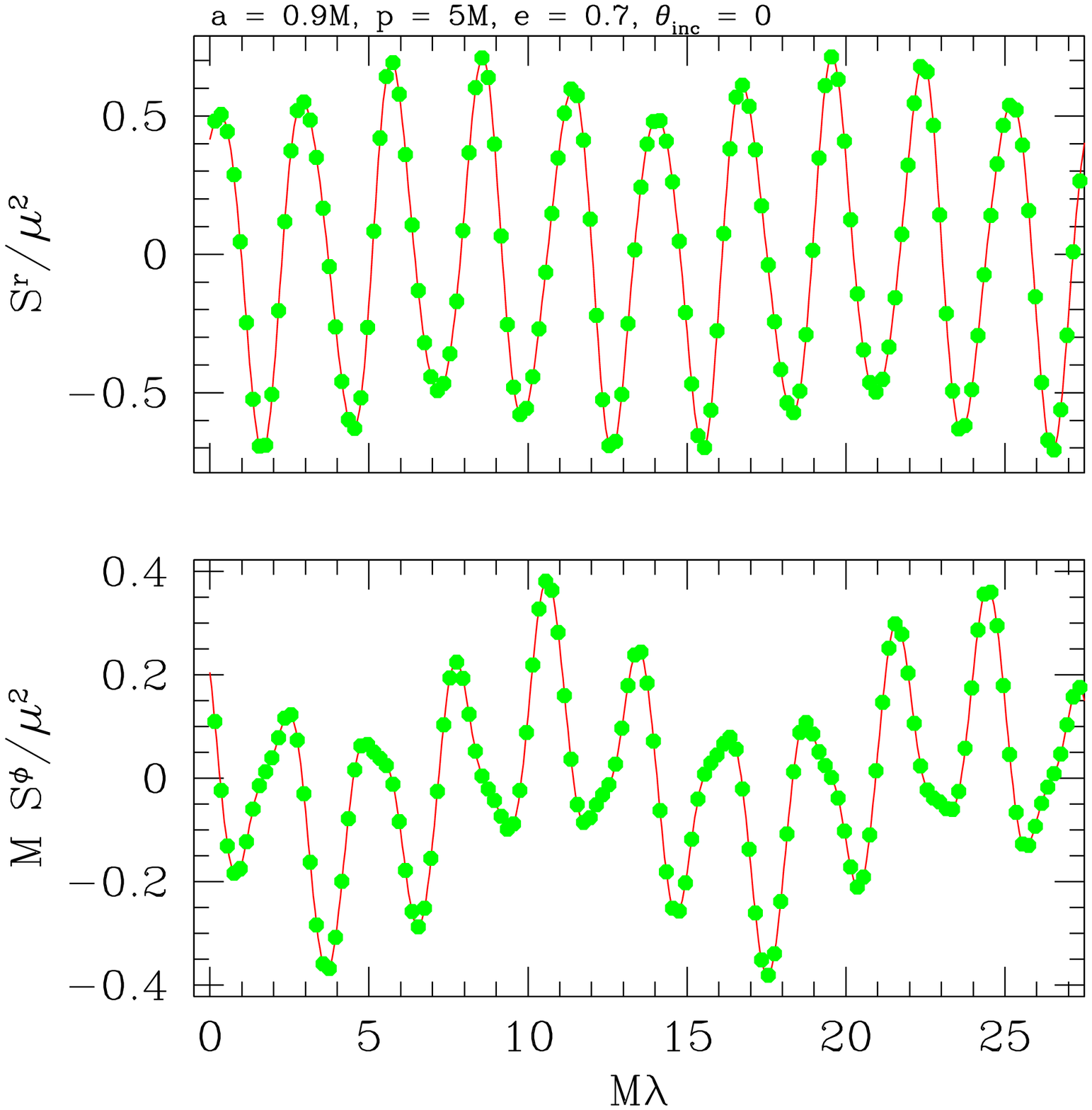}
\includegraphics[width = 0.48\textwidth]{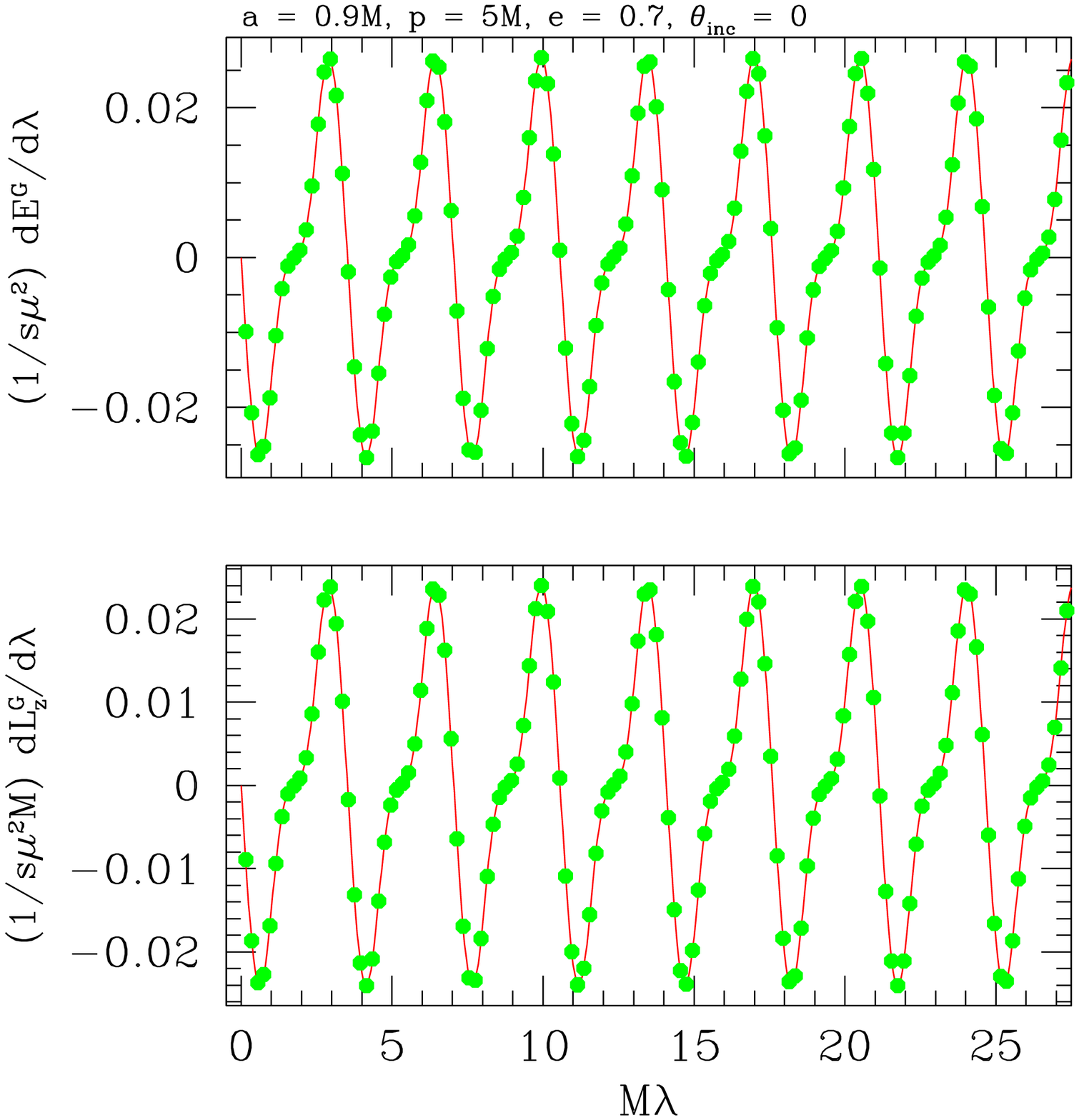}
\caption{Example of spin precession and spin-curvature force for an
  eccentric, equatorial orbit ($a = 0.9M$, $p = 5M$, $e = 0.7$,
  $\theta_{\rm inc} = 0^\circ$).  In the left-hand panel, the solid
  (red) curves show the components of ${\bf S}(\lambda)$ reconstructed
  from the frequency-domain expansion (\ref{eq:spinprecconstrained});
  on the right, the solid (red) curves show the spin-curvature force
  components $dE^{\rm G}/d\lambda$ and $dL^{\rm G}_z/d\lambda$
  reconstructed from the frequency-domain expansion
  (\ref{eq:forcefourier}).  (We don't include the components
  $S^\theta$ or $dK^{\rm G}/d\lambda$; they evolve trivially in this
  case, as discussed in the main text.)  The lowest frequency
  component is a beat between the radial frequency $\Upsilon_r =
  1.7842M$ and the precession frequency $\Upsilon_s = 2.2737M$.  The
  range in $\lambda$ shown is enough for features in the corresponding
  beat period, $2\pi M/(\Upsilon_s - \Upsilon_r) \simeq 12.84$, to be
  seen.  The dots (green) show the same data, but computed by direct
  time-domain integration of the equation of spin precession
  (\ref{eq:dSdlambda}) and direct construction in the time-domain of
  the spin-curvature force (\ref{eq:spinforcelambda}).  The agreement
  between the time- and frequency-domain solutions is outstanding.
  This is typical for equatorial orbits provided the frequency-domain
  expansion is truncated at a sufficiently large value of $n$.  We
  used $n_{\rm max} = 10$ here.  (Note that $S \equiv \sqrt{S^\alpha
    S_\alpha}$ is constant, although this is not apparent from the
  data shown here.)
}
\label{fig:a0.9_p5_e0.7_thi0_results}
\end{figure*}

The constancy of $S^{\hat\theta}$ and $K^{\rm G}$ means that only the
components $S^r$, $S^\phi$, $dE^{\rm G}/d\lambda$, and $dL_z^{\rm
  G}/d\lambda$ are interesting for equatorial orbits.  Figure
{\ref{fig:a0.9_p5_e0.7_thi0_results}} shows the behavior of the spin
components $S^r$ and $S^\phi$, as well as the force components
$dE^{\rm G}/d\lambda$ and $dL^{\rm G}_z/d\lambda$, for the equatorial
eccentric case we examine.  Results for other equatorial cases are
qualitatively similar.  Notice that the frequency-domain solutions
(solid lines) and time-domain solutions (dots) agree quite well.
Detailed analysis shows that the two solutions differ by at most
$10^{-7}$ out to $M\lambda = 500$; see also discussion in
Sec.\ {\ref{sec:converge}}.  This level of agreement is typical for
the equatorial eccentric cases we have examined, provided we include
enough terms in the frequency-domain expansion.  We go to $n = 10$ in
the cases shown in Fig.\ {\ref{fig:a0.9_p5_e0.7_thi0_results}}.

Notice that the harmonic content of the solution is fairly
complicated, especially for $S^\phi$.  One noteworthy feature is a
beat between the radial frequency and the precession frequency, with a
period $\Lambda_{\rm beat} = 2\pi/(\Upsilon_s - \Upsilon_r) \simeq
12.85 M^{-1}$.  We show enough data to see about two full cycles of
this beat.

\subsubsection{Circular and generic orbits}

\begin{figure*}[ht]
\includegraphics[width = 0.48\textwidth]{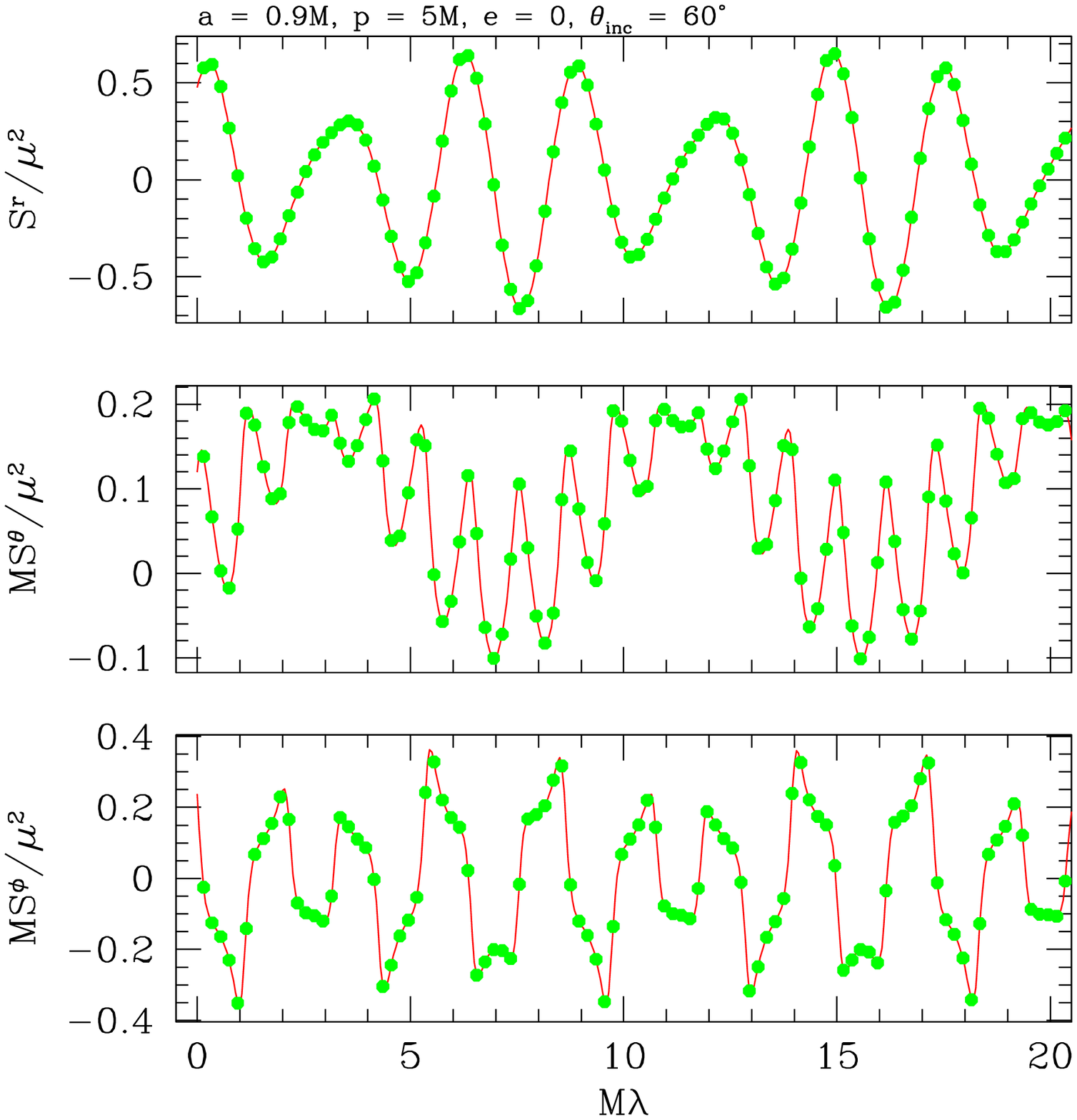}
\includegraphics[width = 0.48\textwidth]{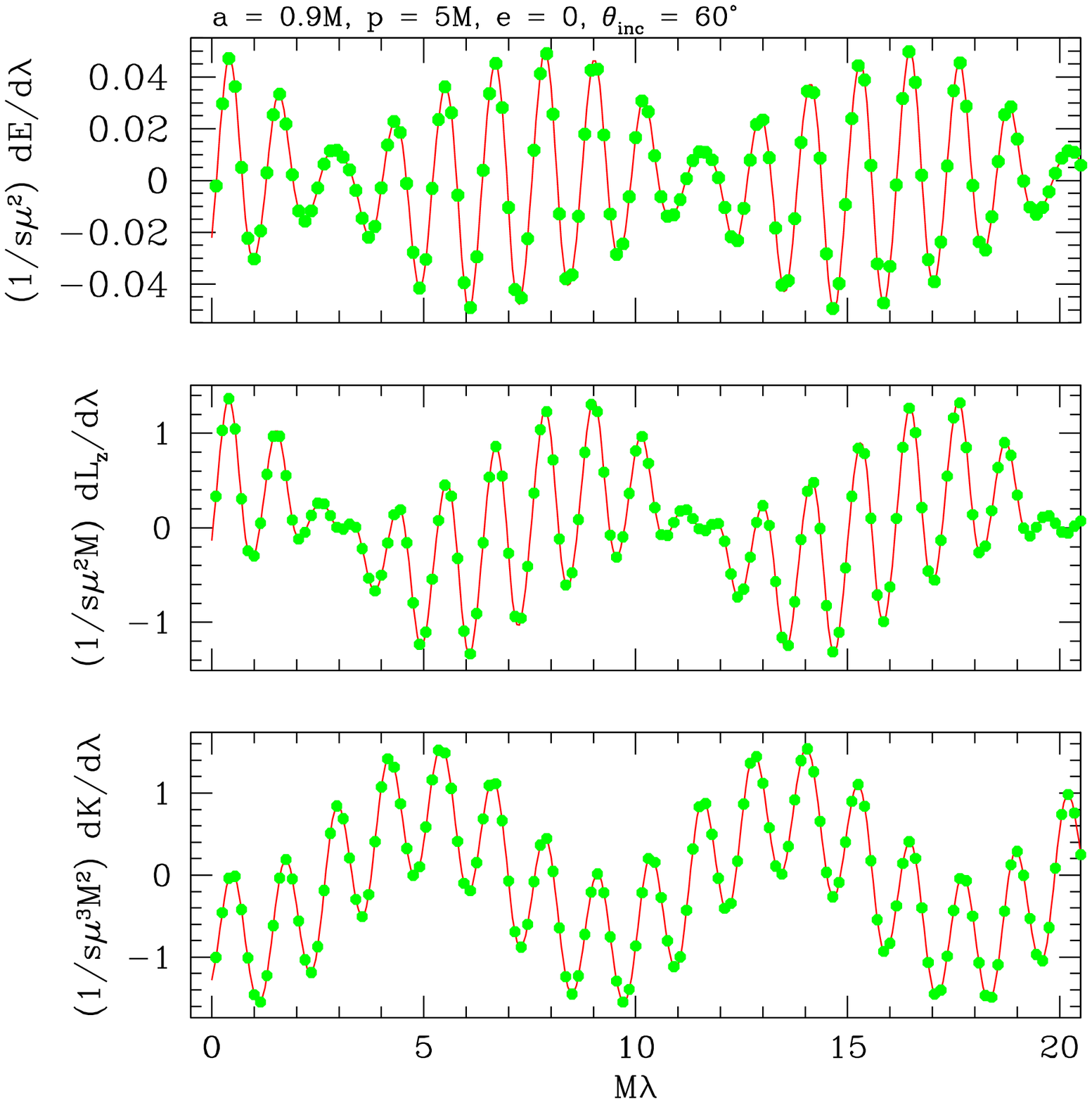}
\caption{Example of spin precession and spin-curvature force for an
  inclined, circular orbit ($a = 0.9M$, $p = 5M$, $e = 0$,
  $\theta_{\rm inc} = 60^\circ$).  As in
  Fig.\ {\ref{fig:a0.9_p5_e0.7_thi0_results}}, the solid (red) curves
  show quantities reconstructed from the frequency-domain expansion.
  Dots (green) show the same data computed by direct time-domain
  integration.  We show the spin-vector components in the left-hand
  panels, and components of the spin-curvature force on the right.
  The lowest frequency component is a beat between the polar frequency
  $\Upsilon_\theta = 2.9230M$ and the precession frequency $\Upsilon_s
  = 2.1833M$; the $\lambda$ range we show is wide enough that features
  in the corresponding period, $2\pi M/(\Upsilon_\theta - \Upsilon_s)
  \simeq 8.49$, can be seen.  We again find outstanding agreement
  between the time- and frequency-domain solutions provided that the
  frequency-domain expansion is truncated at sufficiently large $k$.
  We used $k_{\rm max} = 20$ for the results shown here.}
\label{fig:a0.9_p5_e0_thi60_results}
\end{figure*}

No simplification allows us to disregard components of the spin vector
or the spin-curvature force for circular or generic orbits.  Figure
{\ref{fig:a0.9_p5_e0_thi60_results}} shows the behavior of the three
spin components $S^{r,\theta,\phi}(\lambda)$ and the three force
components $dE^{\rm G}/d\lambda$, $dL_z^{\rm G}/d\lambda$, and
$dK^{\rm G}/d\lambda$ for the inclined circular case.  As in the
equatorial case, we see excellent agreement between the frequency- and
time-domain data in the plot.  We find similar levels of agreement for
other circular inclined cases we have examined, provided we include
enough harmonics.  For this plot, our frequency domain solution
includes terms out to $k_{\rm max} = 20$.  Detailed analysis for this
case shows agreement within $10^{-6}$ for small $\lambda$, drifting to
a disagreement of $\sim 10^{-5}$ at $M \lambda = 500$.  This drift is
due to errors in $\Upsilon_s$, and can be improved by including more
terms in the Fourier expansion; see Sec.\ {\ref{sec:converge}} for
further discussion of this point.  We find that $\Upsilon_s$ is
computed more accurately for smaller values of $\theta_{\rm inc}$.

As in the equatorial case, our solutions have rather ornate harmonic
structure, with complicated beats between the polar frequency and the
precession frequency.  We show enough data to again capture about two
full beat cycles, which have period $\Lambda_{\rm beat} =
2\pi/(\Upsilon_\theta - \Upsilon_s) \simeq 8.49 M^{-1}$ in this case.

\begin{figure*}[ht]
\includegraphics[width = 0.48\textwidth]{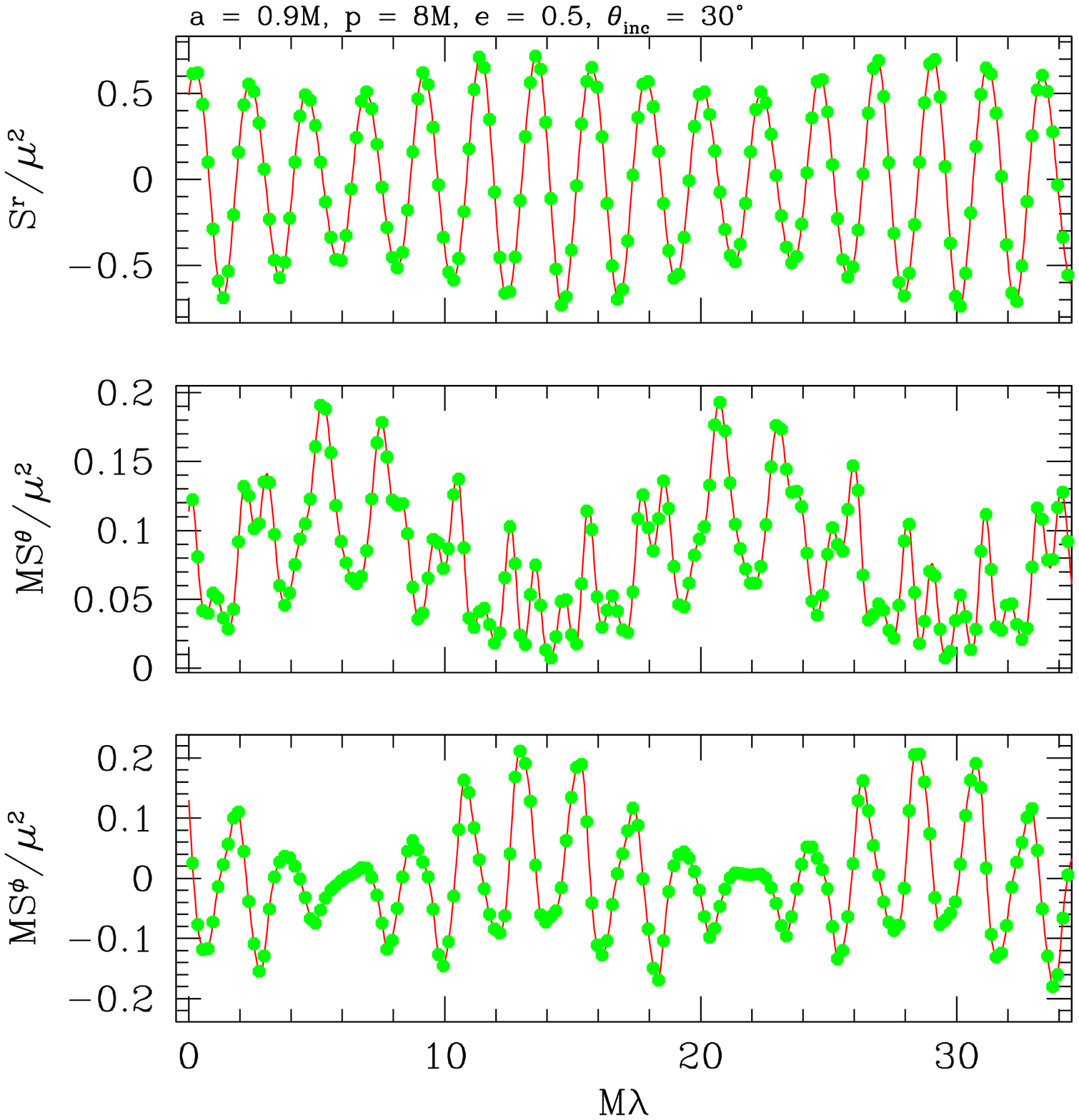}
\includegraphics[width = 0.48\textwidth]{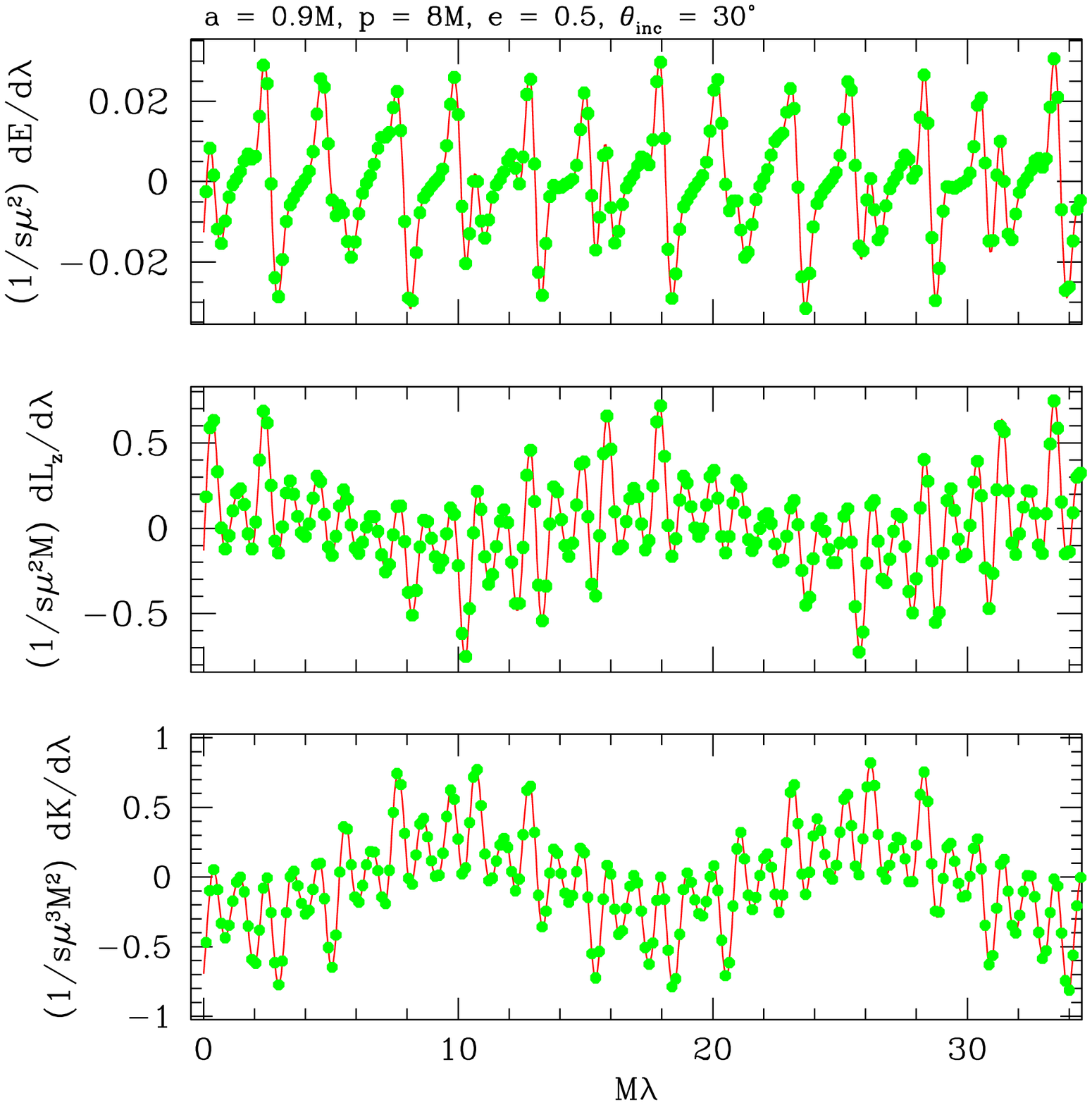}
\caption{Example of spin precession and spin-curvature force for a
  generic orbit ($a = 0.9M$, $p = 8M$, $e = 0.5$, $\theta_{\rm inc} =
  30^\circ$).  As in the previous two figures, the solid (red) curves
  show quantities reconstructed from the frequency-domain expansion.
  Dots (green) show the same data computed by direct time-domain
  integration.  Left-hand panels are the spin-vector components,
  right-hand panels are components of the spin-curvature force.
  Features in these data are present at beats between the radial
  ($\Upsilon_r = 2.4304M$), polar ($\Upsilon_\theta = 3.2367M$), and
  precession ($\Upsilon_s = 2.8429M$) frequencies.  The range in
  $\lambda$ we show allows features at the radial-precession [$2\pi
    M/(\Upsilon_s - \Upsilon_r) \simeq 15.23$] and polar-precession
  [$2\pi M/(\Upsilon_\theta - \Upsilon_s) \simeq 15.96$] beat periods
  to be seen.  As in the previous examples, we find outstanding
  agreement between the solutions provided that the frequency-domain
  expansion is truncated at sufficiently large values of $k$ and $n$.
  We used $k_{\rm max} = n_{\rm max} = 5$ for the results shown
  here.}
\label{fig:a0.9_p8_e0.5_thi30_results}
\end{figure*}

Finally, Fig.\ {\ref{fig:a0.9_p8_e0.5_thi30_results}} shows the spin
and force components for a generic orbit.  We include terms to $k_{\rm
  max} = n_{\rm max} = 10$ here.  The time- and frequency-domain
solutions differ by about $10^{-6}$ for $M\lambda \lesssim 50$, but
the difference drifts to $\mbox{several} \times 10^{-5}$ for $M\lambda
= 500$.  This can be improved by including more terms in the Fourier
sums.

Because two orbital frequencies are available to beat against the
precession frequency, the harmonic structure we find is particularly
ornate for generic orbits.  The span we show is sufficient to show
beat structure at the radial-precession beat period [$\Lambda_{\rm
    beat-rp} = 2\pi/(\Upsilon_s - \Upsilon_r) \simeq 15.23M^{-1}$] and
at the polar-precession beat period [$\Lambda_{\rm beat-pp} =
  2\pi/(\Upsilon_\theta - \Upsilon_s) \simeq 15.96M^{-1}$].

\subsubsection{The radial component}

\begin{figure}[ht]
\includegraphics[width = 0.48\textwidth]{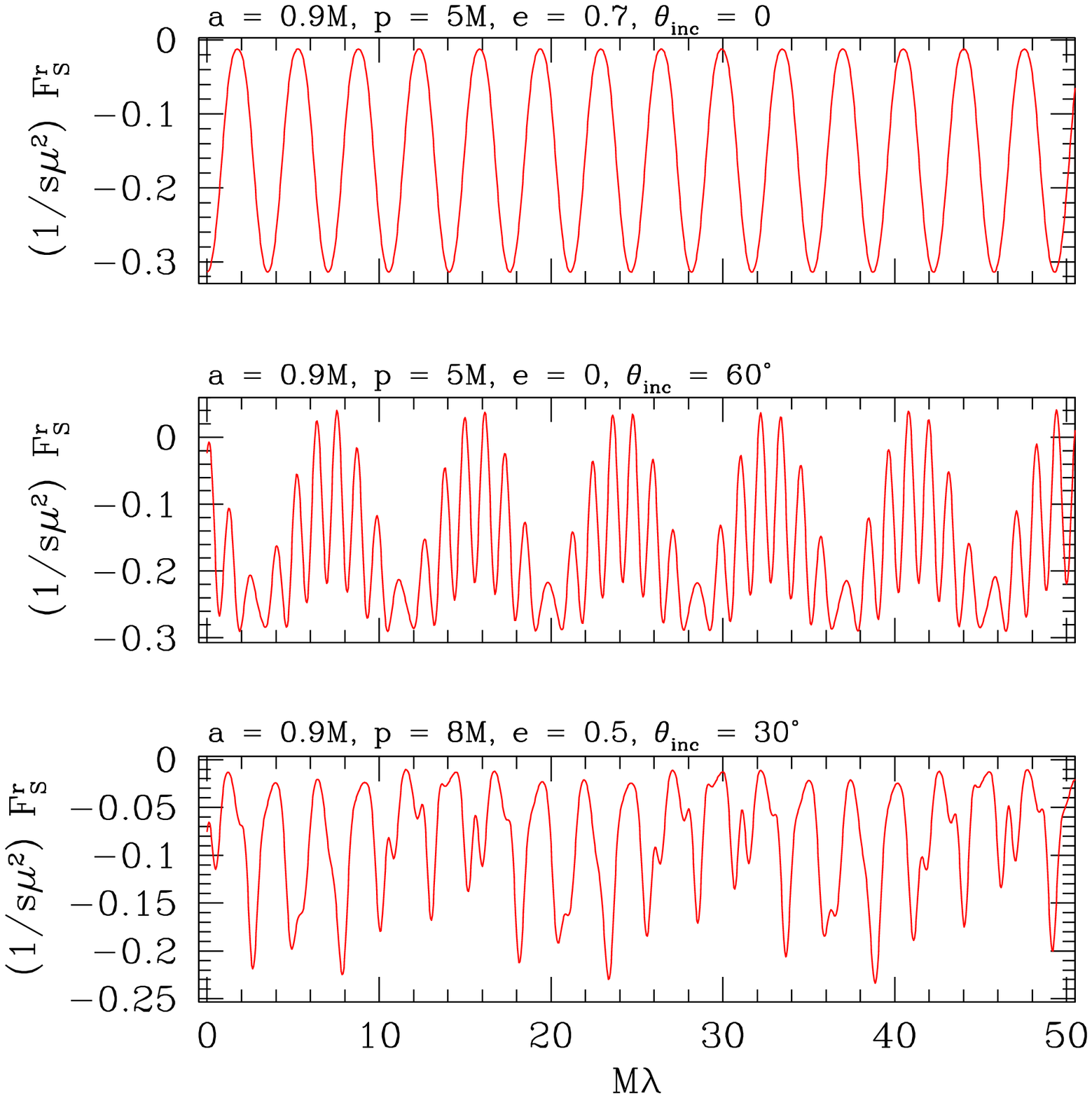}
\caption{The force component $F^r_{\rm S}$ for the configuration used
  to produce Fig.\ {\ref{fig:a0.9_p5_e0.7_thi0_results}} (top panel),
  Fig.\ {\ref{fig:a0.9_p5_e0_thi60_results}} (middle panel), and
  Fig.\ {\ref{fig:a0.9_p8_e0.5_thi30_results}} (bottom panel).  In all
  three cases, this force component has an average value $\langle
  F^r_{\rm S}\rangle < 0$, corresponding to an attractive force
  between the orbiting body and the black hole, consistent with a
  weak-field analysis of orbits with aligned spins.  This component of
  the spin-curvature force will play a role in the kinematics of these
  orbits akin to that played by the conservative self force.}
\label{fig:radialforce}
\end{figure}

As discussed above, we examine the spin-curvature force component
$F^r_{\rm S}$ separately since it decouples from $dE^{\rm
  G}/d\lambda$, $dL_z^{\rm G}/d\lambda$, and $dK^{\rm G}/d\lambda$.
Figure {\ref{fig:radialforce}} shows this component for the three
configurations that were used to produce
Figs.\ {\ref{fig:a0.9_p5_e0.7_thi0_results}},
{\ref{fig:a0.9_p5_e0_thi60_results}}, and
{\ref{fig:a0.9_p8_e0.5_thi30_results}}.  The main new feature that we
see here is that this force component has a non-zero average value:
for these orbits, $\langle F^r_{\rm S}\rangle < 0$, indicating that
there is an attractive force between the body and the black hole that
it orbits.  In retrospect, this is not surprising: on average, the
spin vector in these three configurations has a component that is
parallel to the black hole's spin, and it is well known that the
spin-curvature interaction enhances the gravitational attaction of two
bodies in a way that depends on the relative alignment of their spin
and orbital angular momenta (see, e.g., discussion in Sec.\ VA6 of
Ref.\ {\cite{membrane}}).  To clearly illustrate this, let us examine
Eq.\ (\ref{eq:spinforcelambda}) in the limit of a circular, equatorial
orbit, for an orbiting body whose spin is oriented normal to the
equatorial plane.  For this case, $U^r_{\rm G} = U^\theta_{\rm G} =
0$, $\theta = \pi/2$, and the only non-zero spin vector component is
$S^\theta$.  We find
\begin{equation}
F^r_{\rm S} = \mp\frac{3MS^{\hat\theta}}{r^3}\left(\sqrt{\frac{M}{r}} \mp
\frac{a}{r}\right)\frac{\Delta}{1 \pm 2a\sqrt{M/r^3} - 3M/r}\;,
\label{eq:spinforce_circeq}
\end{equation}
where the upper sign is for prograde orbits and the lower is for
retrograde.  We have written this using the orthonormal form
$S^{\hat\theta} = \sqrt{g_{\theta\theta}} S^\theta$, which is
particularly simple and convenient for a circular, equatorial orbit.
Recall also that our $F^\alpha$ is expressed using Mino time $\lambda$
rather than proper time $\tau$, and hence this differs by a factor
$\Sigma$ from other analyses of this quantity.  Converting to
$d/d\tau$ and considering $r \gg M$, Eq.\ (\ref{eq:spinforce_circeq})
agrees with the ``spin-orbit'' and ``spin-spin'' forces given in
Eqs.\ (5.69b) and (5.69c) of Ref.\ {\cite{membrane}}, which in turn
follows the discussion and derivation given in
Ref.\ {\cite{thornehartle}}.

Because $F^r_{\rm S}$ has an average non-zero value, it will play a
role similar to the conservative self force, causing a shift to
quantities like orbital frequencies, and leading to a secularly
growing term in the orbital phase.  Especially for orbits at small
radius, this could have important implications for gravitational-wave
source modeling.

\subsection{Convergence}
\label{sec:converge}

The Fourier expansions we have introduced,
Eq.\ (\ref{eq:spinprecconstrained}) for the evolution of the spin
vectors and Eq.\ (\ref{eq:forcefourier}) for the spin-curvature force,
are formally correct only when an infinite number of terms are kept in
the expansion.  In this section, we examine how well these quantities
converge when the sums are truncated at a finite number of terms.

\begin{figure}[ht]
\includegraphics[width = 0.48\textwidth]{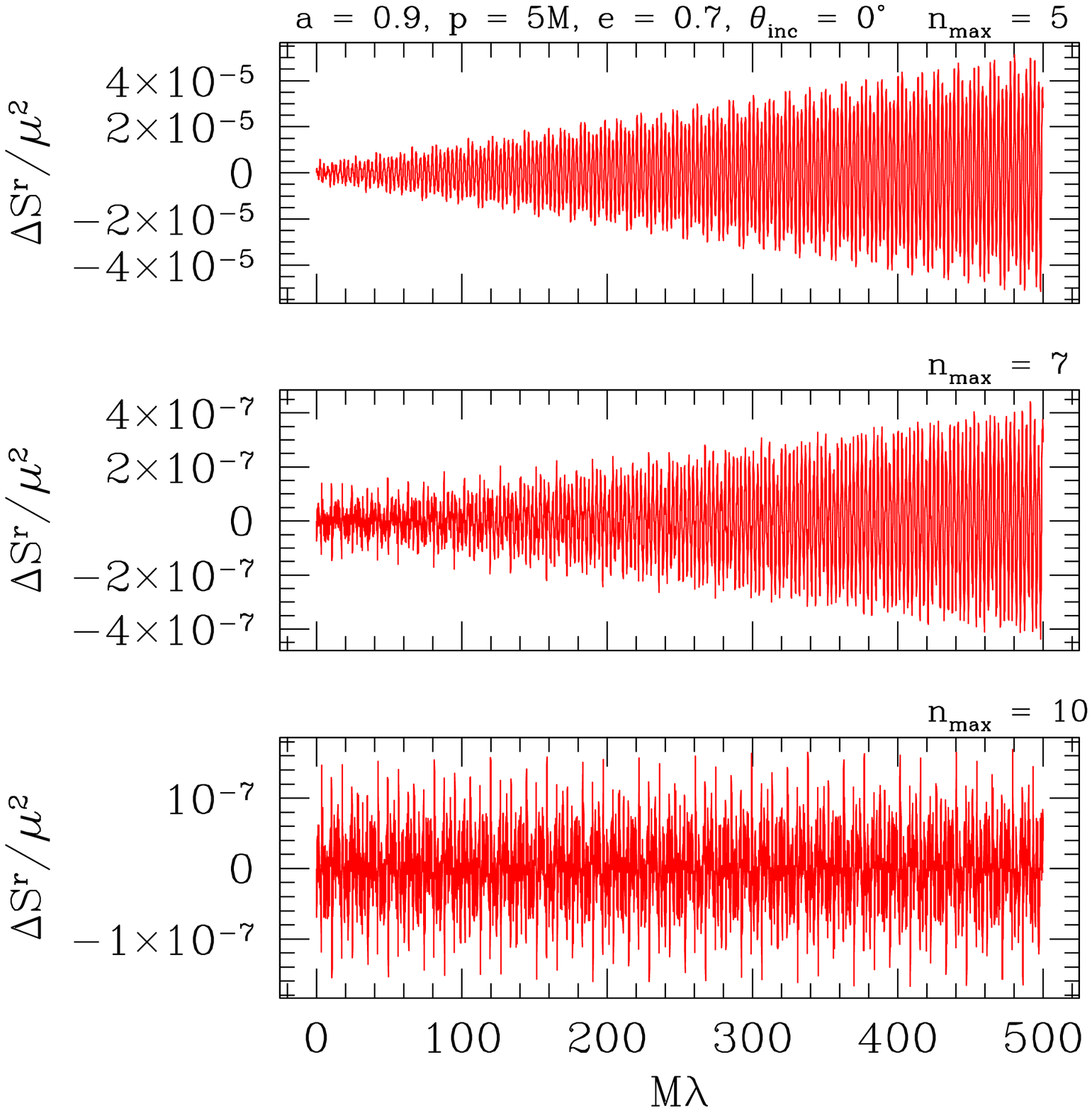}
\caption{Comparison of quantities computed in the frequency domain
  with quantities computed in the time domain.  We show the difference
  between $S^r(\lambda)$ computed in the time domain using
  Eq.\ (\ref{eq:dSdlambda}) and $S^r(\lambda)$ computed in the
  frequency domain using Eq.\ (\ref{eq:spinprecconstrained}).  We
  truncate the Fourier expansion at three different values of $n_{\rm
    max}$ in order to explore how well this expansion converges.  The
  data shown here is for the same orbit we used in
  Fig.\ {\ref{fig:a0.9_p5_e0.7_thi0_results}}, i.e.\ an equatorial
  eccentric orbit with $a = 0.9M$, $p = 5M$, $e = 0.7$.  A major
  source of error is due to inaccuracies in our determination of the
  precession frequency $\Upsilon_s$ This frequency is an eigenvalue of
  Eq.\ (\ref{eq:masterprec}); we cannot solve the eigensystem
  accurately with insufficient Fourier modes.  This accounts for the
  drift with time that we see in the top two panels: when $n_{\rm max}
  = 5$, the two solutions drift by $\mbox{several}\times 10^{-5}$ over
  an integration time $M\lambda = 500$.  When $n_{\rm max} = 7$, there
  is still a drift, but it is reduced by two orders of magnitude.
  When $n_{\rm max} = 10$, the drift has been eliminated, and we find
  a difference $|\Delta S^r/\mu^2| \lesssim 10^{-7}$ at all times.
  The convergence of other spin components and the force components
  behaves in this way for all orbits we have examined, although the
  detailed values of $n_{\rm max}$ and $k_{\rm max}$ needed to
  converge depends on orbit parameters.}
\label{fig:converge}
\end{figure}

Figure {\ref{fig:converge}} shows a typical example of how
frequency-domain expansions converge as more terms are kept in the
Fourier expansion.  We show the convergence of the spin vector
component $S^r$ as a function of Mino-time $\lambda$, for an
equatorial eccentric orbit with parameters $a = 0.9M$, $p = 5M$, $e =
0.7$ (the orbit that was used to generate data for
Fig.\ {\ref{fig:a0.9_p5_e0.7_thi0_results}}).  We examine $\Delta S^r$
as a function of $\lambda$, where
\begin{equation}
\Delta S^r \equiv S^r_{\rm TD} - S^r_{\rm FD,\, n_{\rm max}}\;,
\end{equation}
with $S^r_{\rm TD}$ the result of direct time-domain integration of
Eq.\ (\ref{eq:dSdlambda}) along the geodesic, and $S^r_{\rm FD,\,
  n_{\rm max}}$ from the frequency-domain expansion
(\ref{eq:spinprecconstrained}), truncating at $n = n_{\rm max}$.

To assess convergence, we examine this difference for $n_{\rm max} =
5$, $7$, and $10$.  As expected, the difference decreases as $n_{\rm
  max}$ is increased, but the manner in which it decreases is quite
interesting.  We generally see that, for small $n_{\rm max}$, $\Delta
S^r$ monotonically increases with $\lambda$, meaning that the two
solutions drift away from each other.  This drift is due to errors in
our determination of the precession frequency, $\Upsilon_s$.  Recall
that $\Upsilon_s$ is the eigenvalue of the precession equation
(\ref{eq:masterprec}).  Determining this eigenvalue accurately
requires us to describe the underlying geodesic motion accurately; in
the frequency domain, this description becomes progressively more
accurate as more terms are kept in our Fourier expansion.  When we use
$n_{\rm max} = 5$, our solution drifts by $|\Delta S^r| \simeq
\mbox{several} \times 10^{-5}$ when we integrate to $M\lambda = 500$.
Increasing this to $n_{\rm max} = 7$, the drift is reduced by about
two orders of magnitude.  Increasing still further to $n_{\rm max} =
10$ removes the drift altogether, and the two solutions differ by
roughly $10^{-7}$ at all times.

The convergence behavior shown for $S^r(\lambda)$ is typical: we see
quite similar behavior for the other spin components and for all the
components of the spin-curvature force.  This is true for all the
orbits we have examined, although the values of $n_{\rm max}$ and
$k_{\rm max}$ needed to achieve convergence depends on an orbit's
detailed parameters.  We find it is not difficult to achieve a
convergent frequency-domain representation of all physically important
quantities relating to spin-curvature coupling for astrophysically
interesting orbits.

\section{Are spin-precession resonances interesting?}
\label{sec:res}

Precession introduces a new frequency into the kinematics of an
orbiting body.  This suggests that interesting effects may arise when
the spin frequency $\Upsilon_s$ is commensurate with the orbital
frequencies $\Upsilon_{\theta,r}$, i.e.\ for orbits such that $k$ and
$n$ can be found satisfying
\begin{equation}
j\Upsilon_s + k\Upsilon_\theta + n\Upsilon_r = 0\;,
\label{eq:sc_res}
\end{equation}
with $j \in (-1,0,1)$.  At such spin-orbit resonances, the Fourier
mode corresponding to the harmonics which satisfy (\ref{eq:sc_res})
will be constant with time, suggesting that the system's evolution
could differ substantially from its behavior away from a resonance.
Such behavior has been seen in studies of the dissipative self force
{\cite{fh12,fhr14}} and of gravitational-wave recoil {\cite{vdm2014}}
when the Kerr orbital frequencies $\Omega_r$ and $\Omega_\theta$ are
commensurate.

Although important for the self force problem and for radiation
recoil, one can quickly convince oneself that spin-curvature coupling
cannot produce dynamically important resonances, at least at linear
order in $S$.  Consider Eq.\ (\ref{eq:forcefourier}) for an orbit that
has frequency harmonics which satisfy Eq.\ (\ref{eq:sc_res}).  For
these modes, we would have $d{\cal C}/d\lambda = \mbox{constant}$:
these quantities would grow linearly in $\lambda$ without bound.  Such
growth is inconsistent with the existence of the conserved integrals
$E^{\rm S}$, $L^{\rm S}_z$, and $K^{\rm S}$, defined by
Eqs.\ (\ref{eq:ESdef}), (\ref{eq:LzSdef}), and (\ref{eq:KSdef}).  The
contributions to $E^{\rm S}$, $L^{\rm S}_z$, and $K^{\rm S}$
proportional to the spin tensor are oscillatory, so unbounded growth
of the ``geodesic'' terms would, before long, violate their constant
nature.  To protect the system from these resonances, it must be the
case that $\left(d\mathcal{C}/d\lambda\right)_{jkn} = 0$ for modes and
frequencies which satisfy Eq.\ (\ref{eq:sc_res}).

\begin{figure}[ht]
\includegraphics[width = 0.48\textwidth]{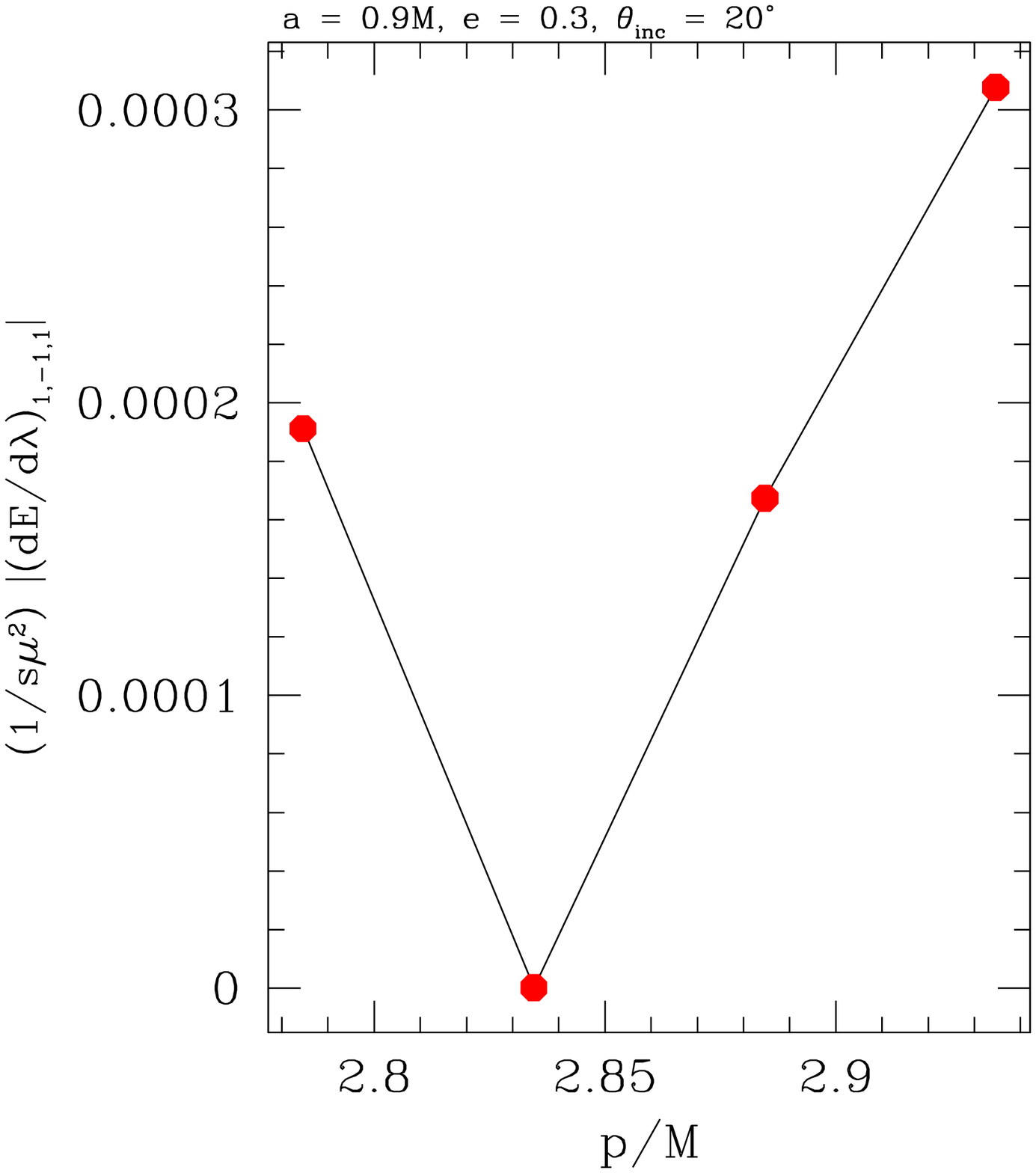}
\caption{Example of the vanishing of spin-curvature coupling on
  resonance.  We examine a set of orbits about a black hole with $a =
  0.9M$; each orbit has eccentricity $e = 0.3$ and inclination
  $\theta_{\rm inc} = 20^\circ$.  We vary the semi-latus rectum: $p/M
  \in (2.78461, 2.83461, 2.88461, 2.98461)$.  The case $p/M = 2.83461$
  is resonant: $\Upsilon_s - \Upsilon_\theta + \Upsilon_r = 0$ for
  that orbit.  We find that the Fourier mode $(dE^{\rm G}/d\lambda)_{j
    = 1, k = -1, n = 1}$ vanishes in the resonant case; the rates of
  change of $L_z^{\rm G}$ and $K^{\rm G}$ likewise vanish on resonance
  for these mode indices.  This behavior protects the conserved
  quantities $E^{\rm S}$, $L_z^{\rm S}$, and $K^{\rm S}$, and
  guarantees that spin-orbit resonances aren't very interesting at
  least to linear order in the equations of motion and precession.
}
\label{fig:resonance}
\end{figure}

Figure {\ref{fig:resonance}} shows an example of
$\left(d\mathcal{C}/d\lambda\right)_{jkn}$ vanishing on resonance.  We
examine a sequence of orbits, each with $a = 0.9M$, $e = 0.2$, and
$\theta_{\rm inc} = 20^\circ$, and varying the semi-latus rectum $p$.
The sequence we examine includes one value of $p$ such that
$\Upsilon_s - \Upsilon_\theta + \Upsilon_r = 0$.  We indeed find that
$\left(d\mathcal{C}/d\lambda\right)_{j = 1, k = -1, n = 1}$ is zero on
resonance, though it is non-zero on nearby orbits.  (We only show
${\cal C} = E^{\rm G}$ in this figure, although our data shows that
this behavior also holds for $L_z^{\rm G}$ and $K^{\rm G}$.)

Although there are no interesting effects due to these resonances at
linear order, there might be interesting effects at higher order.  We
speculate that the behavior near resonance of nonlinear terms (which
we have neglected) may play a role in pushing the dynamical evolution
of spinning bodies from integrable to chaotic motion.  The KAM theorem
{\cite{arnold, moser}} teaches us that an integrable system will
remain integrable under the influence of a weak, non-integrable
perturbing force as long as that force is not resonant with the
integrable motion.  This suggests that circumstances may change when
the perturbing force is in fact in resonance.  If this speculation is
correct, an analysis of spin-curvature forces near these resonances
may help us to understand the onset of chaos that has been seen in
previous analyses of spin-enhanced orbital motion
{\cite{suzukimaeda96, hartl03, levin06}}.

\section{Summary and future work}
\label{sec:conclusion}

We have shown how to compute the coupling between a small body's spin
and the curvature of the black hole spacetime through which it moves
in the frequency domain.  The spin vector precesses along its orbit,
which introduces a new frequency that must be accounted for in a
frequency-domain description of quantities associated with this
motion.  Using the fact that Kerr geodesic motion is itself
characterized by three frequencies, we have shown that all the
quantities relevant to spin-curvature coupling can be computed very
accurately.

Certain aspects of the large mass-ratio limit have proven to be useful
and surprisingly accurate tools for helping to understand the two-body
problem more generally.  In particular, results that have come from
the self force program have been shown to agree very well with results
computed using numerical relativity and have been used to refine
effective one-body binary models
{\cite{letiec2011,abds2012,letiec2013}}.  We speculate that the
results we present here may similarly find use by providing a limit
that can be modeled precisely for understanding spin precessions and
the influence of spin-curvature coupling in the evolution of binary
systems.  A useful starting point may be to check whether the
solutions we find agree with the elegant results found by Gerosa et
al.\ {\cite{gksbo2015}}.  If so, our methods may prove to be a useful
starting point for examining precession and spin-curvature coupling
for more generic binaries than have been considered so far.

We plan to extend this work by using an osculating geodesic integrator
{\cite{poundpoisson08, gfdhb11}} to develop spin-enhanced orbits.  An
osculating geodesic integrator models a non-geodesic worldline as a
sequence of geodesic orbits.  The spin-curvature force
(\ref{eq:spinforcetau}) then acts to move the small body from geodesic
to geodesic in this sequence.  As the orbit evolves, the geodesic
energy $E^{\rm G}$ will oscillate, but will do so in such a way that
the corresponding spin-enhanced energy $E^{\rm S}$ remains fixed;
similar statements hold for $L_z^{\rm G,S}$ and $K^{\rm G,S}$.  The
orbits which we find in this procedure will be very useful tools for
allowing us to understand the importance of the spin on observable
aspects of small body orbits very generally, allowing us to go beyond
the special spin-orbit configurations that have been analyzed in
detail in earlier work.

Of particular interest will be to compare the spin-curvature force to
other non-geodesic effects that have an impact on a binary's
evolution, such as the self force.  We can get a very rough idea of
how such forces compare by simply examining typical spin-curvature
force components that we have computed (e.g., our
Figs.\ {\ref{fig:a0.9_p5_e0.7_thi0_results}} --
{\ref{fig:radialforce}}), along with typical self force components
found by others (e.g., those shown in Figs.\ 6, 7, and 8 of
Ref.\ {\cite{baracksago2010}}).  It should be emphasized that such a
comparison is extremely crude.  Even after correcting for the factor
of $d\tau/d\lambda = \Sigma$ between the two force definitions, and
noting that we examine the forces on rather different orbits, we must
be concerned about gauge.  Best of all would be to develop a
gauge-invariant measure of the influence these forces have on binary
orbits, as was done for simpler spin-orbit configurations in
Refs.\ {\cite{burko2004,burkokhanna2015,sp2012}}.  At present, the
strongest defensible statement we can make is that the spin-curvature
components appear large enough when crudely compared to self force
components that it is very plausible that the small body's spin will
leave an observationally important imprint on a binary's evolution,
especially when we consider motion through the strong field of the
binary's larger black hole.

Once spin-enhanced orbits are fully in hand, we can consider radiation
from such configurations.  As a first pass, it may not be too
difficult to couple these orbits to some kind of ``kludge'' model for
the evolution of orbit constants and wave emission.  For example, a
useful first approximation might be to develop a spin-enhanced
worldline by combining the spin-curvature force with a self force,
allowing us to compute an inspiral with both spin-precession and
spin-coupling effects.  We could then use that worldline as the source
of a time-domain black hole perturbation theory solver (as has been
done in, e.g., {\cite{skhd08}}), or even using a cruder approach based
on some approximate set of radiative multipoles (as in
e.g.\ Refs.\ {\cite{gg2006, bfggh2007}}).  Such a tool would likely be
a useful first cut at building spin-enhanced waveforms to quantify the
role that the small body's spin has the system's waves.  The
frequency-domain description may allow us to go beyond this and
perhaps to extend black hole perturbation theory codes to include the
influence of spin, much as is done in Ref.\ {\cite{harmsetal2015}} but
for general orbits and general spin orientations.

\acknowledgments

This work was supported at MIT by NSF Grant PHY-1403261, and at the
Jet Propulsion Laboratory by an appointment to the NASA Postdoctoral
Program, administered by Oak Ridge Associated Universities through a
contract with NASA.  We are very grateful for feedback and helpful
comments on an earlier draft of this paper from Niels Warburton and
Georgios Loukes-Gerakopoulos, to feedback from L.\ Filipe O.\ Costa
and Jos\'e Nat\'ario on our original arXiv.org posting, and to the
paper's anonymous referee from an extremely thorough and helpful
report.  Many of our calculations were done using the package {\sc
  Mathematica}.  A very early version of this work was supported at
MIT by NASA grant NNX08AL42G, and was published as a chapter in
S.~J.~V.'s Ph.~D.\ thesis {\cite{svthesis}}.

\appendix

\section{$K^{\rm G}$ is constant for an equatorial orbit}
\label{app:eqCarter}

While developing the results we show in Sec.\ {\ref{sec:results}}, we
discovered empirically that for equatorial orbits ($\theta = \pi/2$,
$u_{\rm G}^\theta = 0$), the Carter constant $K^{\rm G}$ remained
constant as the small body moved along its osculating geodesic.  Here
we examine $dK^{\rm G}/d\tau$ for an equatorial orbit analytically.
Begin with the general form for the rate of change of $K^{\rm G}$
under the influence of a force $f^\nu$:
\begin{equation}
\frac{dK^{\rm G}}{d\tau} = 2K_{\mu\nu}p_{\rm G}^\mu f^\nu\;.
\end{equation}
For the force, we use Eq.\ (\ref{eq:spinforcetau}):
\begin{equation}
f^\mu = f^\mu_{\rm S} = -\frac{1}{2} {R^{\mu}}_{\alpha\beta\gamma}
u_{\rm G}^\alpha S^{\beta\gamma}\;.
\end{equation}
Using Eq.\ (\ref{eq:spinvecinvert}) as well as $\theta = \pi/2$ and
$u_{\rm G}^\theta = 0$, we expand this expression and find
\begin{eqnarray}
\frac{dK^{\rm G}}{d\tau} &=& \frac{2M}{r}\frac{S^\theta u_{\rm
    G}^r}{\Delta} \left[(r^2 + a^2) u_{\rm G}^\phi - au_{\rm
    G}^t\right]
\nonumber\\
&\times& \left[(r^2 + a^2)\hat E^{\rm G} - a\hat L^{\rm G}_z +
  \Delta(a u_{\rm G}^\phi - u_{\rm G}^t)\right]\;.
\end{eqnarray}
Plugging in the equatorial values of $u_{\rm G}^t$ and $u_{\rm
  G}^\phi$ [Eqs.\ (\ref{eq:phidot}) and (\ref{eq:tdot}) with $\theta =
  \pi/2$] yields $dK^{\rm G}/d\tau = 0$.

\bibliographystyle{apsrev4-1}
\bibliography{phyjabb,master}

\end{document}